\documentclass[useAMS,usenatbib,a4paper]{mn2e}
\usepackage{amssymb}
\usepackage{amsmath}
\usepackage{graphicx}
\usepackage{subfigure}
\usepackage{txfonts}
\usepackage{mathtools}
\usepackage{ifpdf}
\usepackage[pdftex,colorlinks=true,breaklinks=true,citecolor=blue]{hyperref}
\usepackage{color}
\usepackage{tikz}
\usetikzlibrary{calc,fadings,decorations.pathreplacing}
\usepackage{verbatim}
\usepackage{array}

\newif\ifbw
\bwfalse

\newcommand\LongitudePlane[3][current plane]{%
  \tikzset{#1/.style={cm={cos(#3),sin(#3)*sin(#2),0,cos(#2),(0,0)}}}
}
\newcommand\LatitudePlane[3][current plane]{%
  \pgfmathsetmacro\yshift{cos(#2)*sin(#3)}
  \tikzset{#1/.style={cm={cos(#3),0,0,cos(#3)*sin(#2),(0,\yshift)}}} %
}
\newcommand\DrawLongitudeCircle[2][1]{
  \LongitudePlane{\angEl}{#2}
  \tikzset{current plane/.prefix style={scale=#1}}
  \pgfmathsetmacro\angVis{atan(sin(#2)*cos(\angEl)/sin(\angEl))} %
  \draw[current plane] (\angVis:1) arc (\angVis:\angVis+180:1);
  \draw[current plane,dashed] (\angVis-180:1) arc (\angVis-180:\angVis:1);
}
\newcommand\DrawLatitudeCircle[2][1]{
  \LatitudePlane{\angEl}{#2}
  \tikzset{current plane/.prefix style={scale=#1}}
  \pgfmathsetmacro\sinVis{sin(#2)/cos(#2)*sin(\angEl)/cos(\angEl)}
  \pgfmathsetmacro\angVis{asin(min(1,max(\sinVis,-1)))}
  \draw[current plane] (\angVis:1) arc (\angVis:-\angVis-180:1);
  \draw[current plane,dashed] (180-\angVis:1) arc (180-\angVis:\angVis:1);
}

\newcommand{\bea}{\begin{equation}}
\newcommand{\eea}{\end{equation}}

\newcommand{\eqn}[1]{(#1)}
\newcommand{\Eqn}[1]{(#1)}

\newcommand{\fig}[1]{Fig.~#1}
\newcommand{\Fig}[1]{Fig.~#1}
\newcommand{\sectn}[1]{Sec.~#1}

\newcommand{\appn}[1]{Appendix~#1}

\newcommand{\etal}{\mbox{\it et al.}}
\newcommand{\eg}{\mbox{\it e.g.}}
\newcommand{\ie}{\mbox{\it i.e.}}

\newcommand{\cf}{\mbox{\it cf.}}

\newcommand{\degrees}{\ensuremath{{^\circ}}}

\newcommand{\healpix}{{\tt HEALPix}}

\newcommand{\sshtcode}{{\tt SSHT}}

\newcommand{\spcend}{\ensuremath{\:}}
\newcommand{\img}{\ensuremath{{\rm i}}}
\newcommand{\cconj}{\ensuremath{\ast}}

\newcommand{\integers}{\ensuremath{\mathbb{Z}}}

\newcommand{\sphere}{\ensuremath{{\mathbb{S}^2}}}
\newcommand{\sothree}{\ensuremath{{\mathrm{SO}(3)}}}

\newcommand{\vect}[1]{\ensuremath{\mbox{\boldmath ${#1}$}}}

\newcommand{\dx}{\ensuremath{\mathrm{\,d}}}
\newcommand{\dmu}[1]{\ensuremath{\dx \Omega(#1)}}

\newcommand{\innerp}[2]{\ensuremath{\langle {#1},\: {#2} \rangle}}

\newcommand{\sa}{\ensuremath{\omega}}
\newcommand{\saa}{\ensuremath{\theta}}
\newcommand{\sab}{\ensuremath{\varphi}}
\newcommand{\sas}{\ensuremath{\saa, \sab}}

\newcommand{\eul}{\ensuremath{\mathbf{\rho}}}
\newcommand{\euls}{\ensuremath{\eula, \eulb, \eulc}}
\newcommand{\eula}{\ensuremath{\alpha}}
\newcommand{\eulb}{\ensuremath{\beta}}
\newcommand{\eulc}{\ensuremath{\gamma}}

\newcommand{\el}{\ensuremath{\ell}}
\newcommand{\m}{\ensuremath{m}}
\newcommand{\n}{\ensuremath{n}}

\newcommand{\spin}{\ensuremath{s}}

\newcommand{\elmax}{\ensuremath{{L}}}

\newcommand{\p}{\ensuremath{^\prime}}

\newcommand{\kron}[2]{\ensuremath{\delta_{{#1}{#2}}}}

\renewcommand{\exp}[1]{\ensuremath{{\rm e}^{#1}}}
\newcommand{\shfarg}[3]{\ensuremath{Y_{#1#2}({#3})}}
\newcommand{\shfargc}[3]{\ensuremath{Y_{#1#2}^\cconj({#3})}}

\newcommand{\sshfarg}[4]{\ensuremath{{{}_{#4} Y_{#1#2}({#3})}}}
\newcommand{\sshfargc}[4]{\ensuremath{{{}_{#4} Y_{#1#2}^\cconj({#3})}}}

\newcommand{\shf}[2]{\ensuremath{Y_{#1#2}}}

\newcommand{\shc}[3]{\ensuremath{{#1}_{{#2}{#3}}}}

\newcommand{\sshf}[3]{\ensuremath{{\prescript{}{#3} Y_{#1#2}}}}
\newcommand{\sshfc}[3]{\ensuremath{{\prescript{}{#3} Y_{#1#2}^\cconj}}}
\newcommand{\sshc}[4]{\ensuremath{\prescript{}{#4} {#1}_{{#2}{#3}}}}

\newcommand{\aleg}[3]{\ensuremath{P_{#1}^{#2}({#3})}}

\newcommand{\dmatbig}{\ensuremath{D}}

\newcommand{\rotarg}[1]{\ensuremath{\mathcal{R}_{#1}}}
\newcommand{\rotmat}{\ensuremath{\mathbf{R}}}

\newcommand{\spinup}{\ensuremath{\eth}}
\newcommand{\spindown}{\ensuremath{\bar{\eth}}}

\newcommand{\sumlm}{\ensuremath{\sum_{\el=0}^{\infty} \sum_{\m=-\el}^\el}}

\newcommand{\nside}{\ensuremath{{N_{\rm{side}}}}}

\newcommand{\wlconv}{\ensuremath{\prescript{}{0}{\kappa}}}
\newcommand{\wlshear}{\ensuremath{\prescript{}{2}{\gamma}}}
\newcommand{\wlshearone}{\ensuremath{\gamma_1}}
\newcommand{\wlsheartwo}{\ensuremath{\gamma_2}}
\newcommand{\wlpot}{\ensuremath{\prescript{}{0}{\phi}}}

\newcommand{\wlconvshc}{\ensuremath{\sshc{\hat{\kappa}}{\el}{\m}{0}}}
\newcommand{\wlshearshc}{\ensuremath{\sshc{\hat{\gamma}}{\el}{\m}{2}}}
\newcommand{\wlpotshc}{\ensuremath{\sshc{\hat{\phi}}{\el}{\m}{0}}}

\newcommand{\wlpotf}{\ensuremath{\prescript{}{0}{\hat{\phi}}}}

\newcommand{\wlconvfk}{\ensuremath{\prescript{}{0}{\hat{\kappa}(\wlkx,\wlky)}}}
\newcommand{\wlshearfk}{\ensuremath{\prescript{}{2}{\hat{\gamma}(\wlkx,\wlky)}}}
\newcommand{\wlshearonefk}{\ensuremath{{\hat{\gamma}_1(\wlkx,\wlky)}}}
\newcommand{\wlsheartwofk}{\ensuremath{{\hat{\gamma}_2(\wlkx,\wlky)}}}
\newcommand{\wlpotfk}{\ensuremath{\prescript{}{0}{\hat{\phi}(\wlkx,\wlky)}}}

\newcommand{\wlkx}{\ensuremath{k_x}}
\newcommand{\wlky}{\ensuremath{k_y}}

\newcommand{\wlfactor}{\ensuremath{\mathcal{D}}}
\newcommand{\wlfactorplane}{\ensuremath{\mathcal{E}}}
\newcommand{\wlkernel}{\ensuremath{\prescript{}{2}{\mathcal{K}}}}

\renewcommand{\elmax}{\ensuremath{{\el_{\rm max}}}}
\renewcommand{\eqn}[1]{Equation~(#1)}
\renewcommand{\Eqn}[1]{Equation~(#1)}
\renewcommand{\sectn}[1]{Section~#1}

\renewcommand{\sa}{\ensuremath{{\omega}}}
\newcolumntype{C}[1]{>{\centering\let\newline\\\arraybackslash\hspace{0pt}}m{#1}}

\renewcommand{\fs}{\ensuremath{{}_\spin f}}

\newcommand{\bmtrx}[1]{\ensuremath{\boldsymbol{\mathsf{#1}}}}
\renewcommand{\rotmat}{\ensuremath{\bmtrx{R}}}

\title[Mapping dark matter on the celestial sphere]{Mapping dark matter on the celestial sphere with weak gravitational lensing}

\author[Wallis \etal]
  {Christopher G.~R.~Wallis$^{1}$, Matthew A.~Price$^1$\thanks{m.price.17@ucl.ac.uk}, Jason D.~McEwen$^1$, Thomas D.~Kitching$^1$, \newauthor Boris Leistedt$^{2,3}$, Antoine Plouviez$^1$ \\
    $^1$Mullard Space Science Laboratory, University College London, Surrey, RH5 6NT, UK\\
    $^2$Center for Cosmology and Particle Physics, Department of Physics, New York University, New York, NY 10003, USA\\
    $^3$NASA Einstein Fellow
}

\date{Accepted ---. Received ---; in original form ---}
\pagerange{\pageref{sec:intro}--\pageref{lastpage}}
\pubyear{2017}

\def\LaTeX{L\kern-.36em\raise.3ex\hbox{a}\kern-.15em
    T\kern-.1667em\lower.7ex\hbox{E}\kern-.125emX}

\tikzset{%
  >=latex, %
  inner sep=0pt,%
  outer sep=2pt,%
  mark coordinate/.style={inner sep=0pt,outer sep=0pt,minimum size=3pt,
    fill=black,circle}%
}

\begin{document}
\maketitle

\begin{abstract}
  Convergence maps of the integrated matter distribution are a key science result from weak gravitational lensing surveys.
  To date, recovering convergence maps has been performed using a planar approximation of the celestial sphere. However, with the increasing area of sky covered
  by dark energy experiments, such as Euclid, the Large Synoptic Survey Telescope (LSST), and the
  Wide Field Infrared Survey Telescope (WFIRST), this assumption will no longer be valid. We recover convergence fields on the celestial sphere using an
    extension of the Kaiser-Squires estimator to the spherical setting. Through simulations we study the error introduced by planar approximations.  Moreover, we examine how best to recover convergence maps in the planar setting, considering a variety of different projections and defining the local rotations that are required when projecting spin fields such as cosmic shear.
  For the sky coverages typical of future surveys, errors introduced by projection effects can be of order tens of percent, exceeding 50\% in some cases.  The stereographic projection, which is conformal and so preserves local angles, is the most effective planar projection.  In any case, these errors can be avoided entirely by recovering convergence fields directly on the celestial sphere.  We apply the spherical Kaiser-Squires
  mass-mapping method presented to the public Dark Energy Survey (DES) science verification data to recover convergence maps directly on the celestial sphere.
\end{abstract}


\section{Introduction}
\label{sec:intro}

Weak gravitational lensing distorts the shape and size of images of distant
galaxies due to the gravitational influence of matter perturbations along the
line of sight \citep[see, \eg,][]{bartelmann:2001,schneider:2005, munshi:2008, heavens:2009}. The amplitude of the distortion -- a change in the ellipticity
(third flattening or third eccentricity) and apparent size of an object --
contains information on the integrated Newtonian potential and can be used to
estimate the integrated mass distribution.  The lensing effect is dependent
on the total mass distribution and therefore, because massive structures are
dominated by dark matter, the mass distributions recovered by weak lensing are colloquially
referred to as mass-maps of the dark matter of the Universe.  The creation of such maps constitutes
one of the main empirical observations that underpins the dark matter paradigm
\citep{clowe:2006}.

The most common approach to extract cosmological information from weak lensing surveys is to compute the two-point correlation
function (\eg\ \citealt{kilbinger:2015}) or power spectrum (\eg\
\citealt{alsing:2016}) from observational data and compare to an expectation from
theory. However, such analyses do not use sufficient statistics and are
sensitive only to the Gaussian component of the underlying field.
To capture the entire information content of the shear field higher order statistics (\eg\
\citealt{munshi:2011}) or phase information (\eg\ \citealt{coles:2000}) must
be considered.  Recovering mass-maps provides the basis for performing a wide
variety of complimentary higher order statistical analyses that probe the non-Gaussian
structure of the dark matter distribution.  For example, properties of dark
matter can then be studied using analyses based on peak and void statistics
(\eg\ \citealt{lin:2015a}; \citealt{lin:2015b}; \citealt{lin:2016};
\citealt{peel:2016}), Minkowski functions (\eg\ \citealt{munshi:2012}; \linebreak \citealt{kratochvil:2012}; \citealt{petri:2013}), or wavelets (\cf\ \citealt{hobson:1999}; \citealt{aghanim:2003}; \citealt{vielva:2004}, \citealt{mcewen:2005:ng}), to name just a few.

Further to this, mass-mapping provides an efficient way to cross-correlate weak
lensing  data with other cosmological data (\eg\ with observations of the
cosmic microwave background;  \citealt{liu:2015}). More directly, dark matter maps are of
interest for galaxy evolution studies: it is known from simulations that the
dark matter structure should exhibit a filamentary or ``cosmic web'' structure 
inference of which can then provide dark matter environmental
information that can then be used in galaxy evolution studies
\citep{brouwer:2016}. Finally, mass-mapping is a continuation of cartography
onto the cosmic scale -- the making of such maps is therefore laudable in its
own right.

Recovering mass-maps requires solving an inverse problem to recover the
underlying mass distribution from the observable cosmic shear.  There are a
number of approaches to estimating mass-maps from weak lensing data. The
method mostly commonly used on large scales is colloquially known as ``Kaiser-Squires''
and is named after the paper in which the method was first described
\citep{kaiser:1993}. This approach is based on a direct Fourier inversion of
the equations relating the observed shear field to the convergence field,
which is a scaled version of the integrated mass distribution.  Although it is
widely known that such an approach, based on a direct Fourier inversion, is not
robust to noise, the method remains in widespread use today (in practice, the
resultant mass-map is smoothed to mitigate noise).  Indeed, the Kaiser-Squires
method has been used to recover mass-maps from data from by a number of recent
weak lensing surveys, including data from the Cosmic Evolution Survey (COSMOS;
\citealt{scoville:2007}), the Canada-France-Hawaii Telescope
Lensing Survey (CFHTLenS; \citealt{heymans:2012})
and the Dark Energy Survey (DES; \citealt{flaugher:2015}) Science Verification (SV) data \citep[respectively,][]{massey:2007,van_waerbeke:2013,chang:2015}.
Alternative mass-mapping techniques to recover the convergence field have also
been developed, however these are not typically in widespread use and in many
cased are focused on the galaxy cluster scale.  On the galaxy cluster scale
parametric models \citep[\eg][]{jullo:2007}) and non-parametric methods \citep[\eg][]{massey:2015,lanusse:2016, price:2018} have been considered. \citet{szepietowski:2014} have investigated the use of phase information from galaxy number counts
to improve the reconstruction.

While the methods discussed above focus on recovering the two-dimensional
convergence field, which represents the integrated mass distribution along the
line of sight, it is also possible to recover the full three-dimensional
gravitational potential.  Such an approach involves an additional inverse
problem and thus an additional level of complexity. This has been considered
by a number of works (\citealt{bacon:2003, taylor:2004, massey:2004, simon:2009, vanderplas:2011, leonard:2012, simon:2013, leonard:2014})

\begin{figure}
  \subfigure[DES SV]{\includegraphics[width=.24\textwidth]{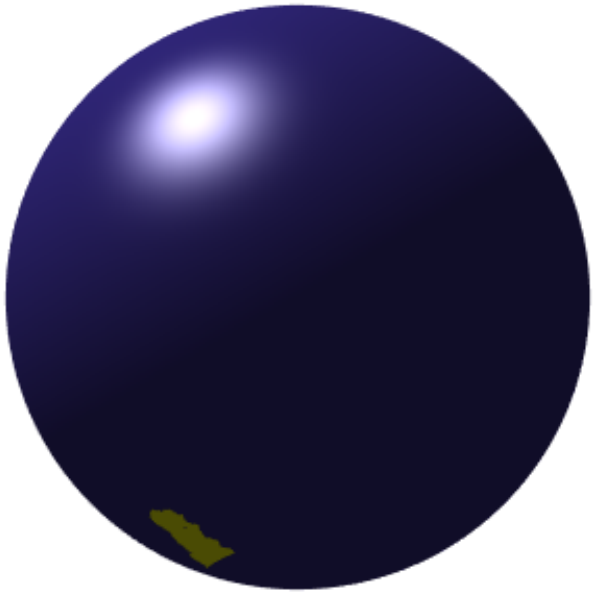}}
  \subfigure[DES full]{\includegraphics[width=.24\textwidth]{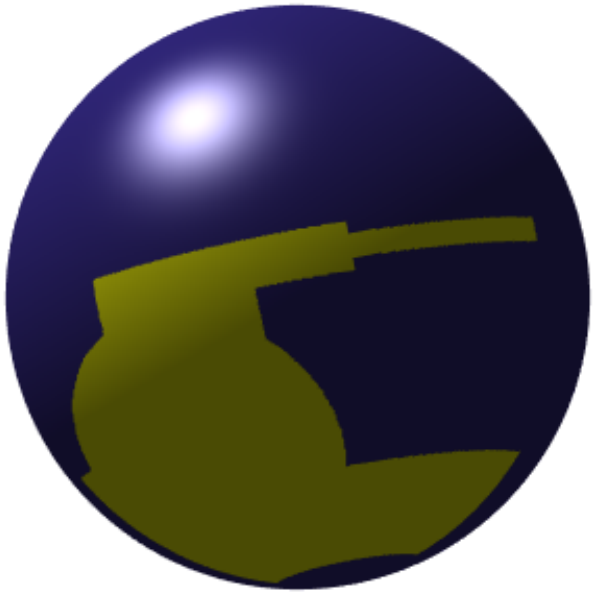}}
  \subfigure[Euclid (front and rear views)]{\includegraphics[width=.24\textwidth]{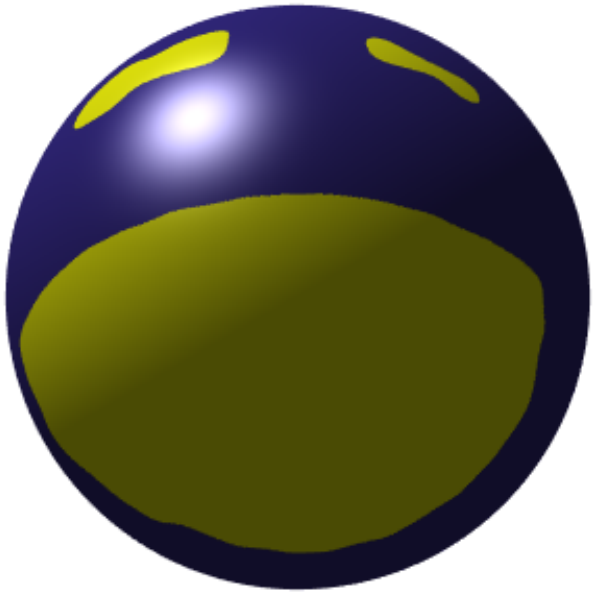}\includegraphics[width=.24\textwidth]{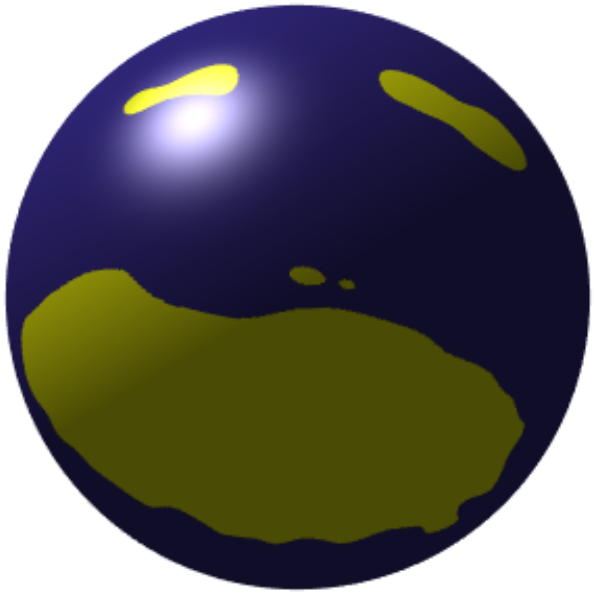}}
  \caption{Approximate coverage area of different weak lensing surveys illustrated on the celestial sphere.
    In particular, the coverage area corresponding to DES SV observations, DES full observations
    and Euclid observations are shown. It is apparent that existing planar mass-mapping techniques will not be appropriate for the large coverage areas of forthcoming surveys.  We extend the Kaiser-Squires technique for mass-mapping to the spherical setting in this article, in order to recover mass-maps on the celestial sphere.}
  \label{fig:sky_coverage}
\end{figure}

In general mass-mapping techniques for weak lensing consider a small field-of-view
of the celestial sphere, which is approximated by a tangent plane.  The
mass-mapping formalism is then developed in a planar setting, where a planar
two-dimensional Fourier transform is adopted.  Such an assumption will not be
appropriate for forthcoming surveys, which will observe significant fractions
of the celestial sphere, such as the Kilo Degree Survey (KiDS\footnote{\url{http://kids.strw.leidenuniv.nl}}; \citealt{de_jong:2013}), the Dark Energy Survey (DES\footnote{\url{http://www.darkenergysurvey.org}};
\citealt{flaugher:2015}), Euclid\footnote{\url{http://euclid-ec.org}}
\citep{laureijs:2011},  the Large Synoptic Survey Telescope (LSST\footnote{\url{https://www.lsst.org}}; \citealt{lsst:2009}), and the
Wide Field Infrared Survey Telescope (WFIRST\footnote{\url{https://wfirst.gsfc.nasa.gov}}; \citealt{spergel:2015}).
\fig{\ref{fig:sky_coverage}} illustrates the approximate sky coverage for DES
SV data, DES full data, and Euclid observations, from which it is apparent
that planar approximations will become increasingly inaccurate as sky coverage
areas grow over time.  Existing mass-mapping techniques that are based on
planar approximations therefore cannot be directly applied to forthcoming
observations.

In this article we consider the Kaiser-Squires approach for recovering mass-maps
  defined on the full celestial sphere. While the harmonic space expressions in the spherical
  setting relating the observed shear field to the convergence field, via the lensing potential,
  have been presented already \citep[\eg][]{taylor:2001,castro:2005,pichon:2010}, to the best of
  our knowledge naive Fourier inversion on the celesital sphere (\ie\ spherical Kaiser-Squires)
  has not been considered previously. We compare the spherical Kaiser-Squires formalism
with the planar case, considering several different spherical projections.\footnote{An
    alternative to recovering mass-maps directly on the sphere is to tile the celestial sphere and
    perform mass-mapping on planar patches, as considered for the lensing of the cosmic
    microwave background by \citet{plaszczynski:2012}.  An extension of this work to galaxy
    lensing, when shear is observed, would be of great interest.}
Spherical mass-mapping techniques have also been
considered by \citet{pichon:2010}, where a \emph{maximum a posteriori} (MAP)
estimator was presented. In addition, the authors consider using a Wiener filter to denoise the
  shear in advance of attempting to recover convergence maps.  However, as far as we are aware these techniques have
not been applied to observational data. The spherical Kaiser-Squires technique that we present here is a
first step towards more sophisticated spherical mass-mapping techniques that
will be the focus of future work. In practice only partial-fields defined on
the celestial sphere are observed. The Kaiser-Squires estimator suffers due to
leakage induced by the masking of the observed region (it is well-known that
the decomposition of a spin field into scalar and pseudo-scalar components,
and consequently mass-mapping, is not unique on a manifold with boundary;
\citealt{bunn:2003}).  Pure mode estimators on the celestial sphere can be developed to remove this
leakage \citep[\eg][]{leistedt:ebsep}.  Furthermore, the impact of noise can be mitigated by the use of regularisation methods adapted to the spherical setting \citep[\eg][]{wallis:s2_recon}.

The remainder of this article is structured as follows. In
\sectn{\ref{sec:background}} we briefly review the mathematical background of
spin fields on the sphere and weak gravitational lensing.  Mass-mapping on
the celestial sphere is presented in \sectn{\ref{sec:mass_mapping}}. In
\sectn{\ref{sec:simulations_tests}} we use simulations to compare the
spherical case to a variety of planar settings for various spherical
projections. In  \sectn{\ref{sec:DES_data}} we present an application of the
spherical Kaiser-Squires technique to DES SV data in order to recover spherical mass-maps.  Concluding remarks are made
in \sectn{\ref{sec:conclusions}}. Throughout we adopt the cubehelix \citep{green:2011}
colour scheme.

\section{Background}
\label{sec:background}

Weak gravitational lensing gives rise to scalar and spin fields defined on the
celestial sphere.   For example, the observed shear field induced by weak
gravitational lensing is a  spin $\pm 2$ field. We therefore review scalar and
spin fields on the sphere and their  harmonic representation, before reviewing
the mathematical details of rotation and the Dirac delta function on the
sphere, which we make use of subsequently when considering mass-mapping on the celestial
sphere.  Weak gravitational lensing in the three-dimensional spherical setting
is then reviewed concisely.

\subsection{Spin fields on the sphere}
Square integrable spin fields on the sphere
$\fs$, with integer spin $\spin\in\integers$, are
defined by their behaviour under local rotations.  By definition, a
spin field transforms as
\begin{equation}
  \label{eqn:spin_rot}
  \fs^\prime(\sa) = \exp{-\img \spin \chi} \: \fs(\sa)
  \spcend ,
\end{equation}
under a local rotation by $\chi \in [0,2\pi)$, where the prime denotes
the rotated
field \citep{newman:1966,goldberg:1967,zaldarriaga:1997,kamionkowski:1996}.\footnote{The sign convention
  adopted for the argument of the complex exponential differs to the original definition \citep{newman:1966} but is
  identical to the convention used typically in astrophysics \citep{zaldarriaga:1997,kamionkowski:1996}.}
It is important to note that the rotation considered here is \emph{not} a global rotation on
the sphere but rather a rotation by $\chi$ in the tangent plane centred on the
spherical coordinates $\sa=(\sas) \in \sphere$, with co-latitude $\saa \in [0,\pi]$ and
longitude $\sab \in [0,2\pi)$. The case $\spin=0$ reduces to the standard scalar setting.

The canonical basis for scalar fields defined on the sphere are given by the (scalar) spherical
harmonics $\shf{\el}{\m}$.  Basis functions for spin fields can be defined
by applying spin lowering and raising operators to the scalar spherical harmonics.  Spin raising and
lowering operators, $\spinup$ and $\spindown$ respectively, increment and decrement the spin order
of a spin-\spin\ field by unity and are defined by
\begin{equation}
  \spinup \equiv
  -\sin^\spin \saa \:
  \Biggl (
  \frac{\partial}{\partial \saa}
  + \frac{\img}{\sin\saa} \frac{\partial}{\partial \sab}
  \Biggr) \:
  \sin^{-\spin}\saa \label{eq:spin_raising}
\end{equation}
and
\begin{equation}
  \spindown \equiv
  -\sin^{-\spin} \saa \:
  \Biggl (
  \frac{\partial}{\partial \saa}
  - \frac{\img}{\sin\saa} \frac{\partial}{\partial \sab}
  \Biggr) \:
  \sin^{\spin}\saa\label{eq:spin_lowering}
  \spcend ,
\end{equation}
respectively \citep{newman:1966,goldberg:1967,zaldarriaga:1997,kamionkowski:1996}.
When applied to spherical harmonics the spin raising and lowering operators take the form:
\begin{equation}
  \spinup \: \sshfarg{\el}{\m}{\sa}{\spin}
  =
  \Bigl[(\el - \spin) (\el + \spin + 1)\Bigr]^{1/2} \:
  \sshfarg{\el}{\m}{\sa}{\spin+1}
  \label{eqn:spinup_harmonic}
\end{equation}
and
\begin{equation}
  \spindown \: \sshfarg{\el}{\m}{\sa}{\spin}
  =
  - \Bigl[(\el + \spin) (\el - \spin + 1)\Bigr]^{1/2} \:
  \sshfarg{\el}{\m}{\sa}{\spin-1}
  \spcend ,
  \label{eqn:spindown_harmonic}
\end{equation}
respectively \citep[see, \eg,][]{zaldarriaga:1997}.
The spin-\spin\ spherical harmonics can thus be expressed in terms of
the scalar (spin-zero) harmonics through the spin raising and lowering
operators by
\begin{equation}
  \sshfarg{\el}{\m}{\sa}{\spin}
  =
  \biggl[ \frac{(\el-\spin)!}{(\el+\spin)!} \biggr]^{1/2}
  \spinup^\spin
  \shfarg{\el}{\m}{\sa}
  \spcend ,
\end{equation}
for $0 \leq \spin \leq \el$, and by
\begin{equation}
  \sshfarg{\el}{\m}{\sa}{\spin}
  =
  (-1)^\spin
  \biggl[ \frac{(\el+\spin)!}{(\el-\spin)!} \biggr]^{1/2}
  \spindown^{-\spin}
  \shfarg{\el}{\m}{\sa}
  \spcend ,
\end{equation}
for $-\el \leq \spin \leq 0$, where $\shf{\el}{\m}$ denote the scalar (spin-zero) spherical harmonics.

Due to the orthogonality and completeness of the spin spherical harmonics, a spin field on the
sphere can be decomposed into its harmonic representation by
\begin{equation}
  \fs(\sa) =
  \sum_{\el=0}^\infty
  \sum_{\m=-\el}^\el
  {}_s\hat{f}_{\ell m} \:
  \sshfarg{\el}{\m}{\sa}{\spin}
  \spcend .
  \label{eq:sh_decomp}
\end{equation}
The harmonic coefficients of $\fs$, denoted by ${}_s\hat{f}$, are given by the usual projection onto the basis functions:
\begin{equation}
  {}_s\hat{f}_{\ell m}
  =
  \innerp{\fs}{\sshf{\el}{\m}{\spin}}
  =
  \int_\sphere
  \dmu{\sa} \:
  \fs(\sa) \:
  \sshfargc{\el}{\m}{\sa}{\spin}
  \spcend ,
\end{equation}
where the rotation invariant measure on the sphere is given by $\dmu{\sa}= \sin \saa \dx \saa \dx \sab$, the
inner product on the sphere is denoted by $\innerp{\cdot}{\cdot}$ and $\cdot^\cconj$ denotes complex conjugation.  In practice we consider harmonic coefficients up to a maximum degree $\elmax$, \ie\ signals on the sphere band-limited at $\elmax$ with $\shc{\fs}{\el}{\m}=0$,
$\forall \el\geq\elmax$, in which case summations over $\el$ can be truncated at $\elmax$.  For notational brevity, we sometimes do not explicitly show the limits of summation where these can be inferred easily.

\subsection{Rotation on the sphere}
\label{sec:background:rotation}

We subsequently consider the rotation of fields on the sphere, defined by
application of the rotation operator $\rotarg{\eul}$, where the rotation is
parameterised by the Euler angles $\eul=(\euls) \in \sothree$.  We adopt the
$zyz$ Euler convention corresponding to the rotation of a physical body in a
\emph{fixed} coordinate system about the $z$, $y$ and $z$ axes by $\eulc$,
$\eulb$ and $\eula$, respectively.  Often we consider rotations with $\eulc=0$ and adopt the shorthand notation $\rotarg{\sa} = \rotarg{(\sab,\saa,0)}$.

The spin spherical harmonic functions are
rotated by \citep[\eg][]{mcewen:s2let_spin}
\begin{equation}
  \label{eqn:spherical_harmonic_rotation}
  (\rotarg{\eul} \: \sshf{\el}{\m}{\spin})(\sa)
  =
  \sum_{\n=-\el}^\el
  \dmatbig_{\n \m}^{\el}(\eul) \:
  \sshfarg{\el}{\n}{\sa}{\spin}
  \spcend ,
\end{equation}
where $\dmatbig_{\n \m}^{\el}$ are the Wigner \dmatbig-functions
\citep{varshalovich:1989}, which follows from the additive property of the
Wigner \dmatbig-functions \citep{marinucci:2011:book}.

The Wigner \dmatbig-functions may also be related to the spin spherical
harmonics by \citep{goldberg:1967}
\begin{equation}
  \label{eqn:ssh_wigner}
  \exp{-\img \spin \eulc}
  \sshfarg{\el}{\m}{\eulb,\eulc}{\spin} = (-1)^\spin
  \sqrt{\frac{2\el+1}{4\pi} } \:
  \dmatbig_{\m,-\spin}^{\el\:\cconj}(\euls)
  \spcend .
\end{equation}

\subsection{Dirac delta on the sphere}
\label{sec:background:dirac}

We subsequently make use of the Dirac delta function on the sphere $\delta^{\rm D}$, defined by
\begin{align}
  (\rotarg{\sa\p} \delta^{\rm D})(\sa)
   & = \frac{1}{\sin\saa} \: \delta^{1{\rm D}}(\saa-\saa\p) \: \delta^{1{\rm D}}(\sab-\sab\p) \\
   & = \sumlm \shfargc{\el}{\m}{\sa\p} \: \shfarg{\el}{\m}{\sa}
  \spcend ,
\end{align}
where $\delta^{1{\rm D}}(\cdot)$ denotes the standard one-dimensional (Euclidean) Dirac delta.  The spherical harmonic coefficients of the Dirac delta defined on the sphere are given by
\begin{equation}
  \shc{\hat{\delta}}{\el}{\m}^{\rm D} = \shfargc{\el}{\m}{\vect{0}} = \sqrt{\frac{2\el+1}{4\pi}} \: \kron{\m}{0}
  \spcend .
  \label{eqn:dirac_harmonic}
\end{equation}

\subsection{Weak gravitational lensing}

We now turn our attention to weak gravitational lensing, concisely reviewing the related mathematical background,
which is covered in more depth in several review articles
\citep[\eg][]{bartelmann:2001,schneider:2005, munshi:2008, heavens:2009}.

The weak gravitational lensing effect is typically expressed in terms of the lensing potential $\phi$,
which depends on the integrated deflection angle along the line of sight, sourced by the local Newtonian potential $\Phi$:
\begin{equation}
  \phi(r,\sa)=\frac{2}{c^2}\int_0^r{\rm d}r'  \frac{f_K(r - r')}{f_K(r)f_K(r')} \ \Phi(r',\sa)\label{eq:grav_potential_to_weak_potential}
  \spcend,
\end{equation}
where $c$ is the speed of light in a vacuum, $r$ and $r^\prime$ are comoving distances,
and $\sa=(\sas)$ denote spherical coordinates, as defined previously.  The angular diameter distance factor reads
\bea
f_K(r) = \left\{ \begin{array}{lll}    \sin(r),    & {\rm if}  \  K = 1 \vspace*{2mm} \\
  r, \quad\quad        & {\rm if} \  K = 0  \vspace*{2mm} \\
  \sinh(r), \quad\quad & {\rm if} \  K = -1\end{array}\right.
\spcend,
\eea
for cosmologies with positive ($K=1$), flat ($K=0$) and negative ($K=-1$) global curvatures.
This expression assumes the Born approximation.
The gravitational potential is related to the density field by Poisson's equation,
\begin{equation}
  \nabla^2 \Phi(r,\sa)
  = \frac{3 \Omega_{\rm M} H_0^2}{2a(r)} \: \delta(r,\sa) \label{eq:poisson}
  \spcend ,
\end{equation}
where $\Omega_{\rm M}$ is the current average matter density of the Universe as a fraction of the critical density,
$H_0$ is the current expansion rate of the Universe, $a(r)$ is the scale factor, and $\delta$ is the fractional
matter over-density.
\Eqn{\ref{eq:grav_potential_to_weak_potential}} and \eqn{\ref{eq:poisson}} relate the matter perturbations $\delta$ to the lensing potential $\phi$.

The lensing potential describes how light from a background source (\eg\ galaxy) at a position $(r,\sa)$ is distorted
by the lensing effect. This deflection, to first order, affects the images of galaxies in two ways. Firstly, images
of background sources are magnified by the convergence $\kappa$, which is related to the lensing potential by
\begin{equation}
  \wlconv(r,\sa)  = \tfrac{1}{4} \bigl( \spinup \spindown + \spindown \spinup \bigr) \wlpot(r,\sa)
  \spcend ,
  \label{eq:kappa_phi}
\end{equation}
through the spin raising and lowering operators introduced in \eqn{\ref{eq:spin_raising}}
and \eqn{\ref{eq:spin_lowering}}.
The convergence is not measured directly in weak lensing experiments because the intrinsic magnitude
of galaxy sizes is unknown.
Here and subsequently we denote the spin of each field explicit with a
proceeding subscript, \ie\ $\wlpot=\phi$ and $\wlconv=\kappa$ are both spin-zero (scalar) fields.
Secondly, images of background sources are sheared by $\wlshear$, which is related to the lensing potential by
\begin{equation}
  \wlshear(r,\sa)  = \tfrac{1}{2} \spinup \spinup \wlpot(r,\sa)
  \spcend , \label{eq:gamma_phi}
\end{equation}
where we make it explicit that the shear is a spin-2 field.
Upon averaging the shapes of many galaxies one would expect the intrinsic shear to average to zero (\ie\
there is no preferred orientation).
Hence, one can  measure shear by averaging the shapes of many galaxies.
In the remainder of this article we do not consider ``tomography'' (the separation of a source galaxy sample
into populations labelled by redshift or time) and so drop the radial dependence shown in the above equations (for notational brevity, henceforth we typically do not show the angular dependence either). For further information see the discussions in \citet{kitching:2016} on spherical-radial and spherical-Bessel representations of the shear field.

In general the potential $\wlpot$ can be decomposed into its parity even and parity odd components, namely the E-mode and B-mode components respectively:
\begin{equation}
  \wlpot = \wlpot^{\rm E} + \img \wlpot^{\rm B}
  \spcend .
\end{equation}
However, the shear induced by gravitational lensing produces an \mbox{E-mode} field only since density (scalar) perturbations cannot induce a parity odd B-mode component.  In the absence of systematic effects, we have $\wlpot^{\rm E} = \wlpot$ and $\wlpot^{\rm B} = 0$.  The convergence can also be decomposed into a parity even E-mode component and a parity odd B-mode component:
\begin{equation}
  \wlconv = \wlconv^{\rm E} + \img \wlconv^{\rm B}
  \spcend ,
  \label{eq:wlconv_eb}
\end{equation}
where the B-mode component is again zero in the absence of systematics effects.  While the E-mode convergence field is of most interest in the standard cosmological model, the B-mode convergence field is important for testing for residual systematics.  Moreover, \mbox{B-modes} are also useful in studying exotic cosmological models that exhibit parity violation \citep[\eg][]{kaufman:2016}. \
Theoretical models of intrinsic alignments of galaxies can create $B$-modes \citep{hirata:2004,crittenden:2001,crittenden:2002}, although the measured level is uncertain \citep{kirk:2015}.

The E-mode convergence field represents a scaled version of the integrated mass distribution and thus
mapping the intervening matter distribution is often performed by estimating the convergence field.
Since the shear is related to the convergence via the lensing potential
through \eqn{\ref{eq:kappa_phi}} and \eqn{\ref{eq:gamma_phi}}, convergence maps can be recovered
from the observable shear field, which amounts to solving an inverse problem.

\section{Mass-mapping on the celestial sphere}
\label{sec:mass_mapping}

In this section we describe the process of estimating a convergence field from an observed shear field in the spherical setting.  Recovering mass-maps by estimating the convergence field involves solving a spherical inverse problem, as discussed above.  First, we define the forward problem in spherical harmonic space and explicitly define the spherical generalisation of the Kaiser-Squires estimator for solving this inverse problem.  Second, we present an equivalent real space representation of the spherical mass-mapping inverse problem, where it can be seen as a deconvolution problem with a spin kernel.  Third, we consider the planar approximation of the full spherical setting, recovering the standard planar Kaiser-Squires estimator.  Finally, we consider iterative refinements to convergence estimators that account for the fact that it is the reduced shear that is observed, rather than the true underlying shear.

\subsection{Harmonic representation}\label{sec:mass_mapping:harmonic_representation}

Using the harmonic representations of the spin raising and lowering operators
it is straightforward to show that the harmonic representations of the convergence and cosmic shear of \eqn{\ref{eq:kappa_phi}} and \eqn{\ref{eq:gamma_phi}} read, respectively,
\begin{equation}
  \wlconvshc = -\tfrac{1}{2} \el (\el+1) \: \wlpotshc
\end{equation}
and
\begin{equation}
  \wlshearshc = \frac{1}{2} \sqrt{\frac{(\el+2)!}{(\el-2)!}} \: \wlpotshc
  \spcend ,
\end{equation}
where $\wlpotshc$ and $\wlconvshc$ are the \emph{scalar} spherical harmonic coefficients of the lensing potential and the converge field respectively, and
$\wlshearshc$ are the \mbox{\emph{spin}-2} spherical harmonic coefficients of the cosmic shear field, \ie\ $\wlpotshc=\innerp{\wlpot}{\shf{\el}{\m}}$, $\wlconvshc=\innerp{\wlconv}{\shf{\el}{\m}}$, and \mbox{$\wlshearshc=\innerp{\wlshear}{\sshf{\el}{\m}{2}}$}.
It follows that the spin-2 harmonic coefficients of the shear are related to the scalar harmonic coefficients of the convergence by
\begin{equation}
  \wlshearshc
  = \wlfactor_\el \wlconvshc
  \spcend ,
  \label{eq:kappa_shear_harmonic}
\end{equation}
where we define the kernel
\begin{equation}
  \wlfactor_\el =  \frac{-1}{\el(\el+1)}
  \sqrt{\frac{(\el+2)!}{(\el-2)!}}
  \spcend .
\end{equation}

Recovering the convergence field from the observable shear field therefore amounts to solving the inverse problem defined by \eqn{\ref{eq:kappa_shear_harmonic}}.  The simplest method to invert this problem is to consider a direct inversion in harmonic space.  In the planar setting, such an approach gives rise to the Kaiser-Squires estimator \citep{kaiser:1993}.  An analogous approach in the full-sky setting leads to the spherical generalisation of the Kaiser-Squires estimator, defined by
\begin{equation}
  _0\hat{\kappa}^{\rm SKS}_{\ell m}
  = \wlfactor_\el^{-1}
  \wlshearshc^{\rm est}
  \spcend ,
  \label{eq:spher_alg_gamma_to_kappa}
\end{equation}
where $\hat{\gamma}^{\rm est}_{\ell m}$ denotes the estimate of the shear harmonic coefficients computed from observational data and
$_0\hat{\kappa}^{\rm SKS}_{\ell m}$ is the spherical Kaiser-Squires (SKS) estimator of the harmonic coefficients of the convergence field.  A spherical convergence map $\wlconv^{\rm SKS}(\sa)$ can then be recovered by an inverse scalar spherical harmonic transform, following \eqn{\ref{eq:sh_decomp}}, from which the E- and B-mode components can be determined by considering the real and complex components, following \eqn{\ref{eq:wlconv_eb}}.

It is well-known that a direct Fourier inversion approach to solving inverse problems, on
which the Kaiser-Squires estimator is based, is susceptible to noise. On large
    scales one typically draws a central limit theory argument for noise Gaussianity, in which case
    a mutlivariate Gaussian noise model is adopted. In such settings the Kaiser-Squires approach
    is straightforwardly given by the maximum likelihood estimator, which implicitly
    assumes a uninformative flat prior. This, combined with the fact that the Kaiser-Squires inversion kernel
    defined by \eqn{\ref{eq:kappa_shear_harmonic}} has a flat frequency response, indicates
    that noise present in the observational dataset propagates unchecked into the convergence
    estimate.

    Typically one may wish to adopt more informative priors, within a Bayesian setting, to regularise
    this noise contribution \citep[see \textit{e.g.}][where Gaussian and wavelet sparsity priors are adopted respectively]{pichon:2010, price:2021}.
    However, for the Kaiser-Squires approach the recovered convergence field is, somewhat
    naively, smoothed with a Gaussian kernel to mitigate the impact of noise. In this paper we
    adopt this post-processing Gaussian smoothing approach and leave more advanced
    alternatives to future research.

\subsection{Real space representation}

It is insightful to express the forward problem connecting the observable cosmic shear and the convergence field in real space.  The differential form of this problem is readily apparent from \eqn{\ref{eq:kappa_phi}} and \eqn{\ref{eq:gamma_phi}}, from which it follows that
\begin{equation}
  \wlshear  = 2 \: \spinup \spinup \:
  \bigl( \spinup \spindown + \spindown \spinup \bigr)^{-1}
  \wlconv
  \spcend .
  \label{eqn:shear_conv_differential}
\end{equation}
An integral form can also be recovered, where the real space spin-2 shear field is related to the scalar convergence by a type of spherical convolution with a spin-2 kernel $\wlkernel$:
\begin{equation}
  \wlshear(\sa)
  =
  \int_\sphere \dmu{\sa\p} \:
  (\rotarg{\sa\p} \wlkernel)(\sa) \: \wlconv(\sa\p)
  \spcend ,
  \label{eqn:shear_conv_integral}
\end{equation}
where the rotation operator $\rotarg{\sa\p}$ is defined in \sectn{\ref{sec:background:rotation}}.  From comparison with \eqn{\ref{eqn:shear_conv_differential}} is it apparent that the kernel is given by
\begin{equation}
  \wlkernel(\sa)
  =
  2 \:
  \spinup \spinup \:
  \bigl( \spinup \spindown + \spindown \spinup \bigr)^{-1} \:
  \delta^{\rm D}(\sa)
  \spcend ,
\end{equation}
where $\delta^{\rm D}(\sa)$ is the Dirac delta function on the sphere defined in \sectn{\ref{sec:background:dirac}}.  Noting the spherical harmonic representation of the Dirac delta function of \eqn{\ref{eqn:dirac_harmonic}} and the harmonic action of the spin raising and lowering operators of \eqn{\ref{eqn:spinup_harmonic}} and \eqn{\ref{eqn:spindown_harmonic}}, it is straightforward to show that the harmonic coefficients of the kernel read
\begin{equation}
  \shc{\wlkernel}{\el}{\m}
  =
  \frac{-1}{\el(\el+1)}
  \sqrt{\frac{(\el+2)!}{(\el-2)!}}
  \sqrt{\frac{2\el+1}{4\pi}} \:
  \kron{\m}{0}
  \spcend .
  \label{eqn:kernel_harmonic}
\end{equation}
An explicit expression for the kernel in real space can then be recovered from its harmonic representation, yielding
\begin{equation}
  \wlkernel(\sa)
  =
  \sum_\el \frac{-1}{\el(\el+1)}  \frac{2 \el+1}{4\pi}\: \aleg{\el}{2}{\cos\theta}
  \spcend ,
\end{equation}
where $\aleg{\el}{2}{\cdot}$ is the associated Legendre function of order two.  The equivalence of the harmonic and real space expressions of the forward problem of \eqn{\ref{eq:kappa_shear_harmonic}} and \eqn{\ref{eqn:shear_conv_integral}}, respectively, can also be seen by the explicit harmonic representation of \eqn{\ref{eqn:shear_conv_integral}}, as shown in \appn{\ref{sec:appendix:harmonic_inverse}}.

\subsection{Planar approximation}\label{sec:mass_mapping:planar_approximation}

We now consider the planar approximation of the spherical mass-mapping estimator presented in \sectn{\ref{sec:mass_mapping:harmonic_representation}}, recovering the standard planar Kaiser-Squires estimator \citep{kaiser:1993}.  Firstly, we note the planar approximations of the spin raising and lowering
operators given by
\begin{equation}
  \spinup \approx -\bigl(\partial_x + \img \partial_y\bigr)
\end{equation}
and
\begin{equation}
  \spindown \approx -\bigl(\partial_x - \img \partial_y\bigr)
  \spcend ,
\end{equation}
respectively \citep[see, \eg,][]{bunn:2003}.  In the planar approximation the convergence and cosmic shear are then related to the lensing potential by
\begin{equation}
  \wlconv =
  \tfrac{1}{4}
  \bigl( \spinup \spindown + \spindown \spinup \bigr) \wlpot
  \approx
  \tfrac{1}{2}
  \bigl( \partial_x^2 + \partial_y^2 \bigr) \wlpot
  \label{eq:kappa_potential_planar}
\end{equation}
and
\begin{equation}
  \wlshear = \tfrac{1}{2} \spinup \spinup \wlpot
  \approx \biggl[ \tfrac{1}{2} \bigl(\partial_x^2 - \img \partial_y^2\bigr)
    + \img \partial_x \partial_y \biggr]
  \wlpot
  \spcend ,
  \label{eq:shear_potential_planar}
\end{equation}
respectively.
It is common to decompose the shear component into its real and imaginary component by
\begin{equation}
  \wlshear = \wlshearone + \img \wlsheartwo
  \spcend .
\end{equation}
The planar Fourier representations of \eqn{\ref{eq:kappa_potential_planar}} and \eqn{\ref{eq:shear_potential_planar}} are then given by
\begin{equation}
  \wlconvfk = -\tfrac{1}{2} \bigl( \wlkx^2 + \wlky^2 \bigr) \wlpotfk
\end{equation}
and
\begin{equation}
  \begin{split}
    \wlshearonefk
    & = -\tfrac{1}{2} \bigl( \wlkx^2 - \wlky^2 \bigr) \wlpotf^{\rm E}(\wlkx,\wlky)
    + \wlkx \wlky \wlpotf^{\rm B}(\wlkx,\wlky) \spcend,\\
    \wlsheartwofk
    & = - \wlkx \wlky \wlpotf^{\rm E}(\wlkx,\wlky)
    -\tfrac{1}{2} \bigl( \wlkx^2 - \wlky^2 \bigr) \wlpotf^{\rm B}(\wlkx,\wlky) \spcend ,
  \end{split}
\end{equation}
respectively, where $\hat{\cdot}$ denotes the Fourier transform and $\wlkx$ and $\wlky$ denote the Fourier coordinates, and we make use of the Fourier derivative property $\widehat{\partial_x f} = \img \wlkx \hat{f}$.
It follows that under the planar approximation the shear can be related to the convergence in Fourier space by
\begin{equation}
  \wlshearfk
  = \wlfactorplane_{\wlkx,\wlky} \wlconvfk
  \spcend ,
  \label{eq:kappa_shear_harmonic_planar}
\end{equation}
where
\begin{equation}
  \wlfactorplane_{\wlkx,\wlky}  = \frac{\wlkx^2 - \wlky^2 + \img 2\wlkx \wlky}{\wlkx^2 + \wlky^2}
  \spcend .
\end{equation}

Analogous to the spherical setting considered in \sectn{\ref{sec:mass_mapping:harmonic_representation}}, in the planar setting recovering the convergence field from the shear amounts to solving the inverse problem defined by \eqn{\ref{eq:kappa_shear_harmonic_planar}}.  Again, the simplest method to invert this problem is to perform a direct inversion in harmonic space, which gives rises to the standard planar Kaiser-Squires (KS) estimator \citep{kaiser:1993} of
\begin{equation}
  _0\hat{\kappa}^{\rm KS}(\wlkx,\wlky)
  = \wlfactorplane_{\wlkx,\wlky}^{-1} \: {}_2\hat{\gamma}^{\rm est}(\wlkx,\wlky)
  = \wlfactorplane_{\wlkx,\wlky}^\cconj \: {}_2\hat{\gamma}^{\rm est}(\wlkx,\wlky)
  \spcend ,
  \label{eq:planar_estimator_begin}
\end{equation}
where we have taken
advantage of the fact that $\wlfactorplane_{\wlkx,\wlky}^{-1}=\wlfactorplane_{\wlkx,\wlky}^*$ since $\vert \wlfactorplane_{\wlkx,\wlky} \vert = 1$.  Recall that $_2\hat{\gamma}^{\rm est}(\wlkx,\wlky)$ is the estimate of the planar Fourier coefficients of the shear computed from observational data. Expanding the real and imaginary components, one recovers the familiar KS estimators for the E- and B-mode component of the convergence given by
\begin{equation}
  _0\hat{\kappa}^{\rm E,KS}(\wlkx,\wlky)
  = \frac{(\wlkx^2 - \wlky^2) \: {}_2\hat{\gamma}^{\rm est}_1(\wlkx,\wlky)
    + 2 \wlkx\wlky \: {}_2\hat{\gamma}^{\rm est}_2(\wlkx,\wlky)}
  {\wlkx^2 + \wlky^2}
\end{equation}
and
\begin{equation}
  _0\hat{\kappa}^{\rm B,KS}(\wlkx,\wlky)
  = \frac{-2 \wlkx\wlky \: {}_2\hat{\gamma}^{\rm est}_1(\wlkx,\wlky)
    + (\wlkx^2 - \wlky^2) \: {}_2\hat{\gamma}^{\rm est}_2(\wlkx,\wlky)}
  {\wlkx^2 + \wlky^2}
  \spcend ,
\end{equation}
respectively.  A planar convergence map $\wlconv^{\rm KS}(\sa)$ can then be recovered by an inverse Fourier transform.

In the above derivation we have not considered the practicalities of the projection of the fields considered, which are defined natively on the celestial sphere, onto a planar region.  In practice, one must choose a specific projection, the choice of which can have a large impact on the quality of the
convergence map recovered from the observed shear. We describe a variety of projections in \appn{\ref{sec:appendix:projections}}
and discuss their properties. Care must be taken when projecting a spin-$2$ field such as the cosmic shear
as local rotations must be taken into account, as described in detail in
\appn{\ref{sec:appendix:projections}}.

\subsection{Reduced shear}\label{sec:mass_mapping:reduced_shear}

In deriving the estimators presented in \sectn{\ref{sec:mass_mapping:harmonic_representation}} and \sectn{\ref{sec:mass_mapping:planar_approximation}} we made the assumption that one could observe the pixelised shear field directly.   However, in practice one can only measure the pixelised \emph{reduced} shear $_2{g}$, which is
related to the true underlying shear by
\begin{equation}
  _2{}g = \frac{{}_2\gamma}{1-{}_0\kappa}
  \spcend .
  \label{eq:reduced_shear}
\end{equation}
The problem of recovering the convergence field then becomes non-linear.  However, this non-linear problem can be solved iteratively \citep[][p.153]{seitz:1995, book:Mediavilla:2016}, as discussed below. These techniques and similar are in common use in the literature \citep[\eg][]{jullo:2014,lanusse:2016, price:2018}

The first step is to denoise the map of reduced shear. In this work we use a Gaussian smoothing.
We make an initial estimate of the shear by assuming it is simply
the measured reduced shear. Then an initial estimate of the pixelised convergence
field is made.  The first step of the iterative algorithm is thus:
\bea
\begin{split}
  {}_2\gamma^{(0)} & = {}_2g \spcend,\label{eq:reduced_shear_initial}\\
  {}_0\kappa^{(0)} & = \mathsf{M}\left[{}_2\gamma^{(0)}\right] \spcend ,
\end{split}
\eea
where $\mathsf{M}$ denotes the mass-mapping estimator used to recover the convergence from the shear (in this
article we consider either the spherical or planar Kaiser-Squires estimators described in \sectn{ \ref{sec:mass_mapping:harmonic_representation}} and \sectn{\ref{sec:mass_mapping:planar_approximation}}, respectively) and the superscript denotes iteration number.
We then use our estimate of the convergence to update the estimate of the shear and
repeat.  The $(i+1)$-th iteration is thus:
\bea
  \begin{split}
    {}_2\gamma^{(i+1)} & = {}_2g(1-{}_0\kappa^{(i)}) \spcend,\\
    {}_0\kappa^{(i+1)} & = \mathsf{M}\left[{}_2\gamma^{(i+1)}\right] \spcend.\label{eq:reduced_shear_end}
  \end{split}
  \eea
Iterations are continued until the absolute difference of the convergence between iterations is below some threshold value.  In this work we choose,
\begin{equation}
  \max_j \: \bigl\vert{}_0\kappa_j^{(i)}-{}_0\kappa_j^{(i-1)}\bigr\vert<10^{-10},
\end{equation}
where $j$ runs over all pixels.
Typically, for a convergence field including ellipticity/shot noise, $4$ to $5$ iterations are required before converging.

\subsection{Implementation}

We have written the python package {\tt massmappy}\footnote{\url{http://www.massmappy.org}} to implement the algorithms presented. The package can perform standard mass-mapping on the plane, with the option to perform iterations to account for reduced shear.
We also implement the spherical Kaiser-Squires estimator described above so that mass-mapping can be performed on the celestial sphere. We support the use of two spherical pixelisations schemes. Firstly, we support the use of \healpix\footnote{\url{http://healpix.jpl.nasa.gov}} \citep{gorski:2005}, an equal area pixelisation with an accompanying software package that can perform fast spherical harmonic transforms. We also support the use of the standard equiangular sampling scheme implemented in \sshtcode\footnote{\url{http://www.spinsht.org}} \citep{mcewen:fssht}.  This sampling scheme supports fast spherical harmonic transforms that are theoretically exact and achieve close to floating point precision in practice.  The most recent release of \sshtcode\
includes fast routines to compute the projections of the sphere onto the plane considered in this work.

\section{Evaluation on simulations}\label{sec:simulations_tests}

In this section we evaluate the mass-mapping algorithms presented in
\sectn{\ref{sec:mass_mapping}} on simulations. We study the error introduced by the planar approximation, for a variety of projections and for varying survey coverage area, when compared to the spherical setting.  We also assess the ability of the iterative algorithm described in \sectn{\ref{sec:mass_mapping:reduced_shear}} to deal with the reduced shear that is observed, rather than the underlying true shear.

\subsection{Comparison of planar and spherical mass-mapping}
\label{sec:simulations_tests:low_res}

We study the impact of the flat-sky planar approximation in mass-mapping, compared to the spherical setting, and determine the typical errors induced for the sky coverages of upcoming surveys.
We do this as an idealised situation to focus the study on the effect of projecting the sphere on to the plane. To do so we need to understand how best one
can estimate mass-maps on the plane for large coverage areas.

When creating convergence maps on the plane (\ie\ mass-maps), the exact projection used to map the celestial sphere to the plane can have a large impact on the quality of the reconstructed convergence map. In \appn{\ref{sec:appendix:projections}} we describe a variety of spherical projections that can be considered, which we evaluate on simulations here. One important aspect
when projecting a non-zero spin field, \eg\ shear (or galaxy ellipticities), is to ensure that the correct local rotations are performed, as described in \appn{\ref{sec:appendix:projections:rotations}}.  This is typically neglected in existing mass-mapping works.

\begin{figure*}

  \begin{tabular}{c p{3cm} p{3cm} p{3cm} p{3cm} p{3cm}}
    \rotatebox{90}{Cylindrical}                                                                                                                      & \includegraphics[width=.18\textwidth, trim=2.5cm 5.15cm 2cm 2.15cm, clip=true]{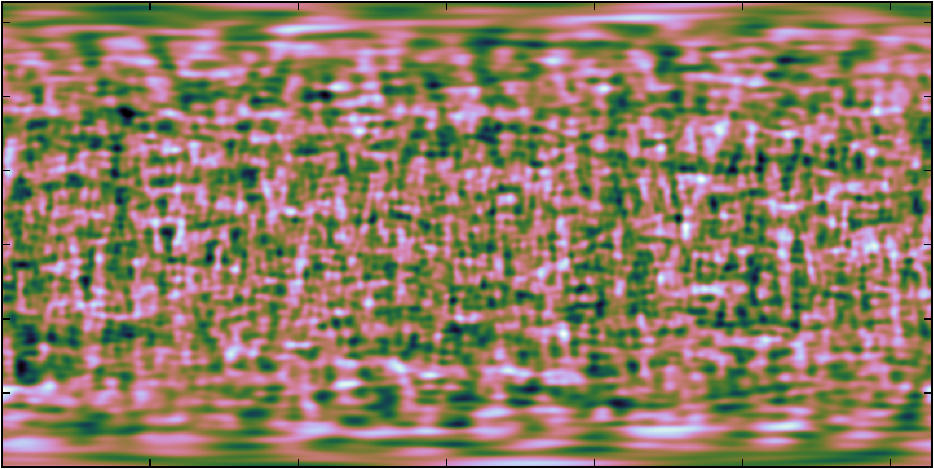}                 &
    \includegraphics[width=.18\textwidth, trim=2.5cm 5.15cm 2cm 2.15cm, clip=true]{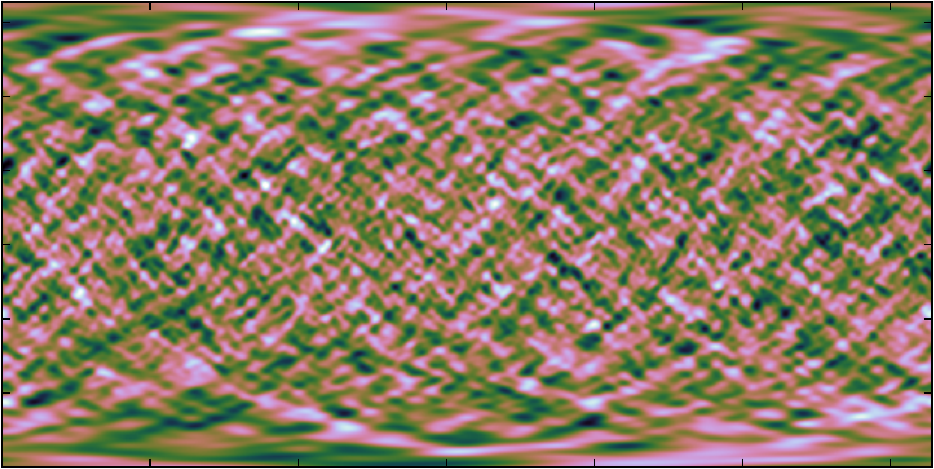}                       &
    \includegraphics[width=.18\textwidth, trim=2.5cm 5.15cm 2cm 2.15cm, clip=true]{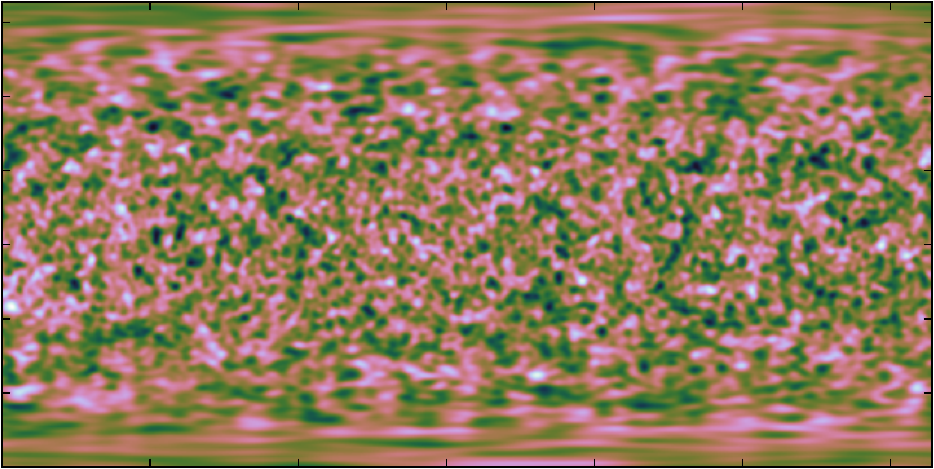}                       &
    \includegraphics[width=.18\textwidth, trim=2.5cm 5.15cm 2cm 2.15cm, clip=true]{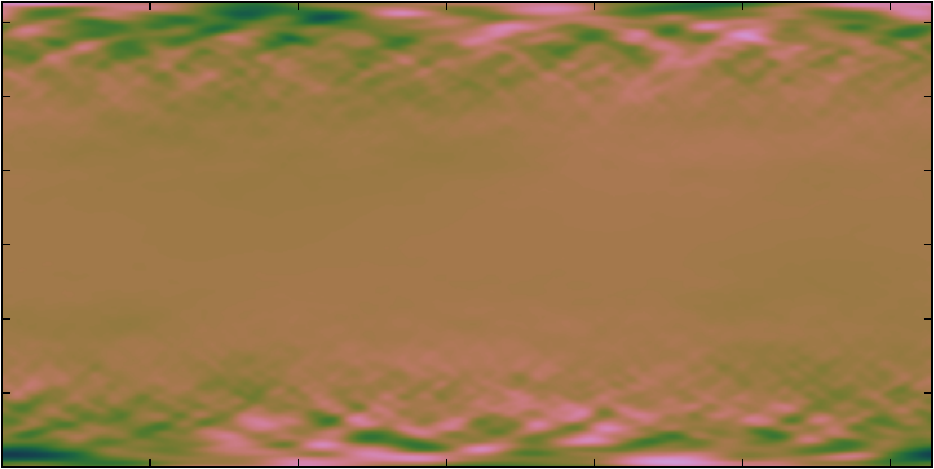}           &
    \includegraphics[width=.18\textwidth, trim=2.5cm 5.15cm 2cm 2.15cm, clip=true]{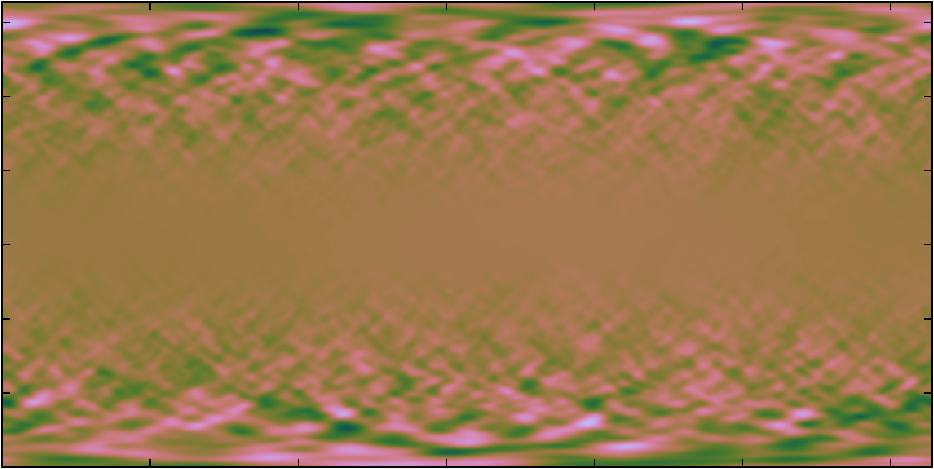}                                                                                                                                                                                                                                                                                                                                                   \\
    \rotatebox{90}{~Mercator}                                                                                                                        & \multicolumn{1}{c}{\includegraphics[width=.18\textwidth, trim=4.6cm 5cm 4.09cm 1.5cm, clip=true]{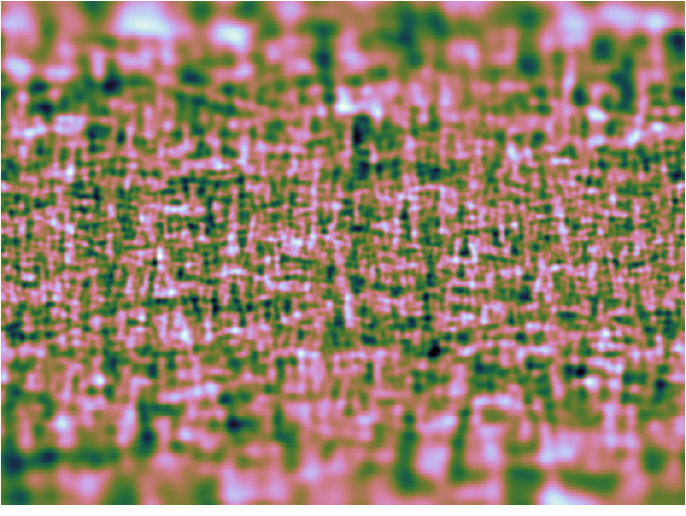}} &
    \multicolumn{1}{c}{\includegraphics[width=.18\textwidth, trim=4.6cm 5cm 4.09cm 1.5cm, clip=true]{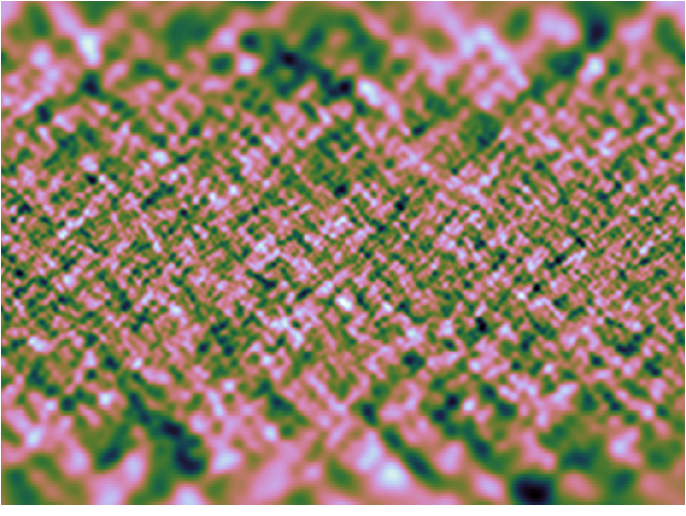}}       &
    \multicolumn{1}{c}{\includegraphics[width=.18\textwidth, trim=4.6cm 5cm 4.09cm 1.5cm, clip=true]{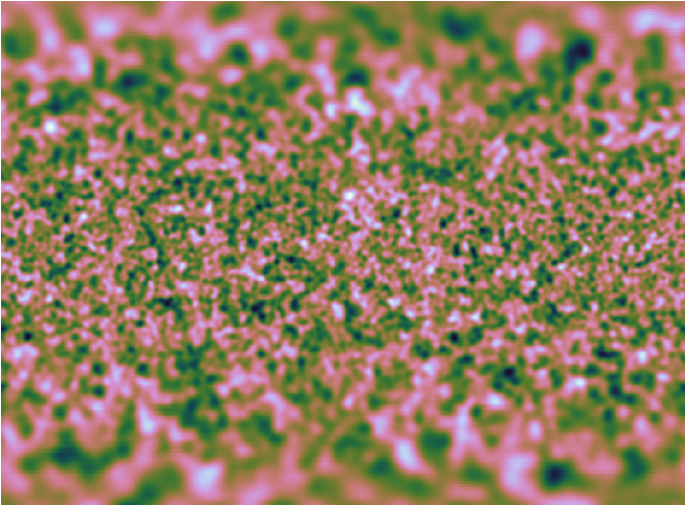}}       &
    \multicolumn{1}{c}{\includegraphics[width=.18\textwidth, trim=4.6cm 5cm 4.09cm 1.5cm, clip=true]{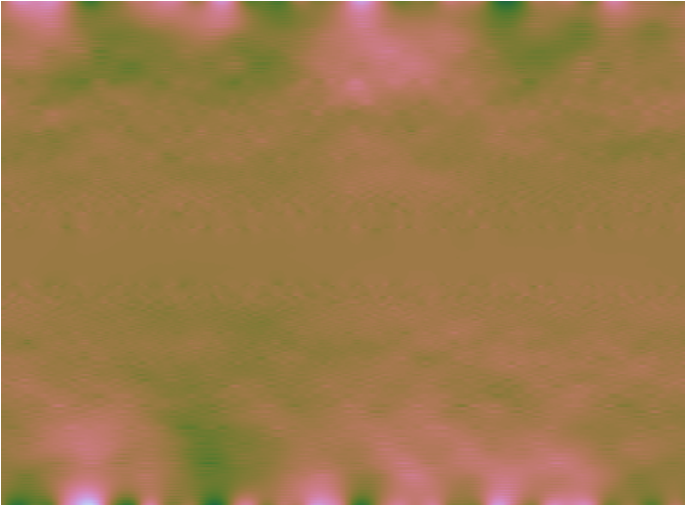}} &
    \multicolumn{1}{c}{\includegraphics[width=.18\textwidth, trim=4.6cm 5cm 4.09cm 1.5cm, clip=true]{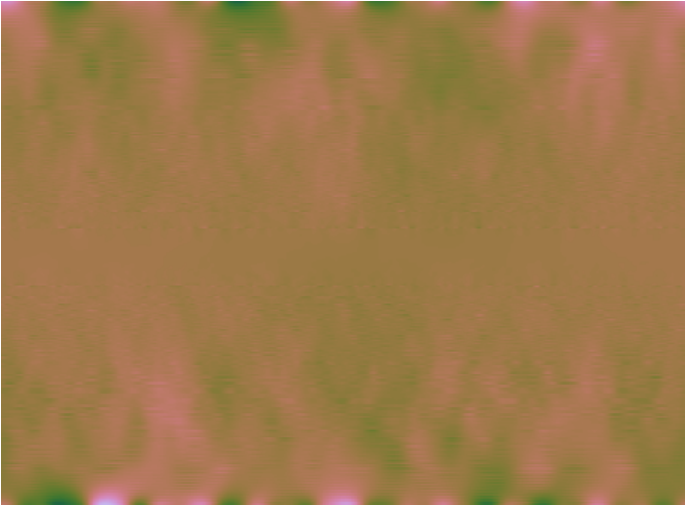}}                                                                                                                                                                                                                                                                                                                                         \\
    \rotatebox{90}{Sinusoidal}                                                                                                                       & \includegraphics[width=.18\textwidth, trim=2.5cm 5.2cm 2cm 2.13cm, clip=true]{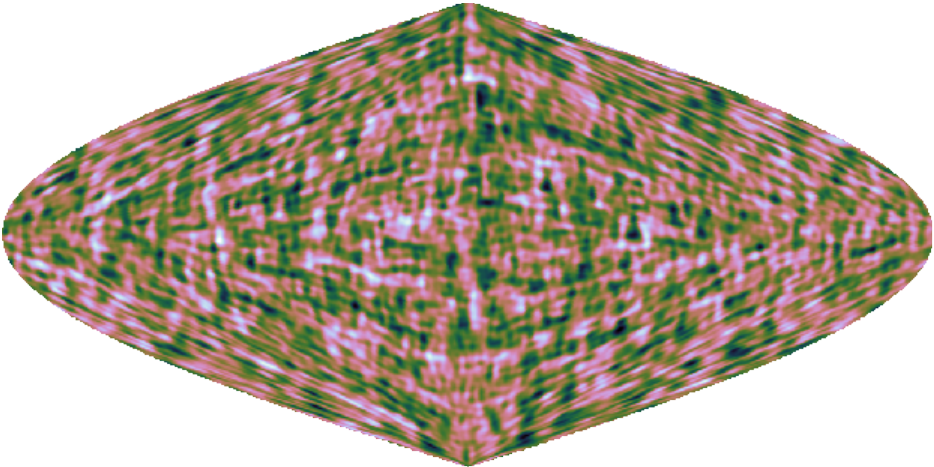}                         &
    \includegraphics[width=.18\textwidth, trim=2.5cm 5.2cm 2cm 2.13cm, clip=true]{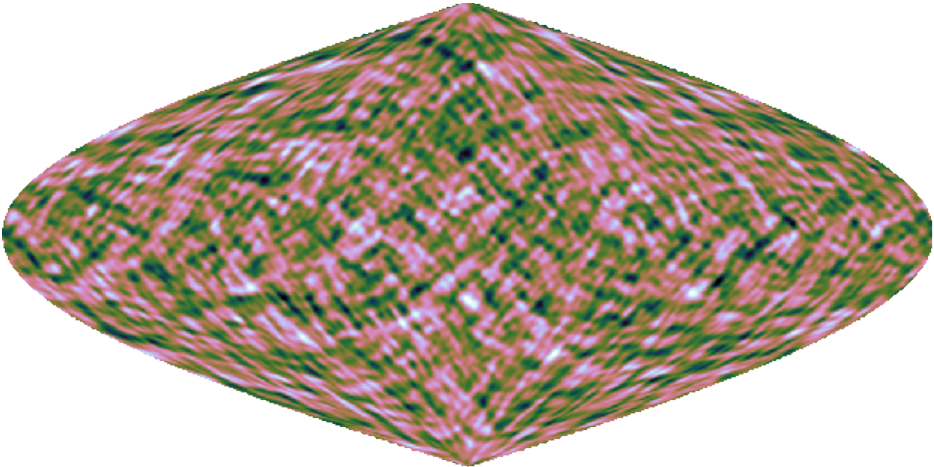}                               &
    \includegraphics[width=.18\textwidth, trim=2.5cm 5.2cm 2cm 2.13cm, clip=true]{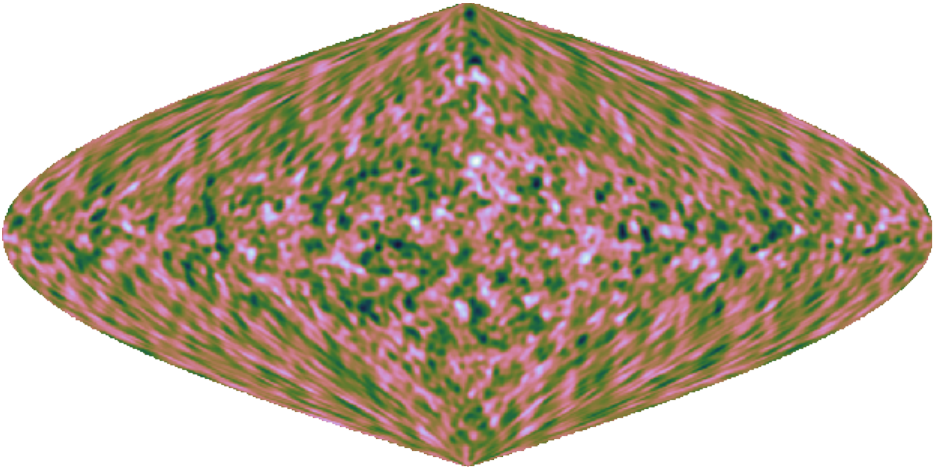}                               &
    \includegraphics[width=.18\textwidth, trim=2.5cm 5.2cm 2cm 2.13cm, clip=true]{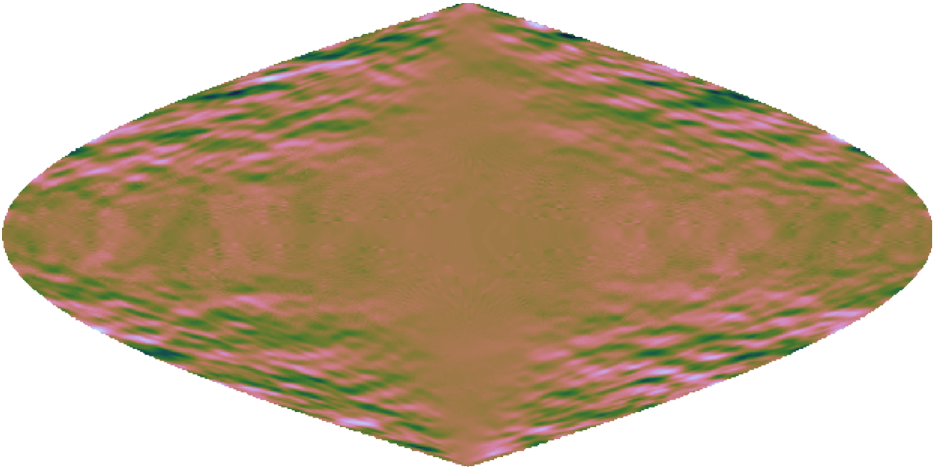}                         &
    \includegraphics[width=.18\textwidth, trim=2.5cm 5.2cm 2cm 2.13cm, clip=true]{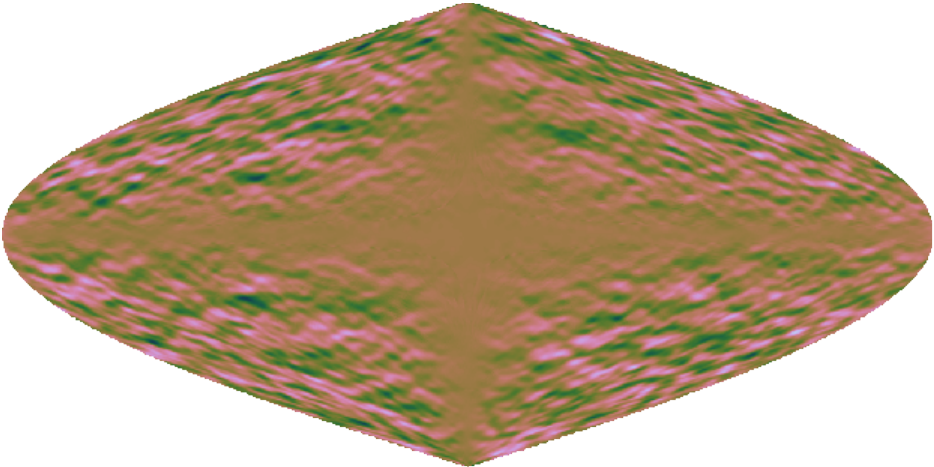}                                                                                                                                                                                                                                                                                                                                                                 \\
                                                                                                                                                     & \multicolumn{1}{c}{(a) $\gamma_1$}                                                                                                         & \multicolumn{1}{c}{(b) $\gamma_2$} & \multicolumn{1}{c}{(c) $\kappa^{\rm E,KS}$} & \multicolumn{1}{c}{(d) $\kappa^{\rm E,KS}$ error} & \multicolumn{1}{c}{(e) $\kappa^{\rm B,KS}$ error} \\
  \end{tabular}

  \includegraphics[width=.5\textwidth, trim=0cm 2cm 0cm 11.2cm, clip=true]{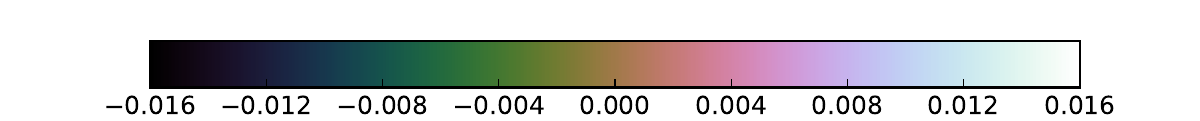}
  \caption{Simulated reconstructions of the convergence field (mass-maps) on large regions of the celestial sphere when using equatorial projections, in order to assess the impact of different planar projections.
    The shear field is shown in the first and second columns (the first showing $\gamma_1$ and the second showing $\gamma_2$).
    The third column shows the reconstructed convergence field ($E$-mode), while the forth and fifth columns shows the error on the $E$-mode and $B$-mode convergence, respectively.
    Each row shows a different projection: the first row shows the simple cylindrical projection; the second shows the Mercator projection; and the final row shows the sinusoidal projection. The
    entire sphere is projected onto the plane, except for the Mercator projection where only $7\pi/16$ radians above and below the equator are considered (as explained in the main text).}
  \label{fig:low_res_equatorial_projections}
\end{figure*}

\begin{figure*}
  \begin{tabular}{c p{3cm} p{3cm} p{3cm} p{3cm} p{3cm}}
    \rotatebox{90}{Orthographic}                                                                                                    &
    \includegraphics[width=.18\textwidth, trim=6.1cm 5.15cm 5.5cm 1.48cm, clip=true]{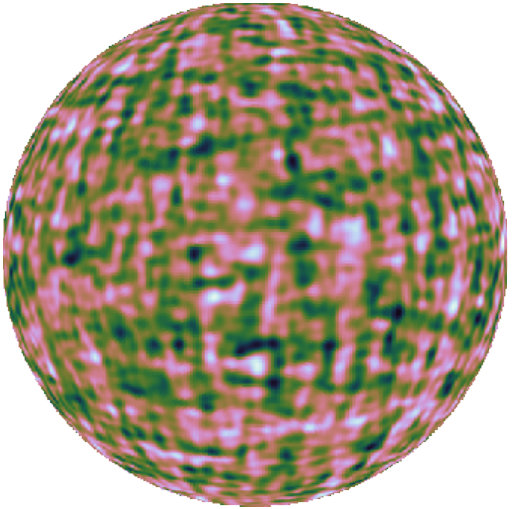}       &
    \includegraphics[width=.18\textwidth, trim=6.1cm 5.15cm 5.5cm 1.48cm, clip=true]{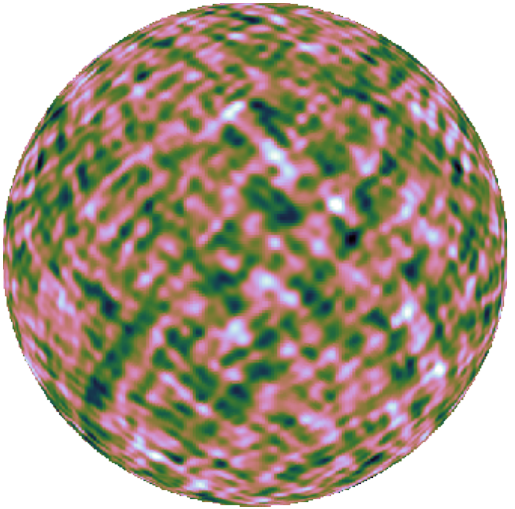}       &
    \includegraphics[width=.18\textwidth, trim=6.1cm 5.15cm 5.5cm 1.48cm, clip=true]{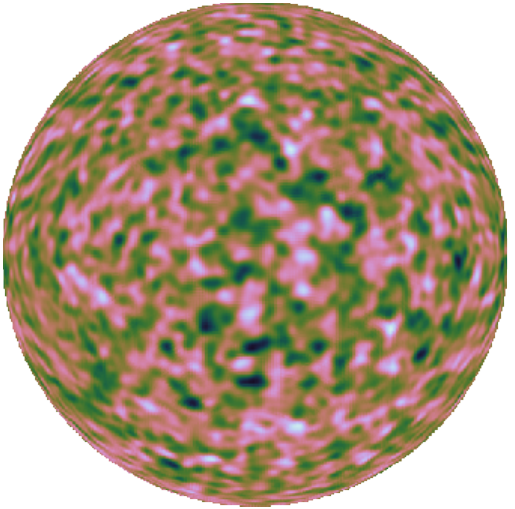}       &
    \includegraphics[width=.18\textwidth, trim=6.1cm 5.15cm 5.5cm 1.48cm, clip=true]{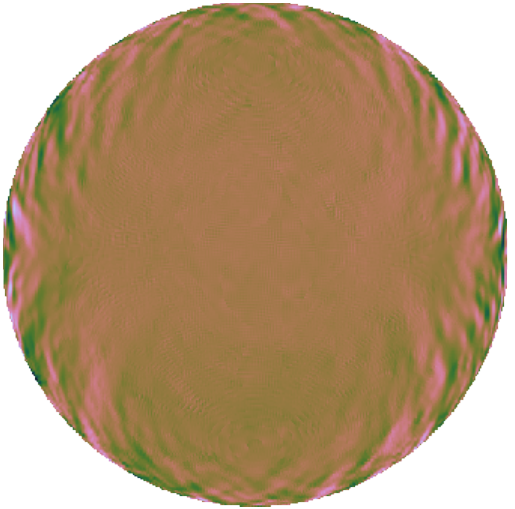} &
    \includegraphics[width=.18\textwidth, trim=6.1cm 5.15cm 5.5cm 1.48cm, clip=true]{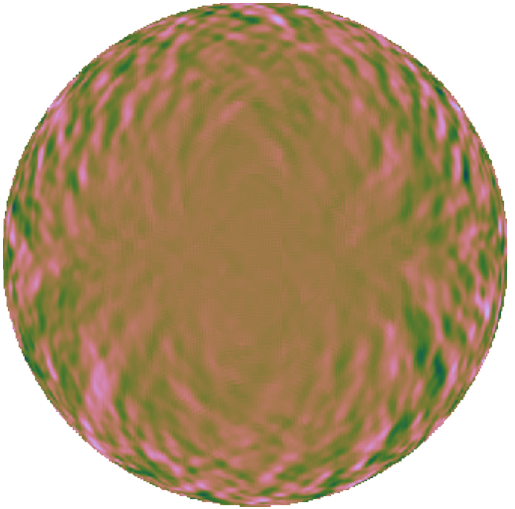}                                                                                                                                                                                                                                 \\

    \rotatebox{90}{Stereographic}                                                                                                   &
    \includegraphics[width=.18\textwidth, trim=6.1cm 5.15cm 5.5cm 1.48cm, clip=true]{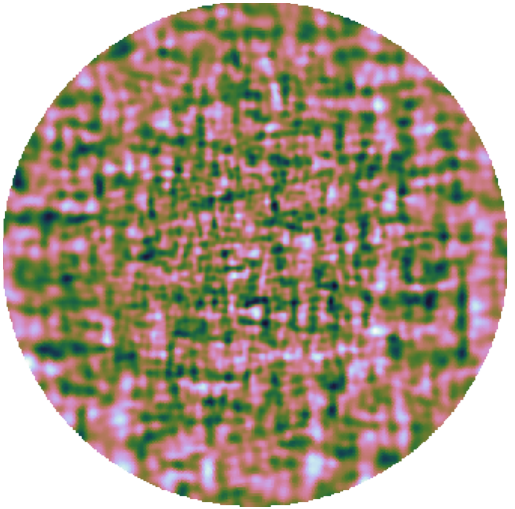}       &
    \includegraphics[width=.18\textwidth, trim=6.1cm 5.15cm 5.5cm 1.48cm, clip=true]{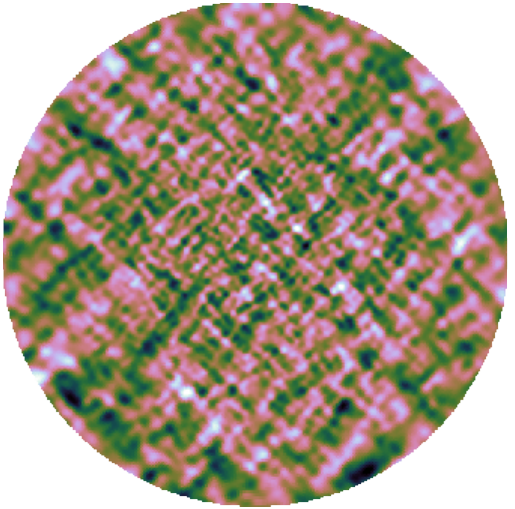}       &
    \includegraphics[width=.18\textwidth, trim=6.1cm 5.15cm 5.5cm 1.48cm, clip=true]{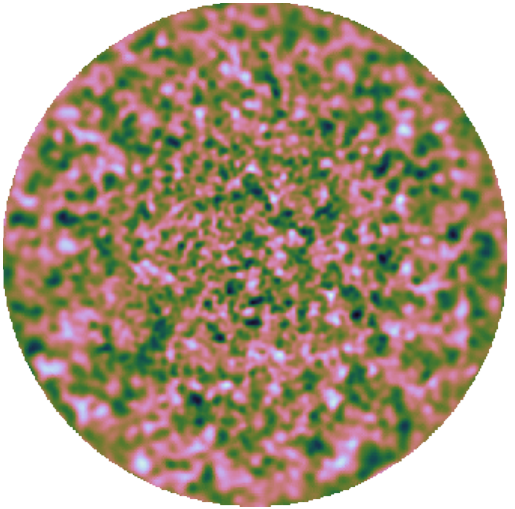}       &
    \includegraphics[width=.18\textwidth, trim=6.1cm 5.15cm 5.5cm 1.48cm, clip=true]{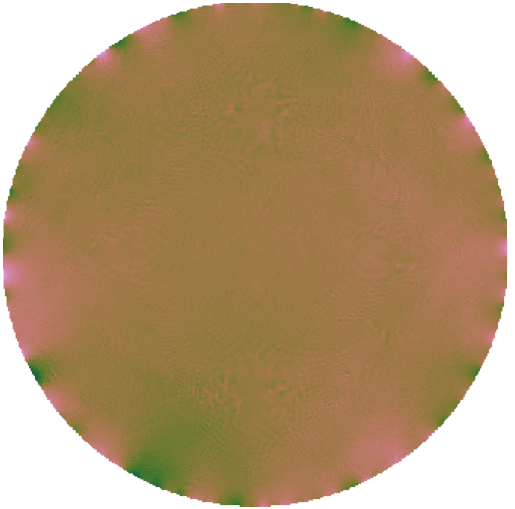} &
    \includegraphics[width=.18\textwidth, trim=6.1cm 5.15cm 5.5cm 1.48cm, clip=true]{figures/low_res_kappa_SP_south_error_imag.pdf}                                                                                                                                                                                                                                 \\

    \rotatebox{90}{Gnomonic}                                                                                                        &
    \includegraphics[width=.18\textwidth, trim=6.1cm 5.15cm 5.5cm 1.48cm, clip=true]{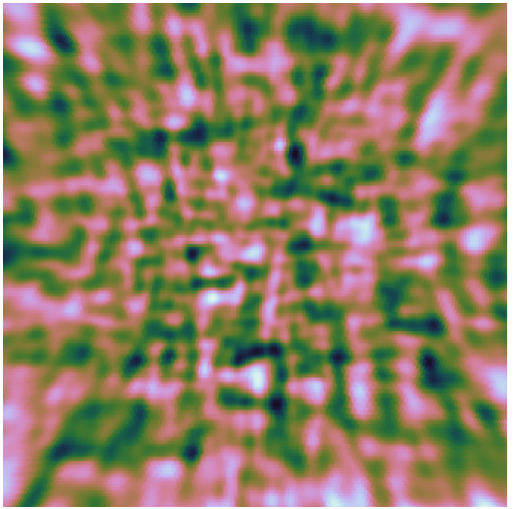}       &
    \includegraphics[width=.18\textwidth, trim=6.1cm 5.15cm 5.5cm 1.48cm, clip=true]{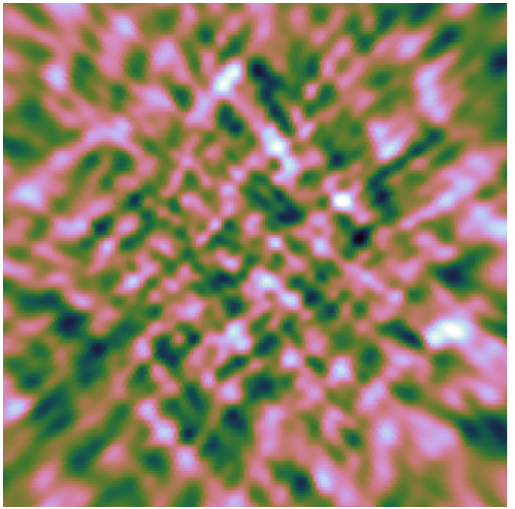}       &
    \includegraphics[width=.18\textwidth, trim=6.1cm 5.15cm 5.5cm 1.48cm, clip=true]{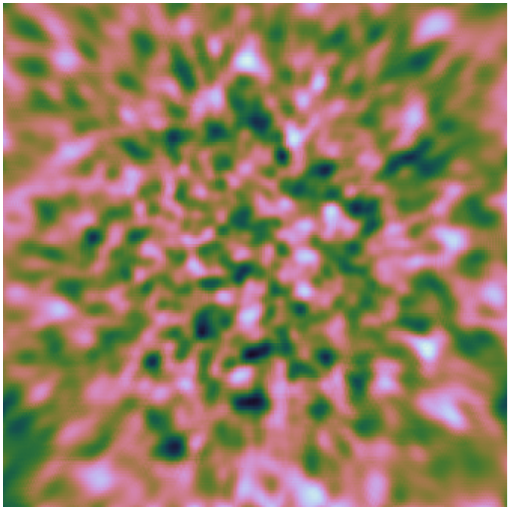}       &
    \includegraphics[width=.18\textwidth, trim=6.1cm 5.15cm 5.5cm 1.48cm, clip=true]{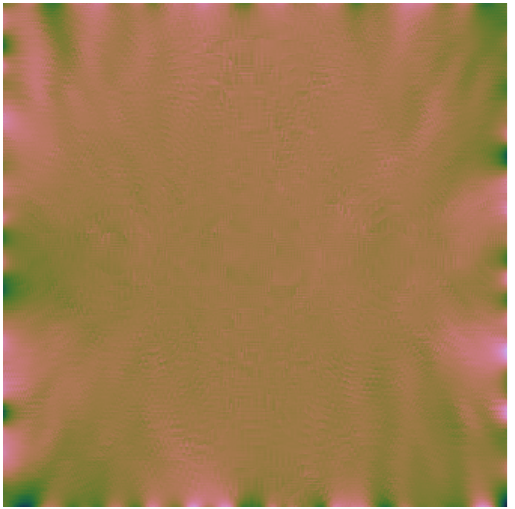} &
    \includegraphics[width=.18\textwidth, trim=6.1cm 5.15cm 5.5cm 1.48cm, clip=true]{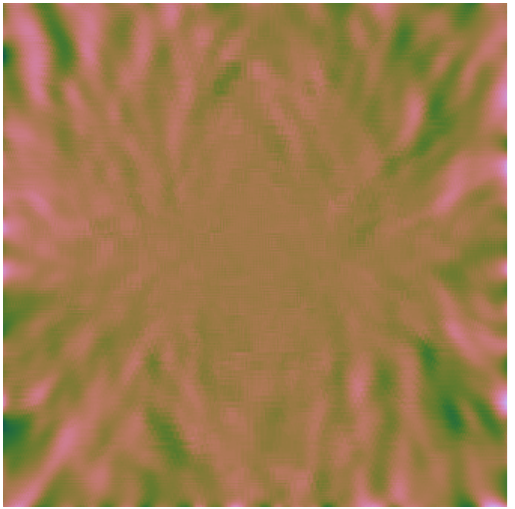}                                                                                                                                                                                                                                 \\
                                                                                                                                    & \multicolumn{1}{c}{(a) $\gamma_1$} & \multicolumn{1}{c}{(b) $\gamma_2$} & \multicolumn{1}{c}{(c) $\kappa^{\rm E,KS}$} & \multicolumn{1}{c}{(d) $\kappa^{\rm E,KS}$ error} & \multicolumn{1}{c}{(e) $\kappa^{\rm B,KS}$ error} \\
  \end{tabular}

  \includegraphics[width=.5\textwidth, trim=0cm 2cm 0cm 11.2cm, clip=true]{figures/low_res_kappa_SP_south_error_imag.pdf}
  \caption{Same as \fig{\ref{fig:low_res_equatorial_projections}} for the polar projections. The first row shows the orthographic projection, the middle row shows the stereographic projection,
    and the third row shows the Gnomonic projection. For these projections we only project one hemisphere onto the sphere, with the pole defined by the $x$-axis. The entire hemisphere is shown except for the
    Gnomonic projection where we project the sphere onto a square where distance from the centre of the square and the edge represents an angle of $\pi/4$ radians (as explained in the main text). 
    Of course, planar approaches are typically restricted to a field of view of $\sim 20$ degrees, these figures simply illustrate why this consensus is adhered to.
  }
  \label{fig:low_res_polar_projections}
\end{figure*}

We now describe the simulations that we use to assess the effect each
projection has on the quality of the reconstruction of convergence maps.  We
simulate Gaussian convergence maps using a convergence power spectrum
generated by the software package \textsc{cosmosis}\footnote{\url{https://bitbucket.org/joezuntz/cosmosis/wiki/Home}} \citep{zuntz:2015}.
The power spectrum was generated with a standard $\Lambda$CDM cosmology with galaxies in high redshift bin $z\gtrsim1$.
We simulate the map up to a
harmonic band-limit of $\elmax=512$ using the sampling of the sphere of
\sshtcode\ \citep{mcewen:fssht}. We consider this spherical sampling scheme
for these numerical experiments since the resulting spherical harmonic
transforms are theoretically exact and the implementations in \sshtcode\ achieve
accuracy close to machine precision (which is not the case for \healpix; see \citet{leistedt:s2let_axisym}
for concise accuracy benchmarks).  Any
errors will therefore be due to projection effects rather than inaccuracies in
harmonic transforms. We smooth the simulated convergence maps with the Gaussian
kernel $G_\ell = e^{-\ell^2\sigma^2}$,
with $\sigma=\pi/256$, to mitigate pixelisation issues. The shear field is simulated
by transforming the scalar convergence field to harmonic space and then
applying \eqn{\ref{eq:kappa_shear_harmonic}}, before transforming back to real
space to recover a spin-2 shear field on the celestial sphere. In these
simulations we aim to understand the effect of the projections so we do not
consider the effects of reduced shear or noise.

To evaluate the accuracy of planar mass-mapping we first project the simulated
shear field from the celestial sphere to the plane, using a particular
projection.  We estimate the convergence field from the planar shear field
using the planar KS estimator of \eqn{\ref{eq:planar_estimator_begin}}.  We
then compare this recovered planar convergence to a planar projection of the
convergence simulated initially on the celestial sphere.
A number of different projections are considered, as defined in \appn{\ref{sec:appendix:projections}}.  In general, we consider two classes of spherical projection: namely, equatorial and polar projections.

In \fig{\ref{fig:low_res_equatorial_projections}} we show example planar
reconstructions and errors for a variety of equatorial projections. These
projections are highly accurate on the equator, with distortion due to the
projection typically increasing with distance from the equator. We consider,
firstly, a simple cylindrical projection, where the $(\sas)$ angles are taken
to be Cartesian coordinates $(x,y)$. We also consider the Mercator projection,
which is often used for geographical maps.  The Mercator projection is a
conformal projection, in that it preserves local angles. The poles in this
projection would be at infinity, so we limit the projection to $7\pi/16$
radians above and below the equator. Finally,
\fig{\ref{fig:low_res_equatorial_projections}} shows results using the
sinusoidal projection, a simple equal area projection used by the DES
collaboration for the convergence map generated from DES SV data
\citep{vikram:2015}.

In \fig{\ref{fig:low_res_polar_projections}} we show example planar
reconstructions and errors for a variety of polar projections. These
projections are highly accurate around the pole defining the centre of the
projection, with distortion increasing as one moves away from this point. For
these projections we project one hemisphere around a pole defined by the
$x$-axis only; hence, two projections (one for each hemisphere) are required
to cover the entire sphere.\footnote{For the stereographic projection, a
  single projection can be applied to map the sphere to the plane.  However, the
  opposite pole is mapped to the point at infinity.  Moreover, the size of the
  planar regions grows considerably as the full coverage of the celestial sphere
  is approached.  Consequently, for practical purposes the two hemispheres are
  projected separately.}  We consider the orthographic projection, which is a
simple vertical projection, the stereographic projection, which is another
conformal projection, and finally the Gnomonic projection, which has the special
property that the local rotations  required for the projection of spin fields
are zero (if no coordinate rotation is performed). The edge of the hemisphere
for the Gnomonic projection lies at infinity so we only project the sphere onto
the square where the distance from the centre of the square and its edge
represents an angle of $\pi/4$ radians.

For all projections, we show in
\fig{\ref{fig:low_res_equatorial_projections}} and
\fig{\ref{fig:low_res_polar_projections}} the projected shear, the recovered
$E$-mode convergence and the error in the $E$ and $B$-mode convergence.  As
expected the convergence reconstruction is best where the planar approximation
is most accurate and worse as one moves away from this region. We can also see by eye that the conformal projections (the Mercator and stereographic projections)
perform the best. This is due to fact that local angles are preserved by the
projection. What is also clear is that for many of the  projections the
$B$-mode convergence error can be large in certain regions even in the absence
of noise or systematic errors.

\begin{figure*}
  \subfigure[$\kappa^{\rm KS}$ $E$-mode error]{\includegraphics[width=.49\textwidth]{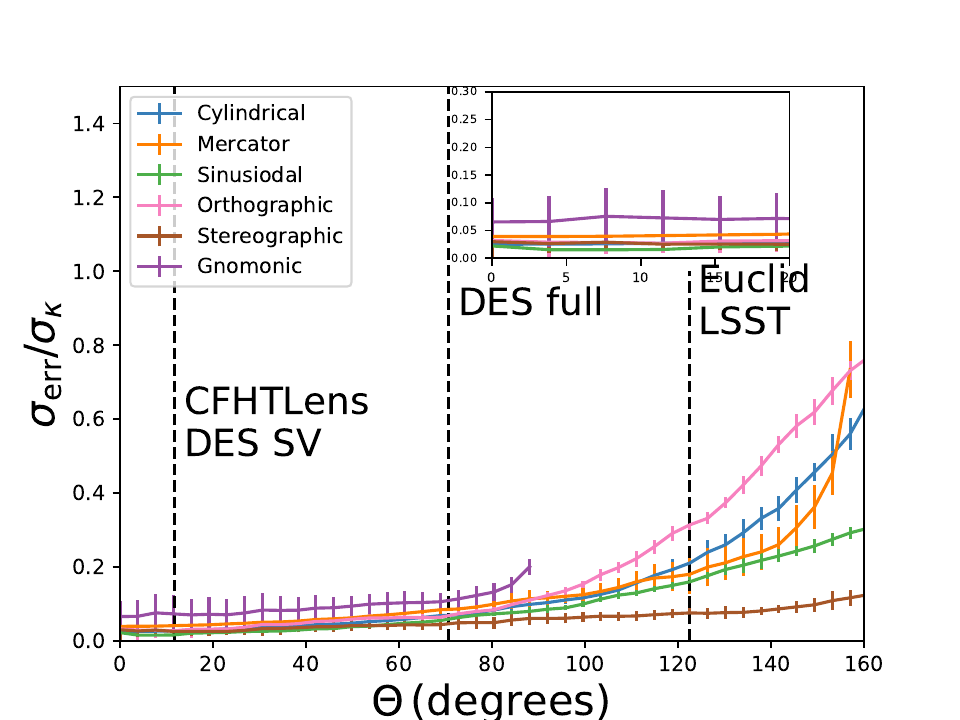}}
  \subfigure[$\kappa^{\rm KS}$ $B$-mode error]{\includegraphics[width=.49\textwidth]{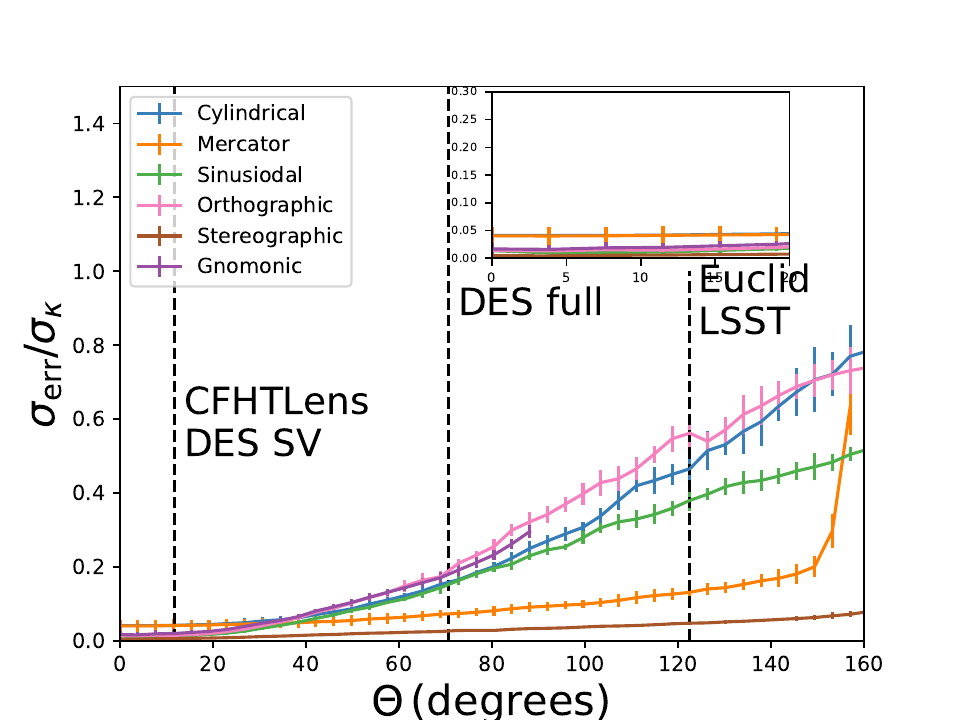}}


  \caption{Relative RMS error of recovered convergence fields (mass-maps) when using
    various planar projections in the standard planar Kaiser-Squires (KS) estimator, as a
    function of angular distance from the centre of the projection. Note that all
        planar projections were significantly zero-padded (to 4 times their original dimension)
        to minimise any contribution to this error from mode mixing at boundaries. Further we note that
        residual boundary effects inevitably exist, thus we clip the figure 20 degrees from any boundary,
        thus restricting the figure to $\Theta$ domains over which it is expected the primary error contribution
        comes from projection effects.  RMS errors are averaged
    over 10 realisations.  Approximate opening angles for the coverages of existing and upcoming
    surveys are overlaid.  For future surveys, such as Euclid and LSST, projection errors can
    be of order tens of percent, exceeding 50\% in some cases.  The conformal projections
    (\ie\ the Mercator and stereographic projections), which preserve local angles, are typically
    superior to the other projections.  In any case, these errors can be avoided entirely by
    recovering convergence fields directly on the celestial sphere.
  }
  \label{fig:angle_error}
\end{figure*}

We can use these simulations to examine the error in the reconstructed
convergence field as a function of angular size. In
\fig{\ref{fig:angle_error}} we show how the accuracy of the recovered
convergence field changes with patch size. We consider a similar simulation
setup as the low resolution experiments described above but now simulate the
convergence field up to a band limit $\elmax=4000$, using the same power
spectrum and smoothing kernel as before. We set a higher band-limit to
eliminate all pixelisation effects (a lower band-limit was sufficient for the previous numerical experiments which were used for visualisation purposes only).
We over-sample on the plane too, again to eliminate all pixelisation effects. For the polar projections we use a square map of $2000\times 2000$ pixels,
capturing the same hemisphere as before. For the equatorial projections we use maps of
size $(2\elmax-1) \times \elmax$ pixels for the cylindrical projection, $(2\elmax-1)\times 5901$ pixels for the Mercator
projection and $(2\elmax-1)\times \elmax$ pixels for the sinusoidal projection. The number of pixels is different of the Mercator projection as it stretches the $\theta$
direction in projection. The equatorial projections, as before,
have the entire sphere projected onto the plane except for the Mercator projection where we
project to $7\pi/16$ radians above and below the equator only as the poles are at infinity in this projection. The exact planar sampling resolutions are not important as we
are intentionally over-sampling to eliminate pixelisation effects.

In a similar way to the other simulations we simulate the convergence and
shear on the sphere,   project the shear on the plane, and recover the
convergence on the plane to compare this to the projected simulated
convergence. We then  calculate the root-mean-square (RMS) error of
$\frac{1}{N}\sum_i^N(\kappa^{\rm KS} -\kappa^{\rm input})^2$ at  different
angular distances from the most accurate region of each projection, where $N$
is the number of  pixels in the region and $\kappa^{\rm input}$ is the input
convergence. The exact angular distances considered for each projection are
defined in \appn{\ref{sec:appendix:projections}}. We calculate the error in annuli
  of constant angular distances away from the centre, defined by the angular metric. The error in the recovered
  convergence will be a result of not only the projection distortion but also a sub-dominant contribution from the
  leakage due to the boundary created by the projection. The leakage due to boundary effects will
  be minimal for small and intermediate scales but will become more significant for the largest scales considered -- \textit{i.e.} as the annuli approach the boundary region.
  Both projection and boundary effects are intrinsic to the projection when using KS inversion and are therefore included here.
  To minimise the contribution of such boundary effects we zero-pad planar projections to 4 times
  their original dimensions, and restrict any analysis to annuli separated by at least 20 degrees from any boundaries.

\fig{\ref{fig:angle_error}} shows the RMS error, averaged over 10 realisations,
at different angular distances for the various projections considered.  We normalise the RMS error with the RMS of the fluctuations in that region to give a relative error.  Relative error for both the $E$- and $B$-modes fields are shown.  Approximate opening angles for the coverages of existing and upcoming surveys are overlaid on \fig{\ref{fig:angle_error}}.  For future surveys, such as Euclid and LSST, projection errors can be of order tens of percent, exceeding 50\% in some cases.  The conformal projections (\ie\ the Mercator and
stereographic projections), which preserve local angles, are typically superior to the other projections.  In any case, these errors can be avoided entirely by recovering convergence fields directly on the celestial sphere.

\section{Application to DES SV data}\label{sec:DES_data}

In this section we apply the mass-mapping techniques presented in \sectn{\ref{sec:mass_mapping}} to the DES
science verification (SV) data, which are publicly available.\footnote{\url{https://des.ncsa.illinois.edu/releases/sva1/doc/shear}}
We use the galaxy shapes estimated by the \textsc{im3shape} method that lie in the range $60^\degrees< {\rm RA} <95^\degrees$
and $-70^\degrees < {\rm dec} <-40^\degrees$, where ${\rm RA}$ and ${\rm dec}$ are the right ascension
and declination in degrees.  We apply the ${\rm \tt sva1\_flag}=0$ selection to the DES SV catalog in order to select galaxies that have a shape that is measured and calibrated
ready to be used for weak lensing studies. These cuts leave $793,743$ galaxies, with a density of 1.4 galaxies per square arcmin.
We pixelise the data by binning into pixels in various settings.  We always pixelise the galaxy in the space that the convergence map is generated; for example,
when a map is made on the sphere the galaxies are pixelated on the sphere directly. In all cases we
apply the recommended weights and corrections to account for multiplicative and additive biases, as described by \citet{becker:2015}.

We create two spherical maps of the reduced shear using the \sshtcode\ and
\healpix\ sampling schemes, considering resolutions to best match the
$\delta\theta = 5$ arcmin pixels considered by \citet{vikram:2015}, which
corresponds to setting an appropriate bandlimit $\elmax$ for the \sshtcode\
sampling scheme and an appropriate $\nside$ resolution parameter for \healpix.
Explicitly, for \sshtcode, we find $\elmax=\pi/\delta\theta=2160$. For
\healpix, we set $N_{\rm side}$ such that the  area of a pixel is as close as
possible to that of a 5  arcmin pixel, \ie\ $A=4\pi/12N_{\rm side}^2 \approx
  (\delta\theta)^2$, yielding $N_{\rm side}=512$ (with the restriction that
\nside\ is a power of two).  The resulting \sshtcode\ map  has pixels of
size 5 arcmin at the equator, while the resulting \healpix\ map has  pixels of
size $7$ arcmin. For the \healpix\ sampled data we use a maximum multipole
$\ell_{\rm max} = 4N_{\rm side}$. The exact choice of $\ell_{\rm max}$ is not critical
as smoothing removes the power on small scales. We smooth the reduced shear
before reconstructing the mass-map with a Gaussian Kernel $G_\ell = e^{-\ell^2\sigma^2}$,
with $\sigma$ such that the half width at half maxima is 20 arcmin, to best match that of \citet{vikram:2015}.

It is academic to note that interpolation errors are effectively unavoidable when
mapping observations continuous in position onto a finite grid. Furthermore, gridding onto
different sampling schemes inherently introduces different interpolation error. One may  wonder,
quite reasonably, which sampling (or corrective measure) minimizes this interpolation error,
however this is beyond the scope of this paper. To normalise for this effect within this analysis
we first grid onto \healpix\ map which we then convert into a \sshtcode\ sampled map with
the aforementioned dimensions. In this way both maps begin with the same information
contaminated with the same interpolation error.

Further one should note that the noise properties of interpolated spherical maps depend fundamentally
on the sampling scheme adopted. When one considers \healpix\ equal area sampling each pixel
contains roughly the same number of observations, whereas for \sshtcode\ equiangular sampling
pixels have significant variation in the number of observations (due to variability in pixel size).
As such, the assumption of noise Gaussianity is more easily justified for \healpix\ maps.

\begin{figure*}

  \subfigure[$\kappa^{\rm SKS}$ $E$-mode (\sshtcode)]{\includegraphics[trim=40 5 42 35,clip=true,width=.49\textwidth]{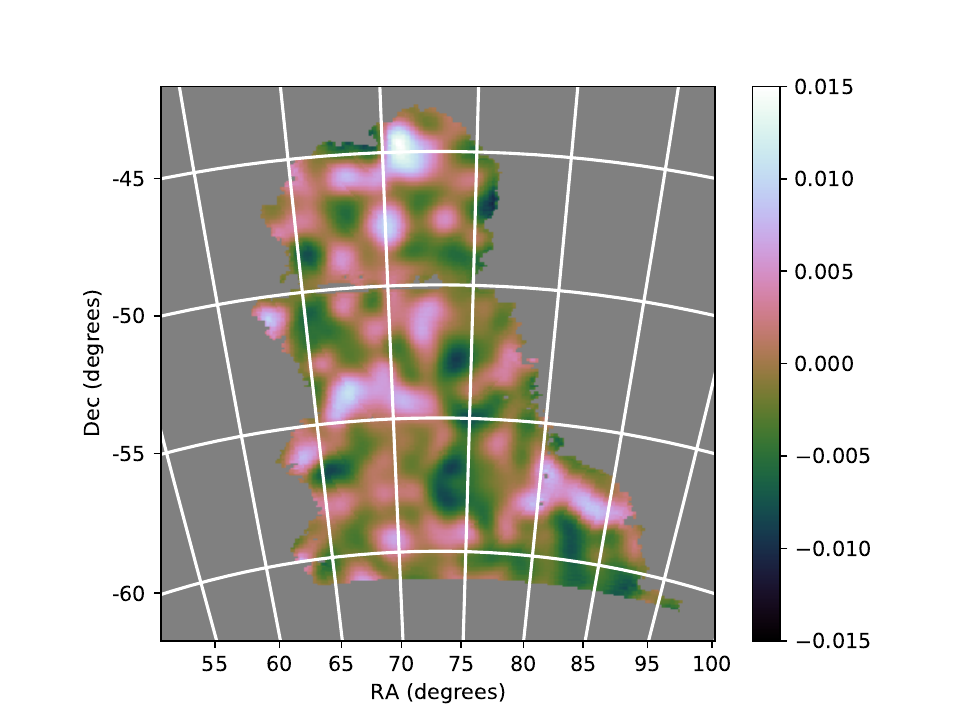}\label{fig:DES_sphere_mw_E}}
  \subfigure[$\kappa^{\rm SKS}$ $B$-mode (\sshtcode)]{\includegraphics[trim=40 5 42 35,clip=true,width=.49\textwidth]{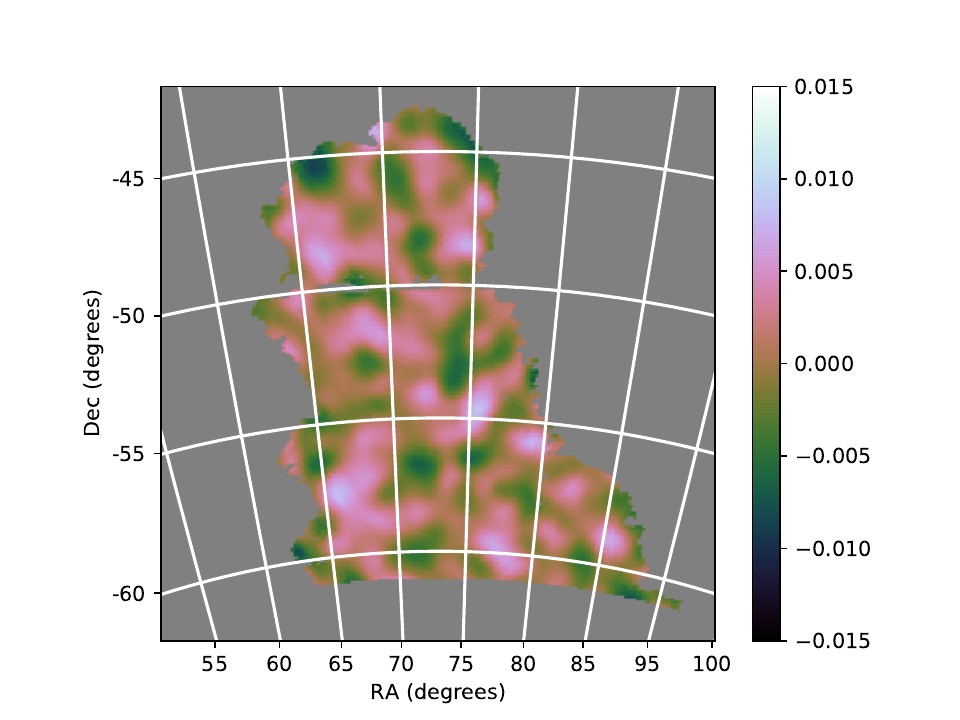}\label{fig:DES_sphere_mw_B}}
  \subfigure[$\kappa^{\rm SKS}$ $E$-mode (\healpix)]   {\includegraphics[trim=40 5 42 35,clip=true,width=.49\textwidth]{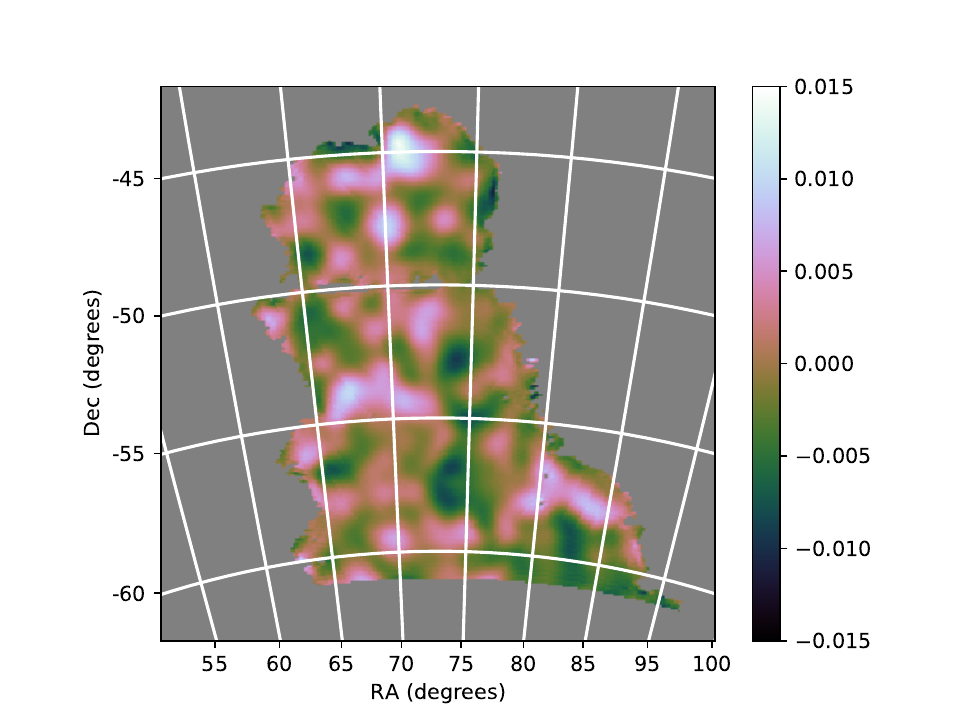}}
  \subfigure[$\kappa^{\rm SKS}$ $B$-mode (\healpix)]   {\includegraphics[trim=40 5 42 35,clip=true,width=.49\textwidth]{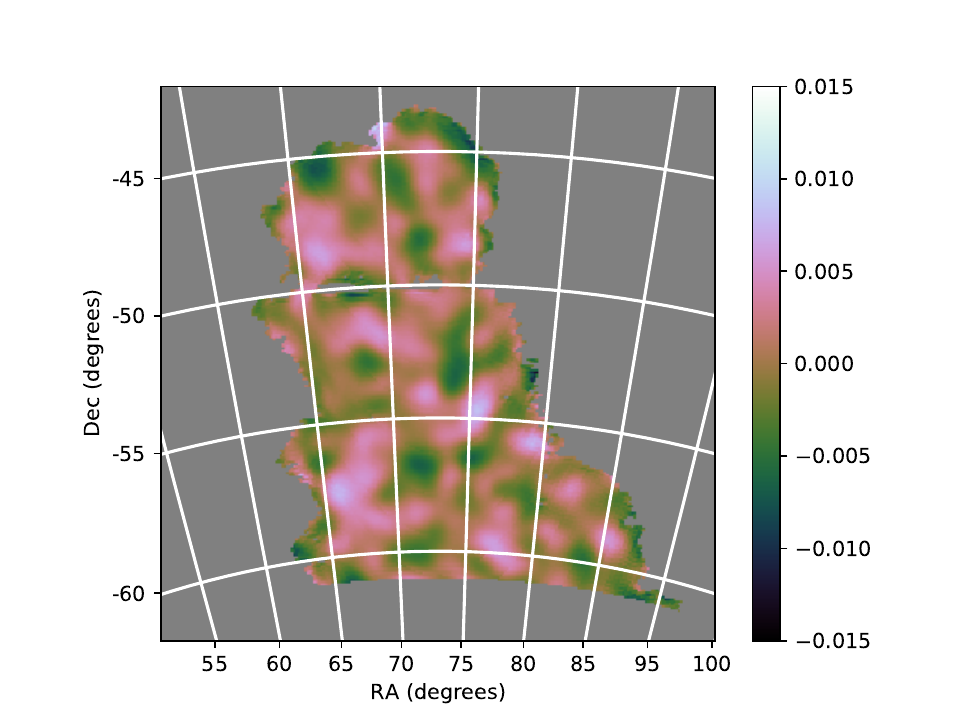}}


  \caption{Spherical convergence maps recovered by the spherical Kaiser-Squires (SKS) estimator applied to spherical maps of the reduced shear created using galaxies from DES SV data.
    The top two plots show stereographic projections of the convergence map recovered on the celestial sphere using the \sshtcode\ sampling, while the
    bottom two plots show Stereographic projections of the convergence maps recovered on the celestial sphere using \healpix\ sampling. The left column shows the recovered $E$-mode convergence, while the right shows the recovered $B$-mode convergence.
      To generate these maps from the DES observation catalogue we first grid onto a \healpix\ sampling scheme, then convert this to a \sshtcode\ sampling scheme
        through harmonic space, thus both reconstructions are working with the same information 
        which mitigates any discrepancies due to the initial catalogue projection. }
  \label{fig:DES_sphere}
\end{figure*}

\fig{\ref{fig:DES_sphere}} shows the $E$- and $B$-mode convergence maps
recovered from the DES SV data using the spherical Kaiser-Squires (SKS)
estimators.  We apply the  iterative algorithm described in
\sectn{\ref{sec:mass_mapping:reduced_shear}} to estimate the underlying shear
from the observed reduced shear. The recovered convergence maps show near perfect
agreement with each other and reasonable agreement with the maps recovered by the DES collaboration
for a similar choice of galaxies \citep[][\fig{2}]{vikram:2015}. It should be
noted that the galaxies used here are not the exact same galaxies used in
estimating the convergence maps recovered by \linebreak \citet{vikram:2015}
due to small differences between the  private and public DES catalogs
(C.~Chang \& J.~Zuntz, private communication).  Therefore, exact equivalence
is not excepted, however, through private communication C.~Chang has provided
convergence maps recovered by the DES map making pipeline when using the
public catalog and in this case there is good agreement between the two
convergence maps.

\begin{figure*}

  \subfigure[$\kappa^{\rm KS}$ $E$-mode for sinusoidal projection]{\includegraphics[trim=40 10 58 35,clip=true,width=.49\textwidth]{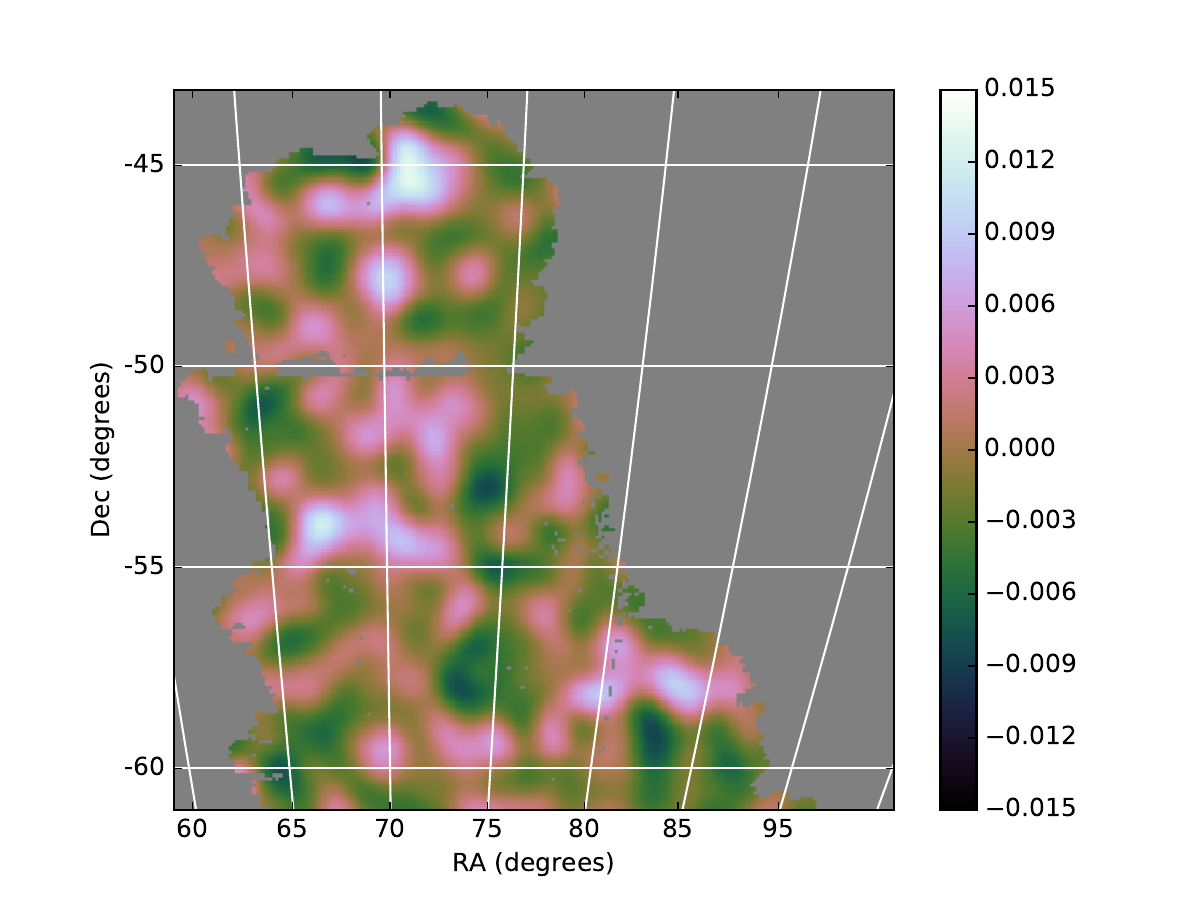}\label{fig:DES_planar:sine_E}}
  \subfigure[$\kappa^{\rm KS}$ $B$-mode  for sinusoidal projection]{\includegraphics[trim=40 10 58 35,clip=true,width=.49\textwidth]{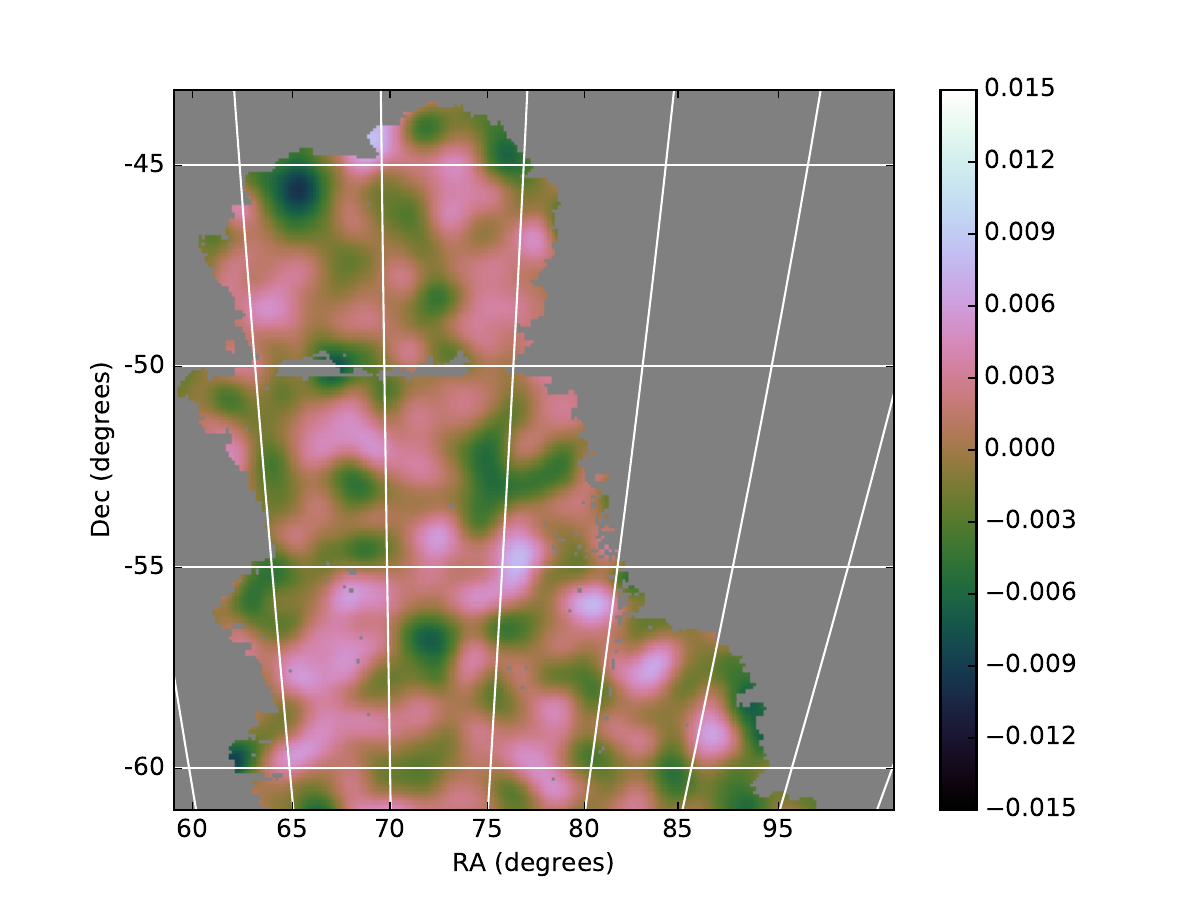}\label{fig:DES_planar:sine_B}}
  \subfigure[$\kappa^{\rm KS}$ $E$-mode  for stereographic projection]{\includegraphics[trim=40 10 58 35,clip=true,width=.49\textwidth]{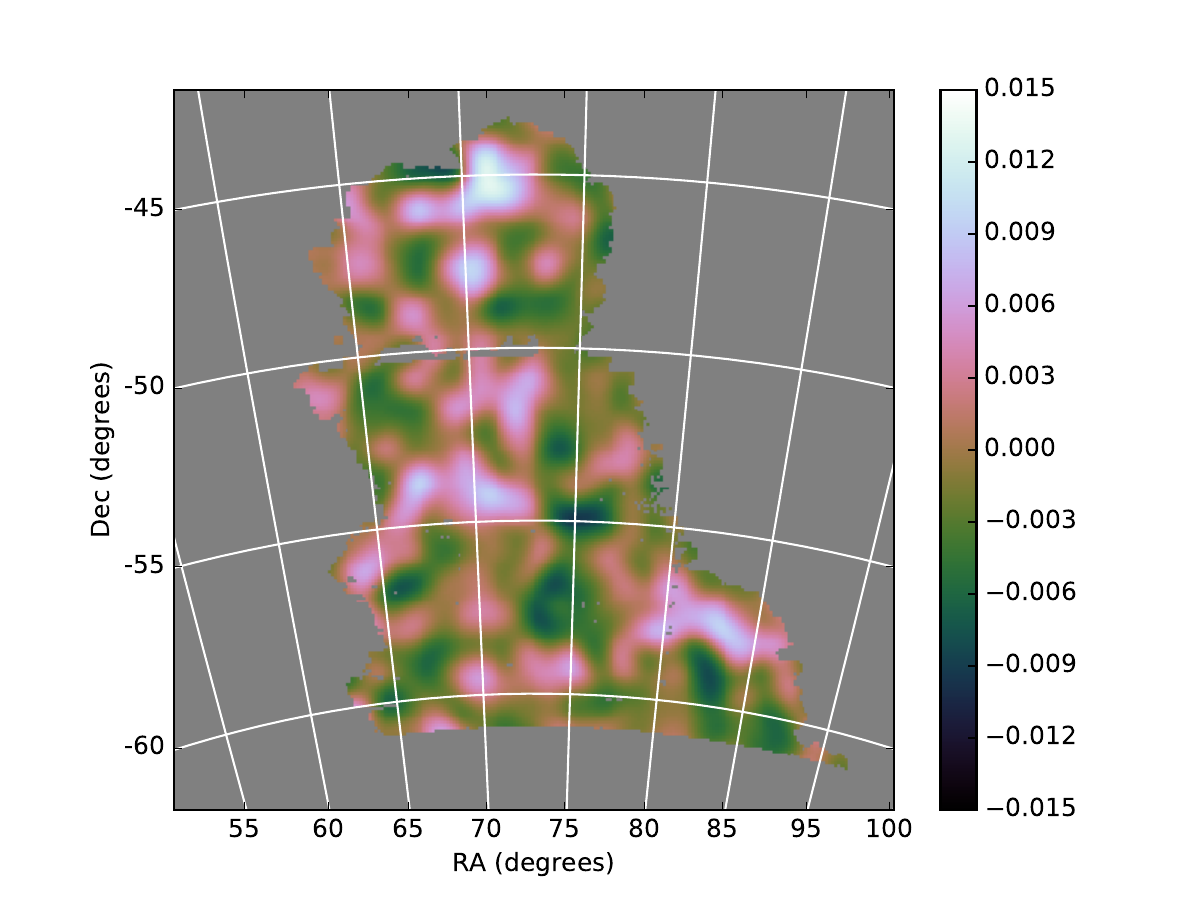}}
  \subfigure[$\kappa^{\rm KS}$ $B$-mode  for stereographic projection]{\includegraphics[trim=40 10 58 35,clip=true,width=.49\textwidth]{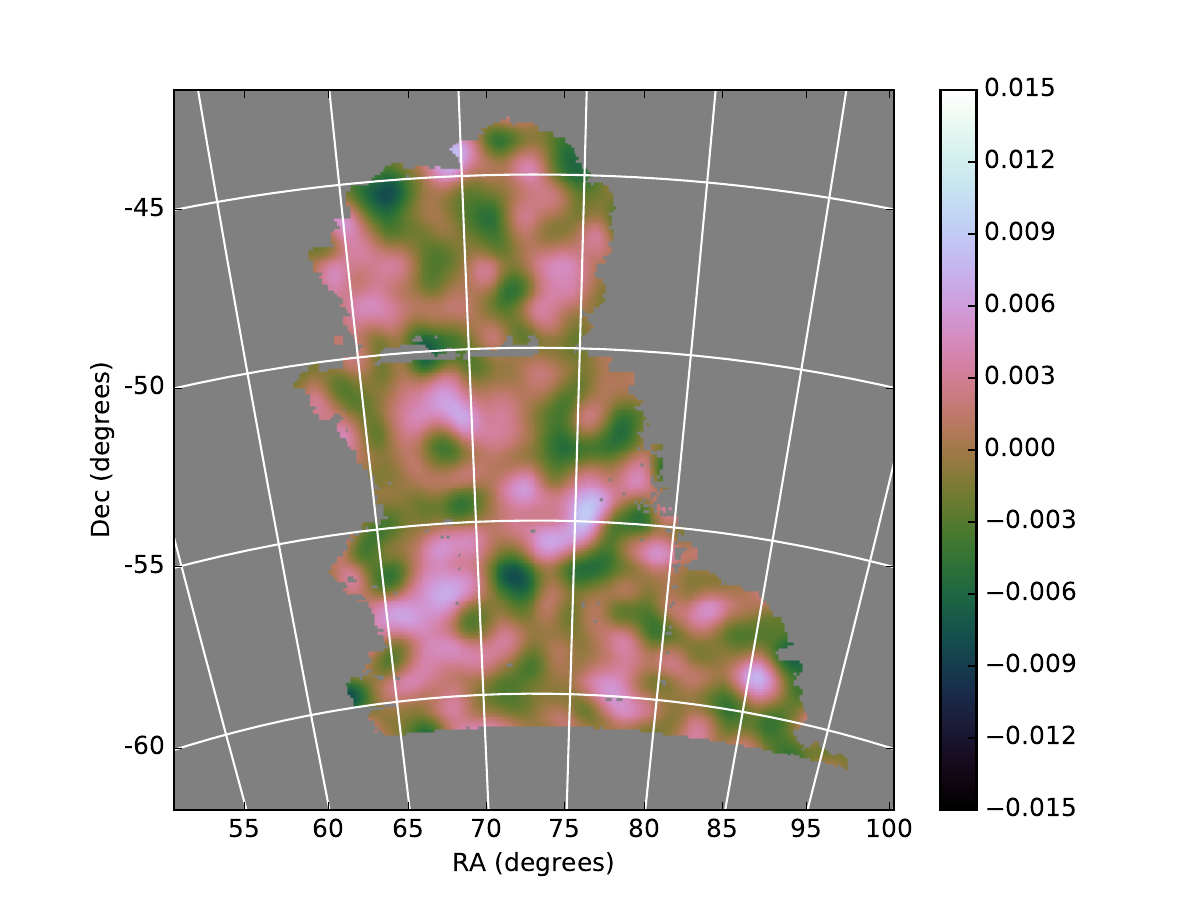}}

  \caption{Planar convergence maps recovered by the planar Kaiser-Squires (KS) estimator applied to planar maps of the reduced shear created using galaxies from the DES SV data.
    The top row of plots show the results where the sinusoidal projection is used, while the bottom row shows the results when the stereographic projection is used. These projections were chosen since the the sinusoidal projection is used by the DES collaboration \citep{vikram:2015}, while the stereographic projection was shown in \fig{\ref{fig:angle_error}} to minimise RMS error. The left column shows the recovered $E$-mode convergence, while the right shows the recovered $B$-mode convergence.}
  \label{fig:DES_planar}
\end{figure*}

\begin{figure*}

  \subfigure[$\kappa^{\rm KS}-\kappa^{\rm SKS}$ $E$-mode for sinusoidal projection]{\includegraphics[trim=40 10 58 35,clip=true,width=.49\textwidth]{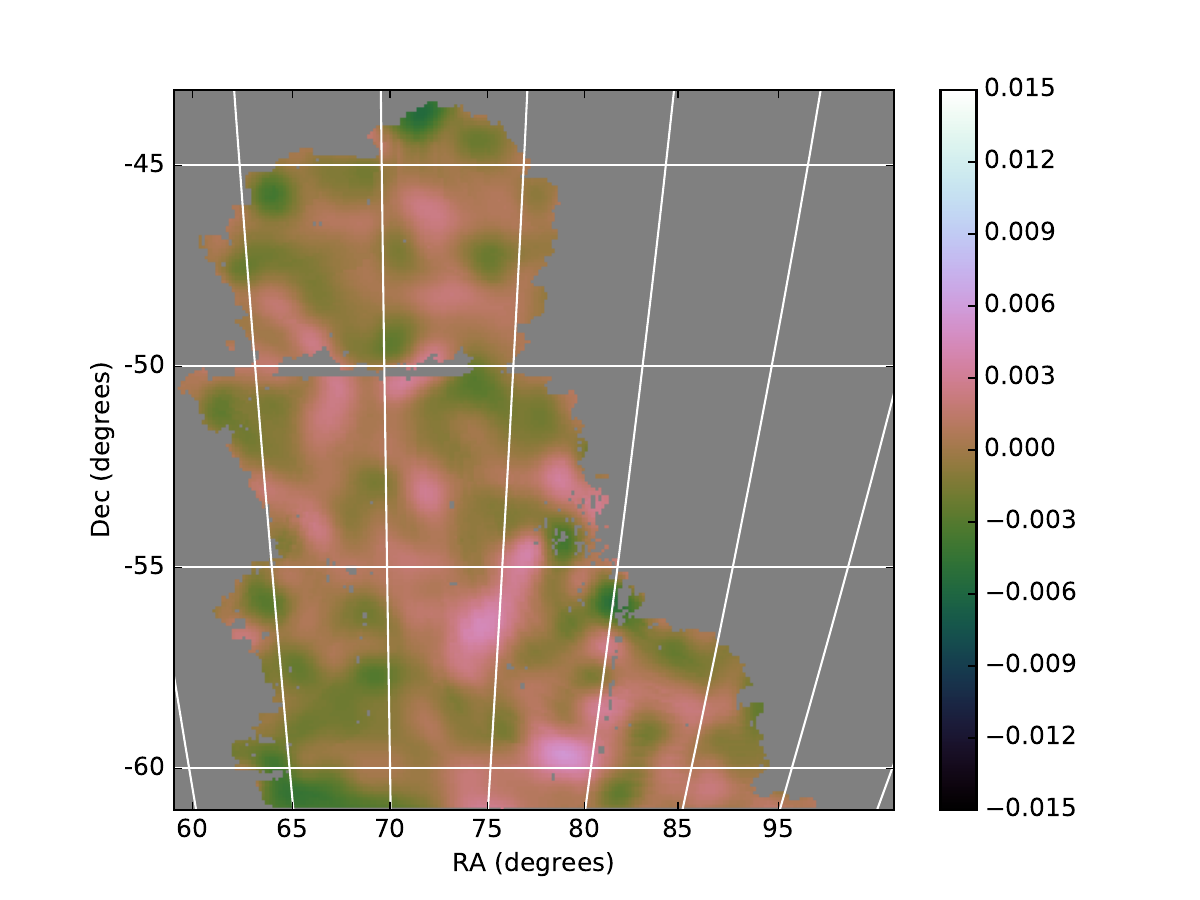}}
  \subfigure[$\kappa^{\rm KS}-\kappa^{\rm SKS}$ $B$-mode for sinusoidal projection]{\includegraphics[trim=40 10 58 35,clip=true,width=.49\textwidth]{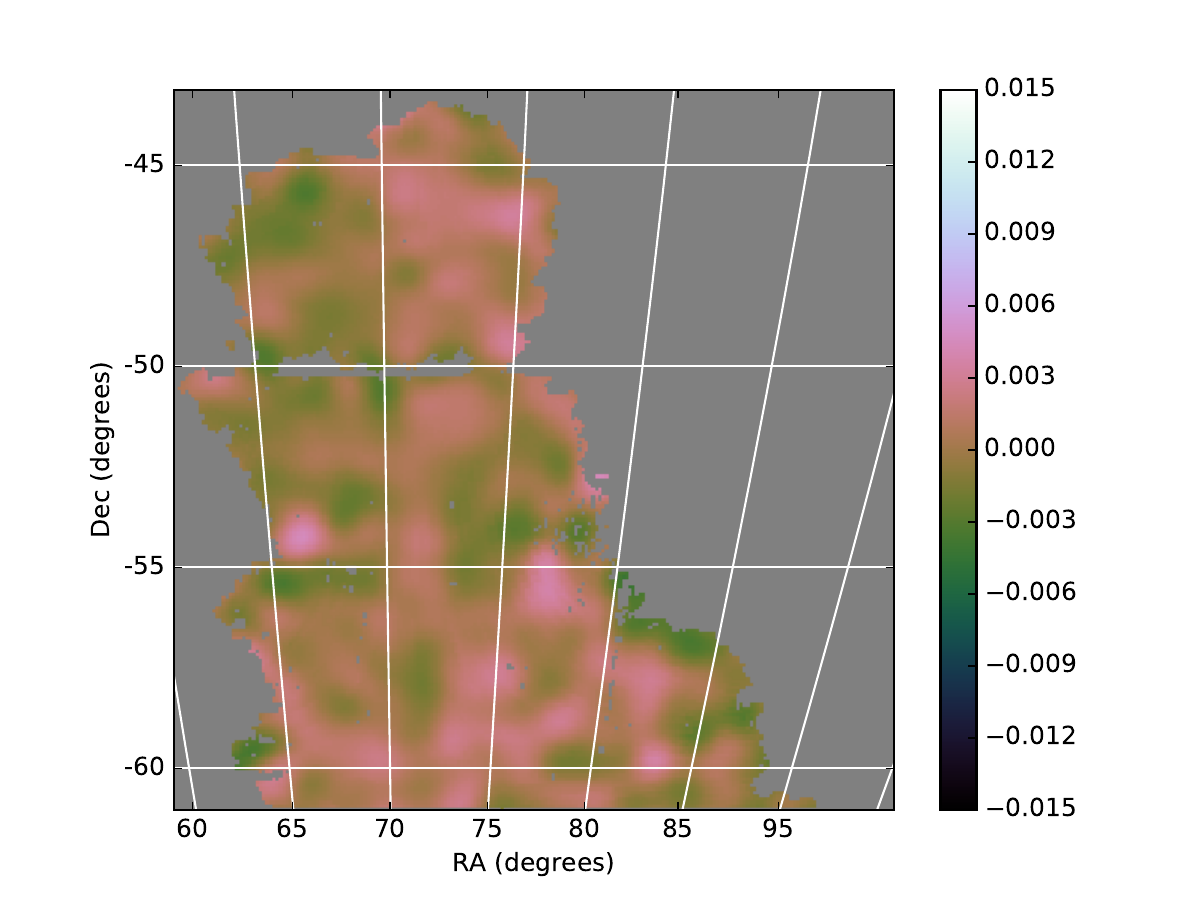}}
  \subfigure[$\kappa^{\rm KS}-\kappa^{\rm SKS}$ $E$-mode for stereographic projection]{\includegraphics[trim=40 10 58 35,clip=true,width=.49\textwidth]{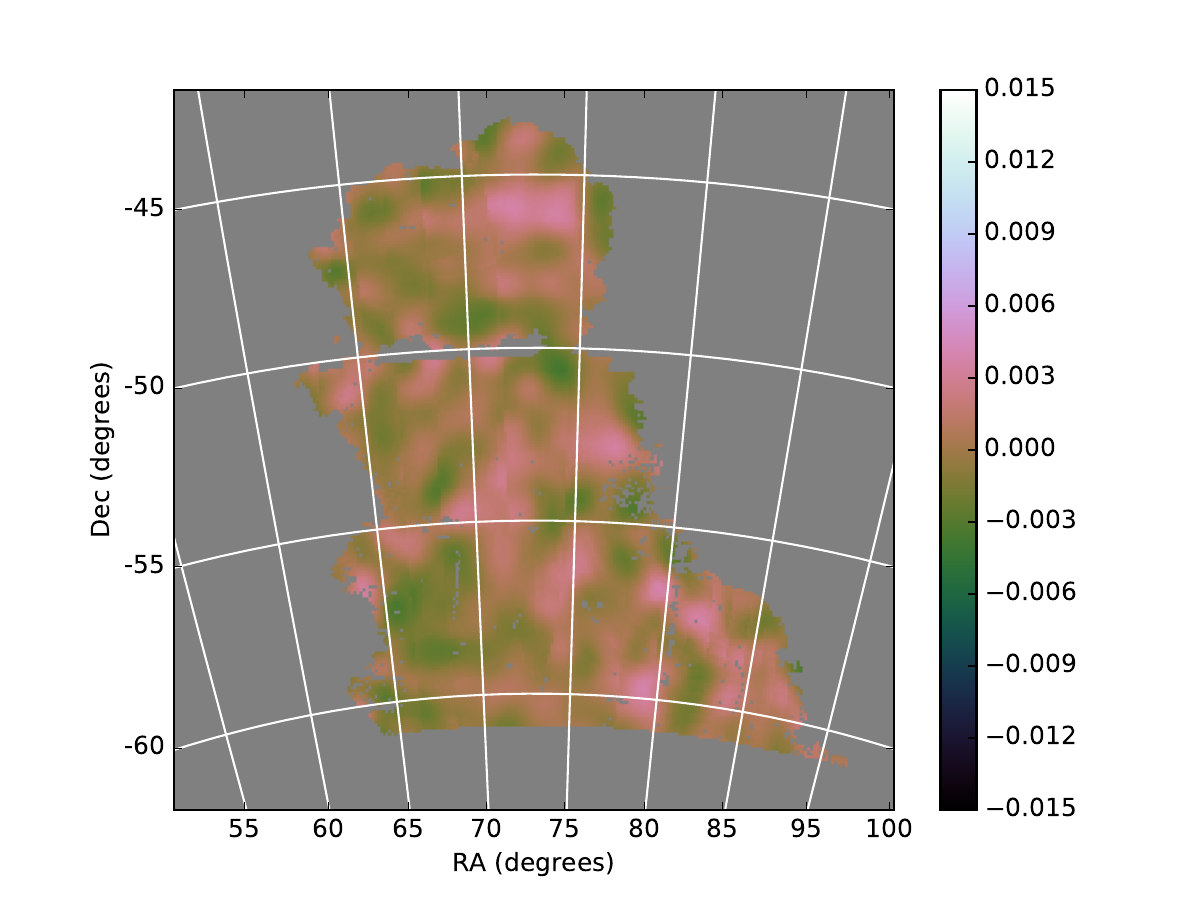}}
  \subfigure[$\kappa^{\rm KS}-\kappa^{\rm SKS}$ $B$-mode for stereographic projection]{\includegraphics[trim=40 10 58 35,clip=true,width=.49\textwidth]{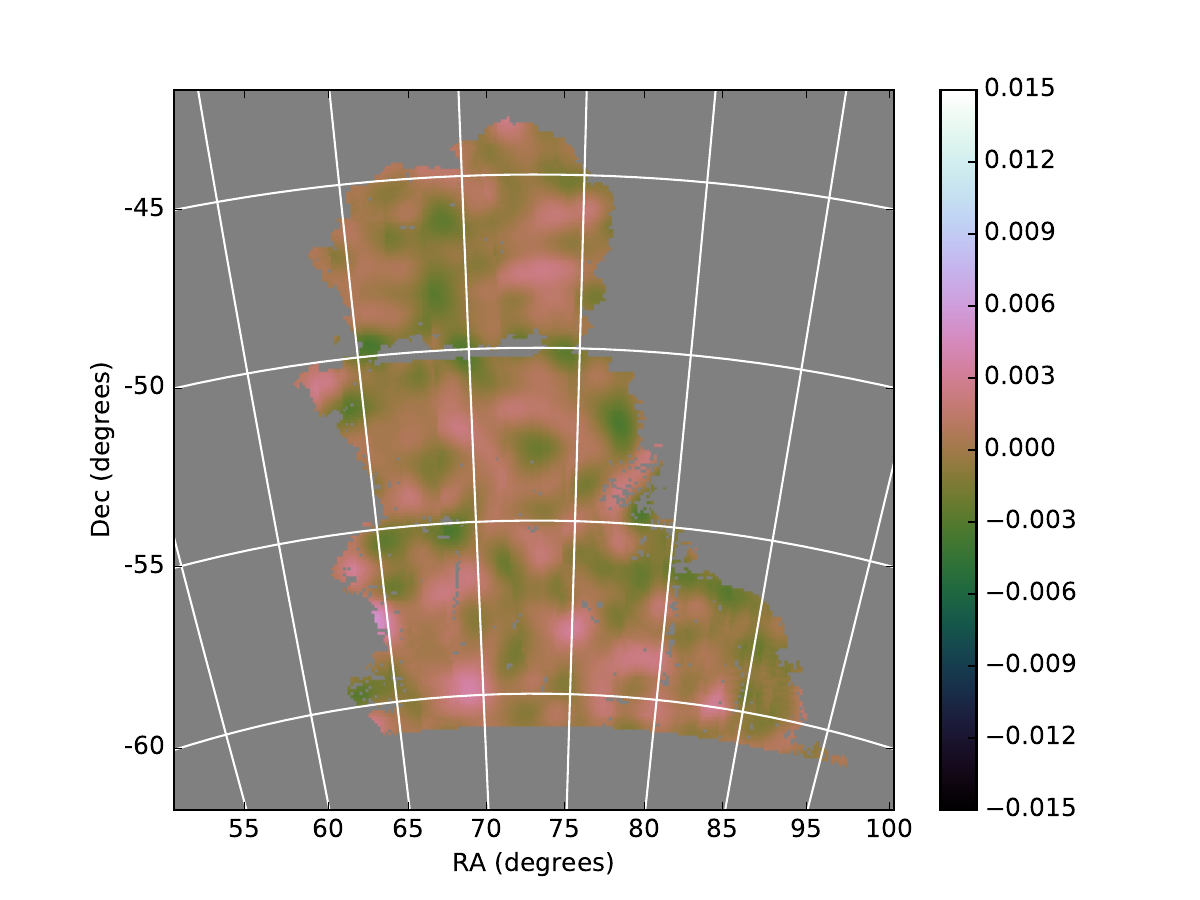}}

  \caption{Similar to \fig{\ref{fig:DES_planar}}, where here we plot the difference between the convergence recovered on the plane by the planar Kaiser-Squires (KS) estimator and the convergence recovered on the sphere by the spherical Kaiser-Squires (SKS) estimator. The purpose of this figure is to compare the planar and spherical results. For the spherical case we consider the \sshtcode\ sampling only, \ie\ differences are relative to \fig{\ref{fig:DES_sphere_mw_E}} and \fig{\ref{fig:DES_sphere_mw_B}}. }
  \label{fig:DES_planar_diff}
\end{figure*}

For comparison purposes, in \fig{\ref{fig:DES_planar}} we show the results when we bin galaxies
onto two planar maps. The top row show the results when using a sinusoidal projection, as also used by the DES collaboration \citep{vikram:2015}. We
rotate the projection such that the central line of the projection corresponds
to ${\rm RA}=70^\degrees$, as also done by \citet{vikram:2015}. No other rotation is
applied to fully centre the region of interest. In the second row we show results using a
stereographic projection that has been rotated by the Euler angles $\alpha =
  159^\degrees$, $\beta  = -37^\degrees$ and $\gamma   = 90^\degrees$, to fully
centre the area of interest to the South pole about which the projection is
then performed. We choose to also show results using the stereographic projection as the results from
\fig{\ref{fig:angle_error}} suggest that this is the best projection to use. In
both cases we use 5 arcmin pixels and apply a 20 arcmin smoothing as they do in \citet{vikram:2015}.
We apply the required local rotations as
described in \appn{\ref{sec:appendix:projections}} (in \appn{\ref{sec:appendix:projections}} we also
examine the effect of not applying such rotations). For these planar results we
also use the reduced shear algorithm described  in
\sectn{\ref{sec:mass_mapping:reduced_shear}}.
\fig{\ref{fig:DES_planar_diff}} shows the difference between the convergence
recovered on the plane for these projections and the projected convergence recovered on the sphere
using the \sshtcode\ sampling shown in \fig{\ref{fig:DES_sphere}}. As is common with the KS estimator, both the planar and the spherical mass-maps suffer
from leakage between the $E$- and $B$-mode due to both the effects of the boundary and, perhaps primarily, the significant complex noise contribution.

\section{Conclusions}
\label{sec:conclusions}

We have described how one can recover convergence fields, or mass-maps,
directly on the celestial sphere, adopting the spherical equivalent of
Kaiser-Squires inversion. We demonstrate that the spherical formulation
reduces to the usual flat-sky Kaiser-Squires approach in the planar
approximation.  We study the accuracy of the planar approximation for mass-mapping
and address the important question of whether one needs to recover the
convergence field on the sphere for forthcoming surveys or whether recovery on
the plane would be sufficient.  The comparison between the planar and
spherical settings depends largely on the projection used. In
\appn{\ref{sec:appendix:projections}} we describe a number of projections that
are used in this work and show how to account for the local rotations required
when projecting spin fields, such as shear, onto the plane.  In
\fig{\ref{fig:angle_error}} the relative error introduced by the planar
approximation, for a variety of projections, is presented.  Conformal
projections, for which local angles are conserved, are found to be the most
effective.  Nevertheless, errors in the planar setting are typically tens of
percent and can exceed 50\% in some cases.  These projection errors can be entirely
eliminated by recovering mass-maps directly on the celestial sphere by the
spherical Kaiser-Squires technique presented in this article.


We apply the spherical Kaiser-Squires estimator to the publicly available DES
SV data. We present maps of the convergence field recovered on the celestial
sphere using both the \sshtcode\ and \healpix\ sampling schemes (see
\fig{\ref{fig:DES_sphere}}), accounting for the fact that one measures reduced
shear, rather than the true underlying shear, by applying the iterative
algorithm discussed above. We compare the results to those recovered on the
plane, using the sinusoidal projection adopted by the DES collaboration and
also the stereographic projection since it was found to be most effective
projection for mass-mapping, particularly for large scales (see \fig{\ref{fig:angle_error}}).  In this
setting we demonstrate reasonable agreement between the spherical and planar
reconstructions. While the coverage area of DES SV data is not sufficiently
large for the planar approximation to induce significant errors (see
\fig{\ref{fig:angle_error}}), recovering spherical mass-maps for DES SV data
is nevertheless a useful demonstration of the spherical Kaiser-Squires
estimator on real observational data.

In this article we consider the most naive estimator of the convergence field on
the celestial sphere, namely a direct spherical harmonic inversion of the equations relating
the observed shear field to the underlying convergence field, \ie\ the generalisation of the
Kaiser-Squires estimator from the plane to the sphere.  In practice, the shear field is not
observed over the entire celestial sphere, which induces leakage in the recovered convergence
field for the simple harmonic estimator considered.  In future work we will apply the pure
mode wavelet estimators developed by \citet{leistedt:ebsep} to remove leakage when
recovering spherical mass-maps.  In addition, in future work we also intend to develop
methods to better mitigate the impact of noise and to estimate the statistical uncertainties
associated with recovered mass-maps \citep[see \textit{e.g.}][]{price:2020}.  In all of these
extensions, however, it is clear that for future surveys like Euclid and LSST it will be essential
to recover mass-maps on the celestial sphere, to avoid the significant errors than are otherwise
induced by planar approximations.

\section*{Acknowledgements}
We thank
the DES team for making their shear data public, and we thank C. Chang and J. Zuntz for assistance in
the manipulation of these catalogues. We thank C. Chang for the private communications including providing us with
convergence maps created using the publicly available shear catalogs and DES map making pipeline.
We thank P. Paykari and Z. Vallis for useful discussions.
This work was supported by the Science and Technology Facilities Council (STFC) through a Euclid Science Support Grant and an LSST:UK Phase A grant, the Engineering and
Physical Sciences Research Council (EPSRC; grant number EP/M011852/1), and the Leverhume Trust. TDK
acknowledges support of a Royal Society University Research Fellowship.

\section{Data Availability}
All data, both observational and simulated, utilized throughout 
this paper is publicly available and can be found alongside the open-source 
code-base massmappy\footnote{\url{https://github.com/astro-informatics/massmappy}} 
developed during this work.


\appendix


\section{Equivalence of different representations of spherical mass-mapping inverse problem}
\label{sec:appendix:harmonic_inverse}

The equivalence of the harmonic and integral expressions, \eqn{\ref{eq:kappa_shear_harmonic}} and \eqn{\ref{eqn:shear_conv_integral}} respectively, connecting the observable cosmic shear field to the convergence field can also be shown by considering the harmonic representation of the integral expression.
Consider the integral representation, decomposing the kernel and convergence field into their harmonic expansions:
\begin{align}
  \wlshear(\sa)
   & =
  \int_\sphere \dmu{\sa\p} \:
  (\rotarg{\sa\p} \wlkernel)(\sa) \: \wlconv(\sa\p) \\
   & =
  \int_\sphere \dmu{\sa\p} \:
  \sum_{\el\m}
  \shc{\wlkernel}{\el}{\m} \:
  \bigl(\rotarg{\sa\p} \sshf{\el}{\m}{2}\bigr)(\sa) \:
  \sum_{\el\p\m\p}
  \sshc{\hat{\kappa}}{\el\p}{\m\p}{0} \:
  \sshfarg{\el\p}{\m\p}{\sa\p}{0}
  \spcend .
  \label{eqn:shear_conv_integral_part}
\end{align}
The rotation of the spin spherical harmonic in the above expression is given by
\begin{align}
  \bigl(\rotarg{\sa\p} \sshf{\el}{0}{2}\bigr)(\sa)
   & =
  \sum_{\n}
  \dmatbig_{\n 0}^{\el}(\sa\p) \:
  \sshf{\el}{\n}{2}(\sa) \\
   & =
  \sqrt{\frac{4\pi}{2\el+1}}
  \sum_{\n}
  \sshfc{\el}{\n}{0}(\sa\p) \:
  \sshf{\el}{\n}{2}(\sa)
  \spcend ,
\end{align}
where it is necessary to only consider $\m=0$ due to the Kronecker delta term $\kron{\m}{0}$ appearing in $\shc{\wlkernel}{\el}{\m}$, as shown in \eqn{\ref{eqn:kernel_harmonic}}, and noting \eqn{\ref{eqn:spherical_harmonic_rotation}} and \eqn{\ref{eqn:ssh_wigner}}.
\eqn{\ref{eqn:shear_conv_integral_part}} can then be written as
\begin{align}
  \wlshear(\sa)
   & =
  \sum_{\el\n}
  \frac{-1}{\el(\el+1)}
  \sqrt{\frac{(\el+2)!}{(\el-2)!}} \:
  \sshc{\hat{\kappa}}{\el}{\n}{0} \:
  \sshf{\el}{\n}{2}(\sa)
  \spcend ,
\end{align}
where we have noted the orthogonality of the spherical harmonics, \ie\ $\innerp{\shf{\el}{\m}}{\shf{\el\p}{\m\p}} = \kron{\el}{\el\p} \kron{\m}{\m\p}$.  The resulting harmonic representation of \eqn{\ref{eqn:shear_conv_integral}} is thus identical to \eqn{\ref{eq:kappa_shear_harmonic}}, as expected.


\section{Projections} \label{sec:appendix:projections}

In this appendix we outline the details of each projection considered. We
firstly define each projection and describe its properties.   Each projection
has different beneficial properties, for example whether the projection is
equal-area, has appropriate boundary conditions or conformal. Conformal projections conserve local angles and are
often used for geographical maps. We also describe the distance metric we use
for each projection to define the opening angle of the patch of sky seen by an
experiment, \ie\ the angle considered in \fig{\ref{fig:angle_error}}. We then
detail how to calculate the local rotation angles required when projecting
spin fields, such as shear (without this rotation $E$- and $B$-modes will be
misinterpreted) and finally illustrate the impact of neglecting this local
rotation on DES SV data.

\subsection{Projection definitions}

We consider two general types of projection: equatorial and polar projections.
Equatorial projections are defined relative to the equator, while polar
projections are defined relative to a pole.  The precise definitions of the
different equatorial and polar projections are given in the following
subsections. The equatorial projections considered include: the sinusoidal
projection, which is a simple equal area projection that was used by the DES
collaboration; the Mercator projection that is a conformal projection, often
used in geographical maps as it preserves local angles; and a simple
cylindrical projection. The polar projections considered include: the
orthographic projection, which is a simple vertical projection from the sphere
to a tangent plane; the Gnomonic projection that has the useful property that
the local rotations are trivial to calculate; and the stereographic projection
that is another conformal projection.

\subsubsection{Equatorial projections}

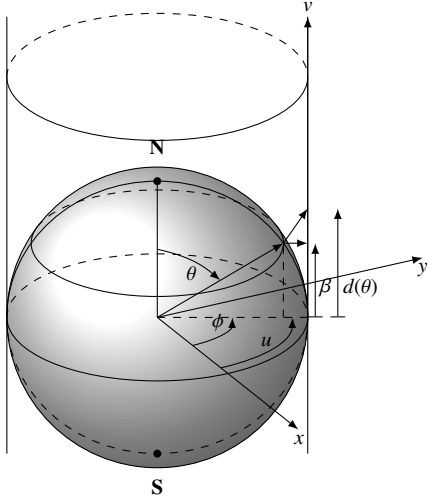
\begin{figure}
\begin{center}
\begin{tikzpicture} 

\def\R{2} 
\def\angEl{25} 
\def\angAz{-60} 
\def\angPhiOne{-50}
\def\angPhiTwo{-35} 
\def\angBeta{33} 
\def\cosBeta{0.838}
\def\sinBeta{0.544639}
\def\angX{-62}
\def\angY{\angX+90}

\pgfmathsetmacro\H{\R*cos(\angEl)} 
\LongitudePlane[xzplane]{\angEl}{\angAz}
\LongitudePlane[pzplane]{\angEl}{0}
\LongitudePlane[qzplane]{\angEl}{\angPhiTwo}
\LatitudePlane[equator]{\angEl}{0}

\fill[ball color=white] (0,0) circle (\R); 
\draw (0,0) circle (\R);


\coordinate (O) at (0,0);
\coordinate[mark coordinate] (N) at (0,\H);
\coordinate[mark coordinate] (S) at (0,-\H);
\path[pzplane] (\R,0) coordinate (XE);
\path[pzplane] (\angBeta:\R) coordinate (P);
\path[pzplane] (\R,0) coordinate (PE);
\path[qzplane] (\angBeta:\R) coordinate (Q);
\path[qzplane] (0,\R) coordinate (QE);

\DrawLongitudeCircle[\R]{-0}
\DrawLatitudeCircle[\R]{\angBeta}
\DrawLatitudeCircle[\R]{0} 

\node at (0,1.6*\R) { \tikz{\DrawLatitudeCircle[\R]{0}} };

\draw (-\R,-\H) -- (-\R,2*\R) (\R,-\H) -- (\R,2*\R);
\draw[->] (XE) -- +(0,2*\R) node[above] {$v$};
\node[above=8pt] at (N) {$\mathbf{N}$};
\node[below=8pt] at (S) {$\mathbf{S}$};
\draw[->] (O) -- (P);
\draw (O) -- (QE);
\draw[pzplane,->,thin] (90:0.5*\R) to[bend left=15]
    node[midway,below] {$\theta$} (\angBeta:0.5*\R);
\draw[equator,->,thin] (\angX:0.5*\R) to[bend right=30]
    node[pos=0.4,above] {$\phi$} (0:0.5*\R);
\draw[equator,->,thin] (\angX:0.9*\R) to[bend right=30]
    node[pos=0.4,above] {$u$} (0:0.9*\R);
\draw[equator,->](0,0) -- (\angX:2*\R) node[below] {$x$};
\draw[equator,->](0,0) -- (\angY:2*\R) node[below] {$y$};
\draw[equator,dashed](0,0) -- (0:\R); 
\draw[pzplane,dashed](\cosBeta*\R,0) -- (\angBeta:\R);
\draw[pzplane,->](\angBeta:\R) -- (\R,\sinBeta*\R);
\draw[pzplane,|->](\R+0.1,0) -- node[pos=0.4, right] {$\beta$} (\R+0.1,\sinBeta*\R);
\draw[pzplane,->](\angBeta:\R) -- (\R,0.8*\R);
\draw[pzplane,|->](\R+0.4,0) -- node[pos=0.4*\sinBeta/0.8, right] {$d(\theta)$} (\R+0.4,0.8*\R);

\end{tikzpicture}
\end{center}
\caption{Diagram to describe graphically the equatorial projections, including the sinusoidal, Mercator and the simple cylindrical projections. These can all be seen as types of cylindrical projections since the sphere is projected onto a cylinder wrapped round the sphere. The $u$ variable simply describes how far round the cylinder a point is and is therefore give by $\phi$ (up to some arbitrary shift), except in the sinusoidal case where the $u$ variable is contracted away from the equator to ensure the projection is equal-area.
The $v$ variable can vary between projections and can be specified by various functions $d(\theta)$. In the Mercator projection this function is chosen to ensure the
projection is conformal.  In the sinusoidal and simple cylindrical projections this function is simply $d(\theta)=\beta=\pi/2-\theta$. }
\label{fig:equatorial_projection_diag} 
\end{figure}

\Fig{\ref{fig:equatorial_projection_diag}} shows graphically how the equatorial projections can be viewed as a projection
onto a cylinder wrapped round the sphere.
Each projection is defined by the relation between the spherical coordinates $(\theta, \phi)$ and the planar coordinates $(u,v)$.

The sinusoidal projection (used by the DES collaboration)
is defined by
\bea
\begin{split}
  u & = (\phi-\pi)\sin(\theta)\spcend,\\
  v & = \theta.
\end{split}
\eea
This projection results in minimal distortion in the central region $(\theta=\pi/2$, $\phi=\pi)$. Moving away from this point in any direction increases the distortion but particularly in a diagonal direction (specifically along the lines $y=x$ or $y=-x$). We define the distance metric for this projection by
\bea
\Theta = \sqrt{(\theta-\pi/2)^2 + (\phi-\pi)^2}
\spcend.
\eea
The sinusoidal projection has the useful property of being equal-area. It is simpler to define than the Mollweide projection, also an equal-area projection, which is
commonly used for plotting in the cosmological community.

The Mercator projection is commonly used for geographical maps and is defined by
\bea
\begin{split}
  u & = \phi-\pi \spcend,\\
  v & = \ln \left[\tan(\pi/2-\theta/2)\right]
  \spcend.
\end{split}
\eea
This projection has the useful property of being conformal, meaning that  local angles on the sphere will not be distorted. The projection introduces minimal distortion at the equator, while the projected image is stretched and distorted as one moves towards the pole.  Since the poles themselves are at infinity the projection cannot completely cover the full sky in practice. The projection is a cylindrical projection and therefore has the correct boundary conditions in the $u$ direction.
The metric used to define the angular distance from the undistorted region is simply given by
\bea
\label{eqn:distance_metric_mercator}
\Theta = \vert \theta-\pi/2 \vert \spcend.
\eea

The final equatorial projection we consider is the simple cylindrical projection defined by
\bea
\begin{split}
  u & = \phi-\pi \spcend ,\\
  v & = \theta-\pi/2\spcend .
\end{split}
\eea
There are no particular properties to inspire us to propose this projection over the more sophisticated cylindrical projection of the Mercator projection. Its attractiveness is in
its simplicity and the ability to map the entire sphere on one plane. The distortions increase away from the equator leading to the same distance metric as the Mercator projection, \ie\ \eqn{\ref{eqn:distance_metric_mercator}}.

\subsubsection{Polar projections}

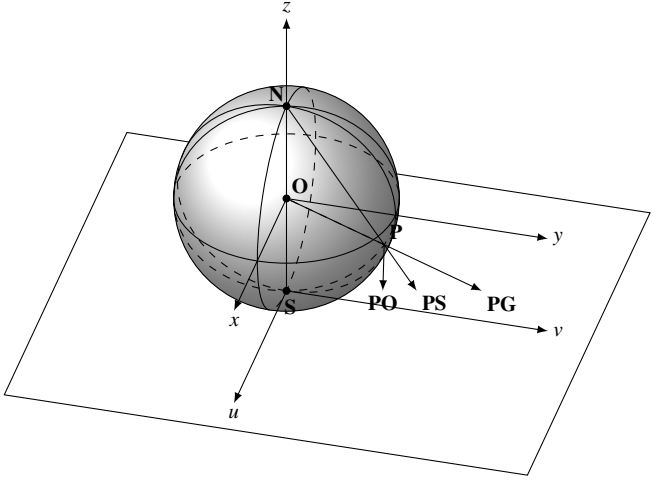
\begin{figure}

\begin{tikzpicture}

\def\R{1.5} 
\def\angEl{35} 
\def\angAz{-105} 
\def\angPhi{0} 
\def\angBeta{-30}

\pgfmathsetmacro\H{\R*cos(\angEl)} 
\tikzset{xyplane/.style={cm={cos(\angAz),sin(\angAz)*sin(\angEl),-sin(\angAz),
                              cos(\angAz)*sin(\angEl),(0,-\H)}}}
\tikzset{origxyplane/.style={cm={cos(\angAz),sin(\angAz)*sin(\angEl),-sin(\angAz),
                              cos(\angAz)*sin(\angEl),(0,0)}}}
\LongitudePlane[xzplane]{\angEl}{\angAz}
\LongitudePlane[pzplane]{\angEl}{\angPhi}
\LatitudePlane[equator]{\angEl}{0}

\draw[xyplane] (-2*\R,-2*\R) rectangle (2.2*\R,2.8*\R);
\fill[ball color=white] (0,0) circle (\R); 
\draw (0,0) circle (\R);

\coordinate[mark coordinate] (O) at (0,0);
\coordinate[mark coordinate] (N) at (0,\H);
\coordinate[mark coordinate] (S) at (0,-\H);
\coordinate (U) at (0,\angBeta*\R);
\path[pzplane] (\angBeta:\R) coordinate (P);
\path[pzplane] (\R,0) coordinate (PE);
\path[xzplane] (\R,0) coordinate (XE);
\path (PE) ++(0,-\H) coordinate (Paux);
\coordinate (PO) at (intersection cs: first line={(U)--(P)},
                                        second line={(S)--(Paux)});
\path (PE) ++(0,-\H) coordinate (Paux); 
\coordinate (PS) at (intersection cs: first line={(N)--(P)},
                                        second line={(S)--(Paux)});
\path (PE) ++(0,-\H) coordinate (Paux); 
\coordinate (PG) at (intersection cs: first line={(O)--(P)},
                                        second line={(S)--(Paux)});

\DrawLatitudeCircle[\R]{0} 
\DrawLongitudeCircle[\R]{\angAz} 
\DrawLongitudeCircle[\R]{\angAz+90} 
\DrawLongitudeCircle[\R]{\angPhi}

\draw[xyplane,<->] (1.8*\R,0) node[below] {$u$} -- (0,0) -- (0,2.4*\R)
    node[right] {$v$};
\draw[origxyplane,<->] (1.8*\R,0) node[below] {$x$} -- (0,0) -- (0,2.4*\R)
    node[right] {$y$};
\draw[->] (0,-\H) -- (0,1.6*\R) node[above] {$z$};

\draw (P) -- (N) +(0.3ex,0.6ex); 
\draw[dashed] (P) -- (O) +(0.3ex,\H) node[above left] {$\mathbf{N}$};
\draw[->] (P) -- (PO) node[below] {$\mathbf{{PO}}$};
\draw[->] (P) -- (PG) node[below right] {$\mathbf{{PG}}$};
\draw[->] (P) -- (PS) node[below right] {$\mathbf{{PS}}$};
\path[->] (S) +(0.4ex,-0.4ex) node[below] {$\mathbf{S}$};
\draw (O) node[above right] {$\mathbf{O}$} -- (P) node[above right] {$\mathbf{P}$};

\end{tikzpicture}

\caption{Diagram to describe graphically the polar projections, including the orthographic, stereographic and gnomic projections.  In these projections a point on the sphere is projected to the tangent plane at a pole (here chosen to be the South pole).
For projections around the North or South pole the angle $\phi$ is simply taken as the polar coordinate $\varphi$ in the planar space. The radial coordinate $\varrho$ is a function of the angle between the point and the pole whos tangent plane is considered ($\pi-\theta$ for the South pole). The orthographic projection is a vertical projection, giving $\varrho = \sin(\pi-\theta)$ for the tangent plane at the South pole. The gnomic projection casts a ray from the origin to the point on the sphere and through to the tangent plane, giving $\varrho = \tan(\pi-\theta)$ for the tangent plane at the South pole. Finally, the stereographic projection casts a ray from the opposite pole to the point on the sphere and 
through to the tangent plane, giving $\varrho=2\tan[(\pi-\theta)/2]$ for the tangent plane at the South pole.  In the diagram, the point P is projected to PO, PS, and PG by the orthographic, stereographic and gnomic projections, respectively.}
\label{fig:polar_projections_diag}
\end{figure}

\Fig{\ref{fig:polar_projections_diag}} shows a graphical representation of the polar projections, where again the spherical coordinates $(\theta, \phi)$ are projected onto the planar coordinates $(u,v)$.
It is most straightforward to define these projections using polar coordinates on the plane $(\varrho,\varphi)$, which are related to the Cartesian coordinates by
\bea
\begin{split}
  u & = \varrho\cos(\varphi)\spcend,\\
  v & = \varrho\sin(\varphi)\spcend.
  \label{eqn:uv_polar}
\end{split}
\eea
In each of the polar projections we simply have that $\varphi=\phi$. The projections differ in the way \saa\ is mapped to $\varrho$, where each projection has its own mapping function $f$, \ie
\bea
\varrho = f(\theta) \spcend.
\eea
It is a common feature of these projections that the entire sphere cannot be projected to a single plane in practice (since in many cases the opposite pole is mapped to the point at infinity). In that case we often project around the South pole as well as the North pole and consider $\varrho=f(\pi-\theta)$. We define the distance metric for these projections to be
\bea
\Theta = \theta \spcend.
\eea

The orthographic projection is defined by
\bea
\varrho = \sin(\theta) \spcend.
\eea
For this projection a point on the sphere is mapped vertically from the sphere to the tangent plane at the North pole. As a result the whole sphere cannot be projected onto one plane in practice and one must project each hemisphere onto a different plane.

We also consider the Gnomonic projection defined by casting a ray from the centre of the sphere to the point considered and then though to the tangent plane at the North pole. The Gnomonic projection is therefore defined by
\bea
\varrho = \tan(\theta) \spcend.
\eea
For this projection the whole sphere again cannot be projected onto one plane in practice since the equator is projected to infinity. One must again project the sphere into a number of regions, for example considering each hemisphere separately.

The final projection we consider is the stereographic projection. This is defined by casting a ray from the South pole to the point considered on the sphere and then through to the tangent plane at the North pole.  The resulting projection is defined by
\bea
\varrho = 2\tan(\theta/2) \spcend.
\eea
We can project almost all of the sphere with this projection, except near the South pole, as the South pole is mapped to infinity. This projection is conformal, preserving local angles.

\subsection{Rotation angles}\label{sec:appendix:projections:rotations}

Spin fields on the sphere have local directions defined relative to the North pole, whereas on the plane the spin fields have their spin defined relative to some universal direction (usually the ``top" of the planar map). We define this direction on the plane by $\vect{\hat{v}}$, the unit vector in the $v$ direction. On projection, the spin field must be rotated from its original coordinate frame on the sphere to the new coordinate frame on the plane. Here we describe how to calculate this local rotation angle.

When we project from the sphere to the plane it is common to rotate our coordinate system before we project. This is done in order to centre the region of interest so that distortions due to the projection are minimised at this point.  We therefore need to define a number of coordinate systems, including the original sphere, the rotated sphere and the plane.  Firstly, consider a field defined on the original sphere with spherical coordinates $(\theta^\prime, \phi^\prime)$ and corresponding Cartesian coordinates $(x^\prime, y^\prime, z^\prime)$.  Consider then the rotated field, where the spherical coordinates of the  rotated sphere are $(\theta, \phi)$, with corresponding Cartesian coordinates $(x, y, z)$. We define the rotation relating the primed frame to the unprimed frame by $\rotarg{\eul}$, with corresponding 3D rotation matrix $\rotmat$. From the rotated sphere the field is then projected onto the plane defined by Cartesian coordinates $(u,v)$ and polar coordinates $(\varrho,\varphi)$.

We need to find the angle between $\vect{\hat{v}}$ and the projected direction of the North pole of the original sphere. To do this we consider an infinitesimal step North on the sphere and then
find the infinitesimal step this makes on the plane $(\dx u,\dx v)$. The rotation angle $\psi$ required is then the angle between the $\vect{\hat{v}}$ direction and the projected North direction.

\subsubsection{Equatorial projections}

The first step is to construct a vector that is an infinitesimal step North in the original space. This vector is given by
\bea
\vect{{\rm d}x}^\prime =  \left( \begin{array}{c}
  0 \\
  0 \\
  1\end{array} \right) {\rm d}\epsilon \spcend,
\eea
where $\dx \epsilon$ is an infinitesimal element of the real line.
When this infinitesimal element is projected onto the sphere at any point it always points North (with the exception of the poles).  Moving in this direction thus yields a vector that is further North but is not normalised to lie on the unit sphere.
The normalisation of the vector is unimportant as later on in this proof we require the direction of this vector only and not its length. In the unprimed frame this infinitesimal step is given by
\bea
\begin{split}
  \vect{{\rm d}x} & =  \rotmat\vect{{\rm d}x}^\prime \spcend ,\\
  \left( \begin{array}{c}
      {\rm d}x \\
      {\rm d}y \\
      {\rm d}z\end{array} \right) & = \left( \begin{array}{c}
    \rotmat_{1,3} \\
    \rotmat_{2,3} \\
    \rotmat_{3,3}\end{array} \right) {\rm d}\epsilon
  \spcend.
\end{split}
\eea
Now we apply the chain rule twice to calculate the projected infinitesimal step in the plane $(\dx u,\dx v)$.  Firstly we note the relation  between $(x, y, z)$ and $(\theta, \phi)$ of
\bea
\begin{split}
  \theta & = \arctan\left(\frac{\sqrt{x^2+y^2}}{z}\right) \spcend,\\
  \phi & = \arctan\left(\frac{y}{x}\right) \spcend,
\end{split}
\eea
where the normalisation of the vector is unimportant, ensuring the definition of $\vect{{\rm d}x}^\prime$ is acceptable. Applying the chain rule we have
\bea
\begin{split}
  {\rm d}\theta & = \cos(\theta)[\cos(\phi){\rm d}x+\sin(\phi){\rm d}y - \tan(\theta){\rm d}z] \spcend,\\
  {\rm d}\phi & = \csc(\theta)[-\sin(\phi){\rm d}x + \cos(\phi){\rm d}y] \spcend,
  \label{eq:appendix:dphi}
\end{split}
\eea
where a unit vector is assumed without loss of generality.
We now generalise the equatorial projections as
\bea
\begin{split}
  u & = g(\theta,\phi) \spcend,\\
  v & = h(\theta,\phi) \spcend.
\end{split}
\eea
We then apply the chain rule again to give
\bea
\begin{split}
  {\rm d}u & = \frac{\partial g}{\partial \theta}{\rm d}\theta + \frac{\partial g}{\partial \phi}{\rm d}\phi \spcend,\\
  {\rm d}v & = \frac{\partial h}{\partial \theta}{\rm d}\theta + \frac{\partial h}{\partial \phi}{\rm d}\phi \spcend,\\
\end{split}
\eea
from which the rotation angle $\psi$ can be calculated by
\begin{equation}
  \psi  = -\arctan\left(\frac{{\rm d}u}{{\rm d}v}\right) \spcend.
  \label{eq:appendix:psi}
\end{equation}
After substituting all the terms from the above expressions into \eqn{\ref{eq:appendix:psi}}, ${\rm d}\epsilon$ cancels out and the limit ${\rm d}\epsilon\to 0$ follows trivially.

\subsubsection{Polar projections}

For polar projections the calculation begins in the same way as for equatorial projections, up until \eqn{\ref{eq:appendix:dphi}}. Then we apply the chain rule giving
\bea
\begin{split}
  {\rm d}\varrho & = \frac{{\rm d}f}{{\rm d}\theta}{\rm d}\theta \spcend,\\
  {\rm d}\varphi & = {\rm d}\phi \spcend.
\end{split}
\eea
Applying the chain rule again to the relation between $(u,v)$ and $(\varrho,\varphi)$ of \eqn{\ref{eqn:uv_polar}} we have
\bea
\begin{split}
  {\rm d}u & = \cos(\varphi){\rm d}\varrho - \varrho\sin(\varphi){\rm d}\varphi \spcend,\\
  {\rm d}v & = \sin(\varphi){\rm d}\varrho + \varrho\cos(\varphi){\rm d}\varphi \spcend.
\end{split}
\eea
We then compute the local rotation angle $\psi$ in the same manner as above, \ie\ by \eqn{\ref{eq:appendix:psi}}. It is possible to show from this result the special property of the Gnomonic projection: when there is no rotation and $f(\theta)=\tan(\theta)$, as is the case for the Gnomonic projection, the rotation angle is zero everywhere.

\subsection{Application to DES SV data}

\begin{figure*}

  \subfigure[$\kappa^{\rm KS}$ $E$-mode with no local rotations]{\includegraphics[trim=40 10 58 35,clip=true,width=.49\textwidthtrim=40 10 58 35,clip=true,width=.49\textwidth]{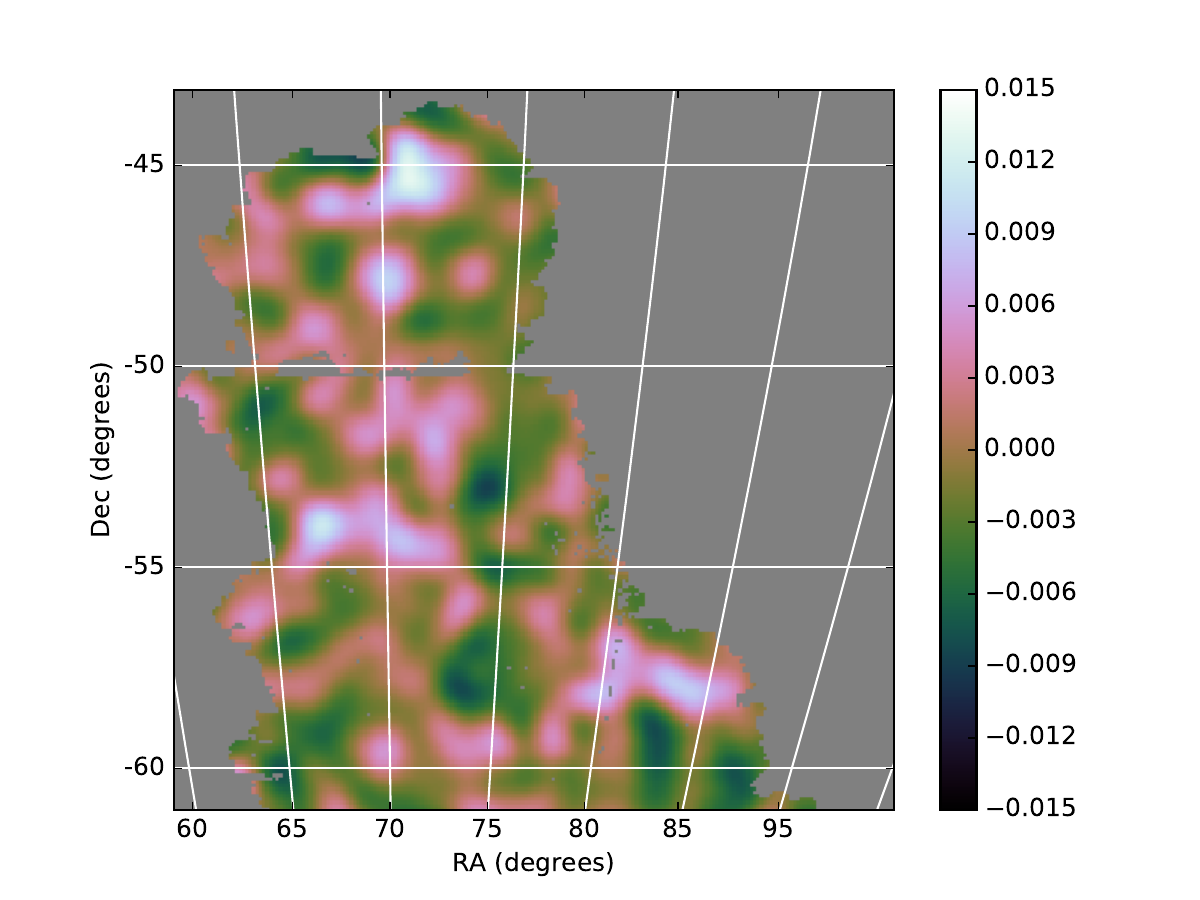}\label{fig:DES_no_rotation:E}}
  \subfigure[$\kappa^{\rm KS}$ $B$-mode with no local rotations]{\includegraphics[trim=40 10 58 35,clip=true,width=.49\textwidthtrim=40 10 58 35,clip=true,width=.49\textwidth]{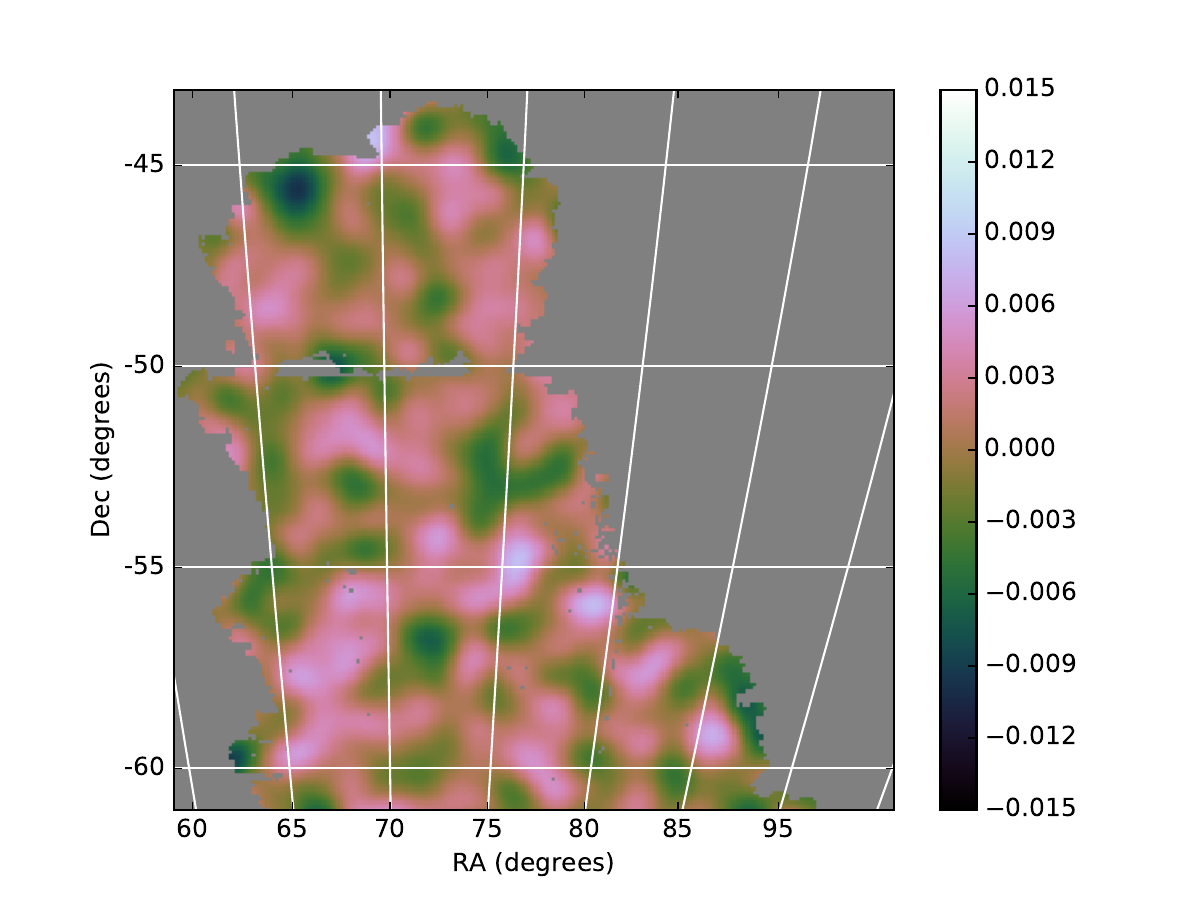}\label{fig:DES_no_rotation:B}}
  \subfigure[Difference between $\kappa^{\rm KS}$ $E$-mode with and without local rotations]{\includegraphics[trim=40 10 58 35,clip=true,width=.49\textwidthtrim=40 10 58 35,clip=true,width=.49\textwidth]{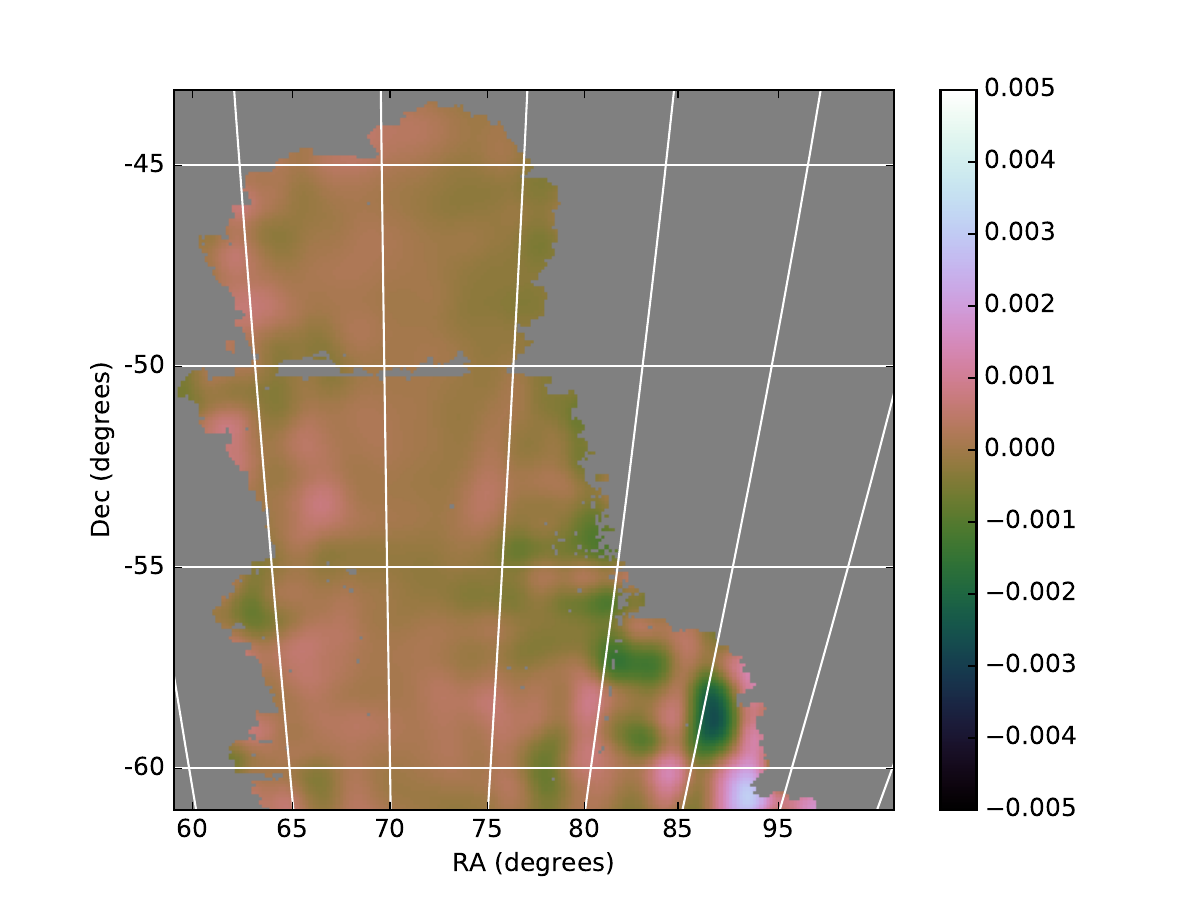}\label{fig:DES_no_rotation:E_dif}}
  \subfigure[Difference between $\kappa^{\rm KS}$ $B$-mode with and without local rotations]{\includegraphics[trim=40 10 58 35,clip=true,width=.49\textwidthtrim=40 10 58 35,clip=true,width=.49\textwidth]{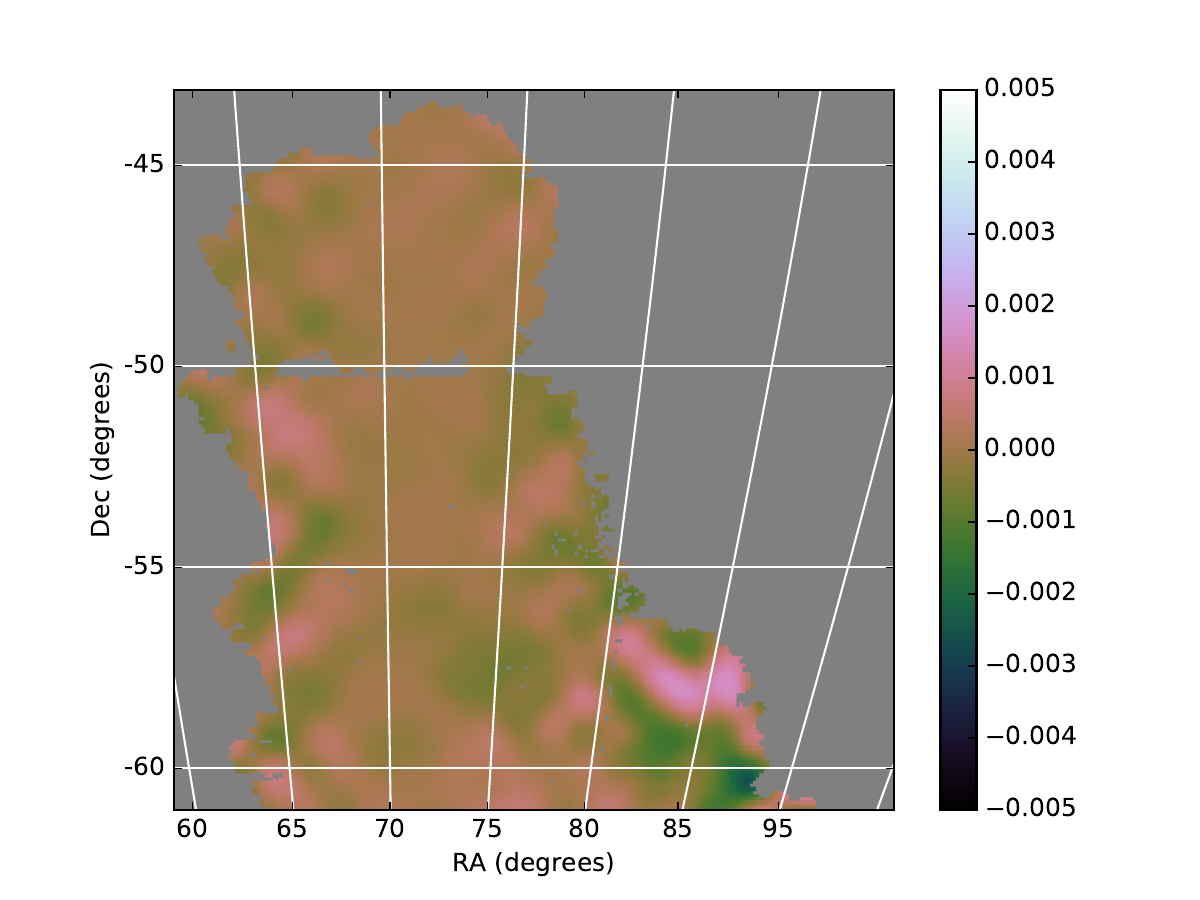}\label{fig:DES_no_rotation:B_dif}}

  \caption{Plot to show the importance of applying the local rotations to real data when performing projections. We project the DES SV data using the sinusoidal projection considered by the DES collaboration. However, in
    this case we do not apply the necessary rotations to the galaxy shapes. Panel (a) and (b) show the results when no rotation is applied, while panels
    (c) and (d) show the error introduced by not applying the local rotations, \ie\ the differences with the maps shown in \fig{\ref{fig:DES_planar:sine_E}}
    and \fig{\ref{fig:DES_planar:sine_B}}.
  }
  \label{fig:DES_no_rotation}
\end{figure*}

Here we demonstrate the importance of applying this rotation in practice, using DES SV data.  As far as we are aware applying these local rotations is not standard practise. We consider the sinusoidal projection also used by the DES collaboration. However, here we do not apply the necessary rotations to the galaxy shapes (as we did in the main body of the article). \fig{\ref{fig:DES_no_rotation}(a)} and \fig{\ref{fig:DES_no_rotation}(b)} show the results when no rotation is applied
and \fig{\ref{fig:DES_no_rotation}(c)} and \fig{\ref{fig:DES_no_rotation}(d)} show the error introduced by not applying the local rotations, \ie\ the differences with the maps shown in \fig{\ref{fig:DES_planar:sine_E}}
and \fig{\ref{fig:DES_planar:sine_B}}. While the effect is not large for DES SV data, it is not insignificant.  Furthermore, if considering planar mass-mapping techniques for larger survey coverages this effect becomes increasingly important.

\begin{thebibliography}{78}
\providecommand{\natexlab}[1]{#1}
\providecommand{\url}[1]{\texttt{#1}}
\providecommand{\urlprefix}{URL }
\providecommand{\eprint}[2][]{\url{#2}}

\bibitem[{{Aghanim} et~al.(2003){Aghanim}, {Kunz}, {Castro} \&
  {Forni}}]{aghanim:2003}
{Aghanim} N., {Kunz} M., {Castro} P.G., {Forni} O., 2003, A\&A, 406, 797,
  \href{http://arXiv.org/abs/astro-ph/0301220}{{\tt astro-ph/0301220}}

\bibitem[{{Alsing} et~al.(2016){Alsing}, {Heavens}, {Jaffe}, {Kiessling},
  {Wandelt} \& {Hoffmann}}]{alsing:2016}
{Alsing} J., {Heavens} A., {Jaffe} A.H., {Kiessling} A., {Wandelt} B.,
  {Hoffmann} T., 2016, MNRAS, 455, 4452,
  \href{http://arXiv.org/abs/arXiv:1505.07840}{{\tt arXiv:1505.07840}}

\bibitem[{{Bacon} \& {Taylor}(2003)}]{bacon:2003}
{Bacon} D.J., {Taylor} A.N., 2003, MNRAS, 344, 1307,
  \href{http://arXiv.org/abs/astro-ph/0212266}{{\tt astro-ph/0212266}}

\bibitem[{{Bartelmann} \& {Schneider}(2001)}]{bartelmann:2001}
{Bartelmann} M., {Schneider} P., 2001, Phys.\ Rep, 340, 291,
  \href{http://arXiv.org/abs/astro-ph/9912508}{{\tt astro-ph/9912508}}

\bibitem[{{Becker} et~al.(2016)}]{becker:2015}
{Becker} M.R., et~al., 2016, PRD, 94, 2, 022002,
  \href{http://arXiv.org/abs/arXiv:1507.05598}{{\tt arXiv:1507.05598}}

\bibitem[{{Brouwer} et~al.(2016)}]{brouwer:2016}
{Brouwer} M.M., et~al., 2016, MNRAS, 462, 4451,
  \href{http://arXiv.org/abs/arXiv:1604.07233}{{\tt arXiv:1604.07233}}

\bibitem[{{Bunn} et~al.(2003){Bunn}, {Zaldarriaga}, {Tegmark} \& {de
  Oliveira-Costa}}]{bunn:2003}
{Bunn} E.F., {Zaldarriaga} M., {Tegmark} M., {de Oliveira-Costa} A., 2003, PRD,
  67, 2, 023501, \href{http://arXiv.org/abs/astro-ph/0207338}{{\tt
  astro-ph/0207338}}

\bibitem[{{Castro} et~al.(2005){Castro}, {Heavens} \& {Kitching}}]{castro:2005}
{Castro} P.G., {Heavens} A.F., {Kitching} T.D., 2005, PRD, 72, 2, 023516,
  \href{http://arXiv.org/abs/astro-ph/0503479}{{\tt astro-ph/0503479}}

\bibitem[{{Chang} et~al.(2015)}]{chang:2015}
{Chang} C., et~al., 2015, Physical Review Letters, 115, 5, 051301,
  \href{http://arXiv.org/abs/arXiv:1505.01871}{{\tt arXiv:1505.01871}}

\bibitem[{{Clowe} et~al.(2006){Clowe}, {Brada{\v c}}, {Gonzalez}, {Markevitch},
  {Randall}, {Jones} \& {Zaritsky}}]{clowe:2006}
{Clowe} D., {Brada{\v c}} M., {Gonzalez} A.H., {Markevitch} M., {Randall} S.W.,
  {Jones} C., {Zaritsky} D., 2006, ApJ L., 648, L109,
  \href{http://arXiv.org/abs/astro-ph/0608407}{{\tt astro-ph/0608407}}

\bibitem[{{Coles} \& {Chiang}(2000)}]{coles:2000}
{Coles} P., {Chiang} L.Y., 2000, Nature, 406, 376,
  \href{http://arXiv.org/abs/astro-ph/0006017}{{\tt astro-ph/0006017}}

\bibitem[{{Crittenden} et~al.(2001){Crittenden}, {Natarajan}, {Pen} \&
  {Theuns}}]{crittenden:2001}
{Crittenden} R.G., {Natarajan} P., {Pen} U.L., {Theuns} T., 2001, ApJ, 559,
  552, \href{http://arXiv.org/abs/astro-ph/0009052}{{\tt astro-ph/0009052}}

\bibitem[{{Crittenden} et~al.(2002){Crittenden}, {Natarajan}, {Pen} \&
  {Theuns}}]{crittenden:2002}
{Crittenden} R.G., {Natarajan} P., {Pen} U.L., {Theuns} T., 2002, ApJ, 568, 20,
  \href{http://arXiv.org/abs/astro-ph/0012336}{{\tt astro-ph/0012336}}

\bibitem[{{de Jong} et~al.(2013){de Jong}, {Verdoes Kleijn}, {Kuijken} \&
  {Valentijn}}]{de_jong:2013}
{de Jong} J.T.A., {Verdoes Kleijn} G.A., {Kuijken} K.H., {Valentijn} E.A.,
  2013, Experimental Astronomy, 35, 25,
  \href{http://arXiv.org/abs/arXiv:1206.1254}{{\tt arXiv:1206.1254}}

\bibitem[{{Flaugher} et~al.(2015)}]{flaugher:2015}
{Flaugher} B., et~al., 2015, AJ, 150, 150,
  \href{http://arXiv.org/abs/arXiv:1504.02900}{{\tt arXiv:1504.02900}}

\bibitem[{Goldberg et~al.(1967)Goldberg, Macfarlane, Newman, Rohrlich \&
  Sudarshan}]{goldberg:1967}
Goldberg J.N., Macfarlane A.J., Newman E.T., Rohrlich F., Sudarshan E.C.G.,
  1967, J.\ Math.\ Phys., 8, 11, 2155

\bibitem[{G\'{o}rski et~al.(2005)G\'{o}rski, Hivon, Banday, Wandelt, Hansen,
  Reinecke \& Bartelmann}]{gorski:2005}
G\'{o}rski K.M., Hivon E., Banday A.J., Wandelt B.D., Hansen F.K., Reinecke M.,
  Bartelmann M., 2005, ApJ, 622, 759,
  \href{http://arXiv.org/abs/astro-ph/0409513}{{\tt astro-ph/0409513}}

\bibitem[{{Green}(2011)}]{green:2011}
{Green} D.A., 2011, Bulletin of the Astronomical Society of India, 39, 289,
  \href{http://arXiv.org/abs/1108.5083}{{\tt 1108.5083}}

\bibitem[{{Heavens}(2009)}]{heavens:2009}
{Heavens} A., 2009, Nuclear Physics B Proceedings Supplements, 194, 76,
  \href{http://arXiv.org/abs/arXiv:0911.0350}{{\tt arXiv:0911.0350}}

\bibitem[{{Heymans} et~al.(2012)}]{heymans:2012}
{Heymans} C., et~al., 2012, MNRAS, 427, 146,
  \href{http://arXiv.org/abs/arXiv:1210.0032}{{\tt arXiv:1210.0032}}

\bibitem[{{Hirata} \& {Seljak}(2004)}]{hirata:2004}
{Hirata} C.M., {Seljak} U., 2004, PRD, 70, 6, 063526,
  \href{http://arXiv.org/abs/astro-ph/0406275}{{\tt astro-ph/0406275}}

\bibitem[{Hobson et~al.(1998)Hobson, Jones \& Lasenby}]{hobson:1999}
Hobson M., Jones A., Lasenby A., 1998, MNRAS, 309, 125,
  \href{http://arXiv.org/abs/astro-ph/9810200}{{\tt astro-ph/9810200}}

\bibitem[{{Jullo} et~al.(2007){Jullo}, {Kneib}, {Limousin},
  {El{\'{\i}}asd{\'o}ttir}, {Marshall} \& {Verdugo}}]{jullo:2007}
{Jullo} E., {Kneib} J.P., {Limousin} M., {El{\'{\i}}asd{\'o}ttir} {\'A}.,
  {Marshall} P.J., {Verdugo} T., 2007, New Journal of Physics, 9, 447,
  \href{http://arXiv.org/abs/arXiv:0706.0048}{{\tt arXiv:0706.0048}}

\bibitem[{{Jullo} et~al.(2014){Jullo}, {Pires}, {Jauzac} \&
  {Kneib}}]{jullo:2014}
{Jullo} E., {Pires} S., {Jauzac} M., {Kneib} J.P., 2014, MNRAS, 437, 3969,
  \href{http://arXiv.org/abs/1309.5718}{{\tt 1309.5718}}

\bibitem[{{Kaiser} \& {Squires}(1993)}]{kaiser:1993}
{Kaiser} N., {Squires} G., 1993, ApJ, 404, 441

\bibitem[{Kamionkowski et~al.(1997)Kamionkowski, Kosowsky \&
  Stebbins}]{kamionkowski:1996}
Kamionkowski M., Kosowsky A., Stebbins A., 1997, PRD, D55, 7368,
  \href{http://arXiv.org/abs/astro-ph/9611125}{{\tt astro-ph/9611125}}

\bibitem[{{Kaufman} et~al.(2016){Kaufman}, {Keating} \&
  {Johnson}}]{kaufman:2016}
{Kaufman} J.P., {Keating} B.G., {Johnson} B.R., 2016, MNRAS, 455, 1981,
  \href{http://arXiv.org/abs/arXiv:1409.8242}{{\tt arXiv:1409.8242}}

\bibitem[{{Kilbinger}(2015)}]{kilbinger:2015}
{Kilbinger} M., 2015, Reports on Progress in Physics, 78, 8, 086901,
  \href{http://arXiv.org/abs/arXiv:1411.0115}{{\tt arXiv:1411.0115}}

\bibitem[{{Kirk} et~al.(2015)}]{kirk:2015}
{Kirk} D., et~al., 2015, Space Science Reviews, 193, 139,
  \href{http://arXiv.org/abs/1504.05465}{{\tt 1504.05465}}

\bibitem[{{Kitching} et~al.(2016){Kitching}, {Alsing}, {Heavens}, {Jimenez},
  {McEwen} \& {Verde}}]{kitching:2016}
{Kitching} T.D., {Alsing} J., {Heavens} A.F., {Jimenez} R., {McEwen} J.D.,
  {Verde} L., 2016, ArXiv e-prints,
  \href{http://arXiv.org/abs/arXiv:1611.04954}{{\tt arXiv:1611.04954}}

\bibitem[{{Kratochvil} et~al.(2012){Kratochvil}, {Lim}, {Wang}, {Haiman}, {May}
  \& {Huffenberger}}]{kratochvil:2012}
{Kratochvil} J.M., {Lim} E.A., {Wang} S., {Haiman} Z., {May} M., {Huffenberger}
  K., 2012, PRD, 85, 10, 103513,
  \href{http://arXiv.org/abs/arXiv:1109.6334}{{\tt arXiv:1109.6334}}

\bibitem[{{Lanusse} et~al.(2016){Lanusse}, {Starck}, {Leonard} \&
  {Pires}}]{lanusse:2016}
{Lanusse} F., {Starck} J.L., {Leonard} A., {Pires} S., 2016, A\&A, 591, A2,
  \href{http://arXiv.org/abs/arXiv:1603.01599}{{\tt arXiv:1603.01599}}

\bibitem[{{Laureijs} et~al.(2011)}]{laureijs:2011}
{Laureijs} R., et~al., 2011, ArXiv e-prints,
  \href{http://arXiv.org/abs/arXiv:1110.3193}{{\tt arXiv:1110.3193}}

\bibitem[{Leistedt et~al.(2017)Leistedt, McEwen, B{\"u}ttner \&
  Peiris}]{leistedt:ebsep}
Leistedt B., McEwen J.D., B{\"u}ttner M., Peiris H.V., 2017, MNRAS, 466, 3,
  3728, \href{http://arXiv.org/abs/arXiv:1605.01414}{{\tt arXiv:1605.01414}}

\bibitem[{Leistedt et~al.(2013)Leistedt, McEwen, Vandergheynst \&
  Wiaux}]{leistedt:s2let_axisym}
Leistedt B., McEwen J.D., Vandergheynst P., Wiaux Y., 2013, A\&A, 558, A128, 1,
  \href{http://arXiv.org/abs/arXiv:1211.1680}{{\tt arXiv:1211.1680}}

\bibitem[{{Leonard} et~al.(2012){Leonard}, {Dup{\'e}} \&
  {Starck}}]{leonard:2012}
{Leonard} A., {Dup{\'e}} F.X., {Starck} J.L., 2012, A\&A, 539, A85,
  \href{http://arXiv.org/abs/arXiv:1111.6478}{{\tt arXiv:1111.6478}}

\bibitem[{{Leonard} et~al.(2014){Leonard}, {Lanusse} \&
  {Starck}}]{leonard:2014}
{Leonard} A., {Lanusse} F., {Starck} J.L., 2014, MNRAS, 440, 1281,
  \href{http://arXiv.org/abs/arXiv:1308.1353}{{\tt arXiv:1308.1353}}

\bibitem[{{Lin} \& {Kilbinger}(2015{\natexlab{a}})}]{lin:2015a}
{Lin} C.A., {Kilbinger} M., 2015{\natexlab{a}}, A\&A, 576, A24,
  \href{http://arXiv.org/abs/arXiv:1410.6955}{{\tt arXiv:1410.6955}}

\bibitem[{{Lin} \& {Kilbinger}(2015{\natexlab{b}})}]{lin:2015b}
{Lin} C.A., {Kilbinger} M., 2015{\natexlab{b}}, A\&A, 583, A70,
  \href{http://arXiv.org/abs/arXiv:1506.01076}{{\tt arXiv:1506.01076}}

\bibitem[{{Lin} et~al.(2016){Lin}, {Kilbinger} \& {Pires}}]{lin:2016}
{Lin} C.A., {Kilbinger} M., {Pires} S., 2016, A\&A, 593, A88,
  \href{http://arXiv.org/abs/arXiv:1603.06773}{{\tt arXiv:1603.06773}}

\bibitem[{{Liu} \& {Hill}(2015)}]{liu:2015}
{Liu} J., {Hill} J.C., 2015, PRD, 92, 6, 063517,
  \href{http://arXiv.org/abs/arXiv:1504.05598}{{\tt arXiv:1504.05598}}

\bibitem[{{LSST Science Collaboration} et~al.(2009)}]{lsst:2009}
{LSST Science Collaboration}, et~al., 2009, ArXiv e-prints,
  \href{http://arXiv.org/abs/arXiv:0912.0201}{{\tt arXiv:0912.0201}}

\bibitem[{{Marinucci} \& {Peccati}(2011)}]{marinucci:2011:book}
{Marinucci} D., {Peccati} G., 2011, \emph{Random Fields on the Sphere:
  Representation, Limit Theorem and Cosmological Applications}, Cambridge
  University Press

\bibitem[{{Massey} et~al.(2004)}]{massey:2004}
{Massey} R., et~al., 2004, AJ, 127, 3089,
  \href{http://arXiv.org/abs/astro-ph/0304418}{{\tt astro-ph/0304418}}

\bibitem[{{Massey} et~al.(2007)}]{massey:2007}
{Massey} R., et~al., 2007, Nature, 445, 286,
  \href{http://arXiv.org/abs/astro-ph/0701594}{{\tt astro-ph/0701594}}

\bibitem[{{Massey} et~al.(2015)}]{massey:2015}
{Massey} R., et~al., 2015, MNRAS, 449, 3393,
  \href{http://arXiv.org/abs/arXiv:1504.03388}{{\tt arXiv:1504.03388}}

\bibitem[{McEwen et~al.(2005)McEwen, Hobson, Lasenby \&
  Mortlock}]{mcewen:2005:ng}
McEwen J.D., Hobson M.P., Lasenby A.N., Mortlock D.J., 2005, MNRAS, 359, 1583,
  \href{http://arXiv.org/abs/astro-ph/0406604}{{\tt astro-ph/0406604}}

\bibitem[{McEwen et~al.(2015)McEwen, Leistedt, B{\"u}ttner, Peiris \&
  Wiaux}]{mcewen:s2let_spin}
McEwen J.D., Leistedt B., B{\"u}ttner M., Peiris H.V., Wiaux Y., 2015, IEEE
  TSP, submitted, \href{http://arXiv.org/abs/arXiv:1509.06749}{{\tt
  arXiv:1509.06749}}

\bibitem[{McEwen \& Wiaux(2011)}]{mcewen:fssht}
McEwen J.D., Wiaux Y., 2011, IEEE TSP, 59, 12, 5876,
  \href{http://arXiv.org/abs/arXiv:1110.6298}{{\tt arXiv:1110.6298}}

\bibitem[{{Mediavilla} et~al.(2016){Mediavilla}, {Mu{\~n}oz}, {Garz{\'o}n} \&
  {Mahoney}}]{book:Mediavilla:2016}
{Mediavilla} E., {Mu{\~n}oz} J.A., {Garz{\'o}n} F., {Mahoney} T.J., 2016,
  \emph{{Astrophysical Applications of Gravitational Lensing}}, {Cambridge
  University Press}

\bibitem[{{Munshi} et~al.(2011){Munshi}, {Kitching}, {Heavens} \&
  {Coles}}]{munshi:2011}
{Munshi} D., {Kitching} T., {Heavens} A., {Coles} P., 2011, MNRAS, 416, 1629,
  \href{http://arXiv.org/abs/arXiv:1012.3658}{{\tt arXiv:1012.3658}}

\bibitem[{{Munshi} et~al.(2008){Munshi}, {Valageas}, {van Waerbeke} \&
  {Heavens}}]{munshi:2008}
{Munshi} D., {Valageas} P., {van Waerbeke} L., {Heavens} A., 2008, Phys.\ Rep,
  462, 67, \href{http://arXiv.org/abs/astro-ph/0612667}{{\tt astro-ph/0612667}}

\bibitem[{{Munshi} et~al.(2012){Munshi}, {van Waerbeke}, {Smidt} \&
  {Coles}}]{munshi:2012}
{Munshi} D., {van Waerbeke} L., {Smidt} J., {Coles} P., 2012, MNRAS, 419, 536,
  \href{http://arXiv.org/abs/arXiv:1103.1876}{{\tt arXiv:1103.1876}}

\bibitem[{Newman \& Penrose(1966)}]{newman:1966}
Newman E.T., Penrose R., 1966, J.\ Math.\ Phys., 7, 5, 863

\bibitem[{{Peel} et~al.(2016){Peel}, {Lin}, {Lanusse}, {Leonard}, {Starck} \&
  {Kilbinger}}]{peel:2016}
{Peel} A., {Lin} C.A., {Lanusse} F., {Leonard} A., {Starck} J.L., {Kilbinger}
  M., 2016, ArXiv e-prints, \href{http://arXiv.org/abs/arXiv:1612.02264}{{\tt
  arXiv:1612.02264}}

\bibitem[{{Petri} et~al.(2013){Petri}, {Haiman}, {Hui}, {May} \&
  {Kratochvil}}]{petri:2013}
{Petri} A., {Haiman} Z., {Hui} L., {May} M., {Kratochvil} J.M., 2013, PRD, 88,
  12, 123002, \href{http://arXiv.org/abs/arXiv:1309.4460}{{\tt
  arXiv:1309.4460}}

\bibitem[{{Pichon} et~al.(2010){Pichon}, {Thi{\'e}baut}, {Prunet}, {Benabed},
  {Colombi}, {Sousbie} \& {Teyssier}}]{pichon:2010}
{Pichon} C., {Thi{\'e}baut} E., {Prunet} S., {Benabed} K., {Colombi} S.,
  {Sousbie} T., {Teyssier} R., 2010, MNRAS, 401, 705,
  \href{http://arXiv.org/abs/arXiv:0901.2001}{{\tt arXiv:0901.2001}}

\bibitem[{{Plaszczynski} et~al.(2012){Plaszczynski}, {Lavabre}, {Perotto} \&
  {Starck}}]{plaszczynski:2012}
{Plaszczynski} S., {Lavabre} A., {Perotto} L., {Starck} J.L., 2012, A\&A, 544,
  A27, \href{http://arXiv.org/abs/1201.5779}{{\tt 1201.5779}}

\bibitem[{{Price} et~al.(2020){Price}, {Cai}, {McEwen}, {Pereyra}, {Kitching}
  \& {LSST Dark Energy Science Collaboration}}]{price:2020}
{Price} M.A., {Cai} X., {McEwen} J.D., {Pereyra} M., {Kitching} T.D., {LSST
  Dark Energy Science Collaboration}, 2020, MNRAS, 492, 1, 394,
  \href{http://arXiv.org/abs/1812.04017}{{\tt 1812.04017}}

\bibitem[{Price et~al.(2021)Price, McEwen, Cai, Kitching \&
  Wallis}]{price:2018}
Price M.A., McEwen J.D., Cai X., Kitching T.D., Wallis C.G.R., 2021, MNRAS,
  ISSN 0035-8711, stab1983

\bibitem[{{Price} et~al.(2021){Price}, {McEwen}, {Pratley} \&
  {Kitching}}]{price:2021}
{Price} M.A., {McEwen} J.D., {Pratley} L., {Kitching} T.D., 2021, MNRAS, 500,
  4, 5436, \href{http://arXiv.org/abs/2004.07855}{{\tt 2004.07855}}

\bibitem[{{Schneider}(2005)}]{schneider:2005}
{Schneider} P., 2005, ArXiv Astrophysics e-prints,
  \href{http://arXiv.org/abs/astro-ph/0509252}{{\tt astro-ph/0509252}}

\bibitem[{{Scoville} et~al.(2007)}]{scoville:2007}
{Scoville} N., et~al., 2007, ApJ S., 172, 1,
  \href{http://arXiv.org/abs/astro-ph/0612305}{{\tt astro-ph/0612305}}

\bibitem[{{Seitz} \& {Schneider}(1995)}]{seitz:1995}
{Seitz} C., {Schneider} P., 1995, A\&A, 297, 287,
  \href{http://arXiv.org/abs/astro-ph/9408050}{{\tt astro-ph/9408050}}

\bibitem[{{Simon}(2013)}]{simon:2013}
{Simon} P., 2013, A\&A, 560, A33,
  \href{http://arXiv.org/abs/arXiv:1203.6205}{{\tt arXiv:1203.6205}}

\bibitem[{{Simon} et~al.(2009){Simon}, {Taylor} \& {Hartlap}}]{simon:2009}
{Simon} P., {Taylor} A.N., {Hartlap} J., 2009, MNRAS, 399, 48,
  \href{http://arXiv.org/abs/arXiv:0907.0016}{{\tt arXiv:0907.0016}}

\bibitem[{{Spergel} et~al.(2015)}]{spergel:2015}
{Spergel} D., et~al., 2015, ArXiv e-prints,
  \href{http://arXiv.org/abs/arXiv:1503.03757}{{\tt arXiv:1503.03757}}

\bibitem[{{Szepietowski} et~al.(2014){Szepietowski}, {Bacon}, {Dietrich},
  {Busha}, {Wechsler} \& {Melchior}}]{szepietowski:2014}
{Szepietowski} R.M., {Bacon} D.J., {Dietrich} J.P., {Busha} M., {Wechsler} R.,
  {Melchior} P., 2014, MNRAS, 440, 2191,
  \href{http://arXiv.org/abs/arXiv:1306.5324}{{\tt arXiv:1306.5324}}

\bibitem[{{Taylor}(2001)}]{taylor:2001}
{Taylor} A.N., 2001, ArXiv, \href{http://arXiv.org/abs/astro-ph/0111605}{{\tt
  astro-ph/0111605}}

\bibitem[{{Taylor} et~al.(2004)}]{taylor:2004}
{Taylor} A.N., et~al., 2004, MNRAS, 353, 1176,
  \href{http://arXiv.org/abs/astro-ph/0402095}{{\tt astro-ph/0402095}}

\bibitem[{{Van Waerbeke} et~al.(2013)}]{van_waerbeke:2013}
{Van Waerbeke} L., et~al., 2013, MNRAS, 433, 3373,
  \href{http://arXiv.org/abs/arXiv:1303.1806}{{\tt arXiv:1303.1806}}

\bibitem[{{VanderPlas} et~al.(2011){VanderPlas}, {Connolly}, {Jain} \&
  {Jarvis}}]{vanderplas:2011}
{VanderPlas} J.T., {Connolly} A.J., {Jain} B., {Jarvis} M., 2011, ApJ, 727,
  118, \href{http://arXiv.org/abs/arXiv:1008.2396}{{\tt arXiv:1008.2396}}

\bibitem[{Varshalovich et~al.(1989)Varshalovich, Moskalev \&
  Khersonskii}]{varshalovich:1989}
Varshalovich D.A., Moskalev A.N., Khersonskii V.K., 1989, \emph{Quantum theory
  of angular momentum}, World Scientific, Singapore

\bibitem[{Vielva et~al.(2004)Vielva, Mart\'{\i}nez-Gonz\'{a}lez, Barreiro, Sanz
  \& Cay\'{o}n}]{vielva:2004}
Vielva P., Mart\'{\i}nez-Gonz\'{a}lez E., Barreiro R.B., Sanz J.L., Cay\'{o}n
  L., 2004, ApJ, 609, 22, \href{http://arXiv.org/abs/astro-ph/0310273}{{\tt
  astro-ph/0310273}}

\bibitem[{{Vikram} et~al.(2015)}]{vikram:2015}
{Vikram} V., et~al., 2015, PRD, 92, 2, 022006,
  \href{http://arXiv.org/abs/arXiv:1504.03002}{{\tt arXiv:1504.03002}}

\bibitem[{Wallis et~al.(2016)Wallis, Wiaux \& McEwen}]{wallis:s2_recon}
Wallis C.G.R., Wiaux Y., McEwen J.D., 2016, IEEE TIP, submitted,
  \href{http://arXiv.org/abs/arXiv:1608.00553}{{\tt arXiv:1608.00553}}

\bibitem[{Zaldarriaga \& Seljak(1997)}]{zaldarriaga:1997}
Zaldarriaga M., Seljak U., 1997, PRD, 55, 4, 1830,
  \href{http://arXiv.org/abs/astro-ph/9609170}{{\tt astro-ph/9609170}}

\bibitem[{{Zuntz} et~al.(2015)}]{zuntz:2015}
{Zuntz} J., et~al., 2015, Astronomy and Computing, 12, 45,
  \href{http://arXiv.org/abs/arXiv:1409.3409}{{\tt arXiv:1409.3409}}

\end{thebibliography}

\label{lastpage}
\end{document}